%% file: Jpsi_SSA_SIDIS_arxiv.tex
\documentclass[prd,aps,showpacs,notitlepage,superscriptaddress,nofootinbib]{revtex4-1}

\usepackage{amsmath}
\usepackage{multirow}
\usepackage{float}
\usepackage{soul}
\usepackage{caption}
\usepackage {tikz}
\usepackage{gnuplot-lua-tikz}

\def\bea#1\eea{\begin{align}#1\end{align}}

\makeatletter
\def\slash#1{{\mathpalette\c@ncel{#1}}} % TeXbook, bottom of p360
\makeatother

\newcommand\beq{\begin{eqnarray}}
\newcommand\eeq{\end{eqnarray}}

\newcommand\la{\langle}
\newcommand\ra{\rangle}

\begin{document}
\title{The twist-3 gluon contribution to Sivers asymmetry in $J/\psi$ production in semi-inclusive deep inelastic scattering}

\date{\today}

\author{Longjie Chen}
\email{chenlongjieusc@163.com}
\affiliation{Key Laboratory of Atomic and Subatomic Structure and Quantum Control (MOE), Institute of Quantum Matter, South China Normal University, Guangzhou 510006, China} 
\affiliation{Guangdong Provincial Key Laboratory of Nuclear Science, Institute of Quantum Matter, South
China Normal University, Guangzhou 510006, China} 
\affiliation{Guangdong-Hong Kong Joint Laboratory of Quantum Matter, Southern Nuclear Science Computing Center, South China Normal University, Guangzhou 510006, China}

\author{Hongxi Xing}
\email{hxing@m.scnu.edu.cn}
\affiliation{Key Laboratory of Atomic and Subatomic Structure and Quantum Control (MOE), Institute of Quantum Matter, South China Normal University, Guangzhou 510006, China} 
\affiliation{Guangdong Provincial Key Laboratory of Nuclear Science, Institute of Quantum Matter, South
China Normal University, Guangzhou 510006, China}
\affiliation{Guangdong-Hong Kong Joint Laboratory of Quantum Matter, Southern Nuclear Science Computing Center, South China Normal University, Guangzhou 510006, China}

\author{Shinsuke Yoshida}
\email{shinyoshida85@gmail.com}
\affiliation{Key Laboratory of Atomic and Subatomic Structure and Quantum Control (MOE), Institute of Quantum Matter, South China Normal University, Guangzhou 510006, China} 
\affiliation{Guangdong Provincial Key Laboratory of Nuclear Science, Institute of Quantum Matter, South
China Normal University, Guangzhou 510006, China}
\affiliation{Guangdong-Hong Kong Joint Laboratory of Quantum Matter, Southern Nuclear Science Computing Center, South China Normal University, Guangzhou 510006, China}

\begin{abstract}

We carry out the first calculation for the twist-3 gluon contribution to the single 
transverse-spin asymmetry(SSA) in $J/\psi$ production in semi-inclusive deep inelastic scattering. 
Our result shows that the $J/\psi$ SSA is an ideal observable to pin down the $C$-even type twist-3 
gluon distribution that has a direct relationship with 
the gluon transverse-momentum-dependent distribution function. We also perform some numerical simulations of the $J/\psi$ SSA
for the kinematics accessible at the future electron-ion-collider experiment. 
For color-singlet contribution,
the hadronization effect of $J/\psi$ is completely canceled at the level of the SSA and the 
spin-dependent structure functions directly reflect the behavior of the $C$-even twist-3 gluon 
distribution.

\end{abstract}

\maketitle

%----------------------- Section 1: Introduction ----------------------

\section{Introduction}

The investigation of the nucleon internal structure in high energy scatterings has been one of
the central subjects in basic science since quantum chromodynamics(QCD) 
was established as a fundamental theory of the strong interaction. A lot of knowledge have been 
accumulated in the past half century through the perturbative QCD analysis of experimental data.
However, in spite of tremendous theoretical and experimental effort,  a lot of mysteries still lie in the 
nucleon structure. In particular, the role of gluons inside the nucleon is leaving a lot of room for 
research.  The investigation of the gluon spin structure is one of the main subjects of the experiment at
Relativistic Heavy Ion Collider(RHIC) that was launched in 2000. The evaluation of the gluon spin contribution
to the proton spin is a major progress that was made 
in the past couple of decades\cite{PHENIX:2008swq,deFlorian:2014yva}. Further investigation 
will be inherited by the next-generation collider experiment, 
Electron Ion Collider(EIC)\cite{Accardi:2012qut}.
The EIC experiment aims to understand deeper gluon structure like the 3-dimensional orbital motion of 
gluons inside the proton. The importance of the orbital motions of the partons was realized by the emergence
of the large single transverse-spin asymmetry(SSA) in high energy hadron scatterings. The large SSA 
was first observed in the late 70s\cite{Klem:1976ui,Bunce:1976yb} and it turned 
out that the conventional parton picture could not 
describe it at all. The observation of the large SSA motivated the improvement of the conventional
perturbative QCD framework. The transverse-momentum-dependent(TMD) factorization is known as
one of the successful frameworks in describing existing data of the SSA. The nonperturbative functions
in the TMD factorization represents the 3-dimensional motion of the partons and the success of 
this framework made us realize the importance of the orbital motions of the partons. 
Another successful framework is the twist-3 contributions in the collinear factorization. 
This framework describes incoherent multi-parton scattering
in a hard process and this is successful in describing SSAs in single-scale processes like 
$pp\to \pi X$ measured at RHIC. Although these two frameworks basically have different applicable 
conditions, it was found that there is a marginal region where both frameworks are valid 
in some processes\cite{Ji:2006br,Ji:2006vf,Koike:2007dg,Bacchetta:2008xw,Yuan:2009dw,Zhou:2009jm}. 
The equivalence of two frameworks is important for giving a unified picture 
to the origin of the SSA.

Heavy flavored hadron productions are important in the context of the investigation of the gluon structure
because the heavy quark fragmenting into a final state hadron is mainly produced
by a fusion of gluon inside the proton\cite{Qiu:2020xum}. 
The SSA in the heavy flavored hadron production
is one of ideal observables to investigate the orbital motions of gluons. Those SSAs have been well discussed
based on the TMD factorization for $D$-meson 
production\cite{Anselmino:2004nk,Godbole:2016tvq,DAlesio:2017rzj,Godbole:2017fab,DAlesio:2018rnv} 
and $J/\psi$ production\cite{Godbole:2014tha,Mukherjee:2016qxa,DAlesio:2017rzj,Rajesh:2018qks,DAlesio:2018rnv,Sun:2019tuk,DAlesio:2019qpk,Kishore:2019fzb,DAlesio:2019gnu,DAlesio:2020eqo}.
On the other hand, the twist-3 gluon contribution to the SSA
has been calculated only for $D$-meson 
production\cite{Kang:2008qh,Kang:2008ih,Beppu:2010qn,Koike:2011mb,Beppu:2012vi}.
In this paper, we carry out the first calculation for the twist-3 gluon contribution to the 
SSA in $J/\psi$ production in semi-inclusive deep inelastic scattering(SIDIS).
This is required to deal with the data of the $J/\psi$ SSA in a full kinematic range of 
the EIC experiment together with the TMD 
framework and give a unified interpretation to the data. We will also perform some numerical simulations 
for the $J/\psi$ SSA. Our result clarifies the role of the $J/\psi$ SSA in the determination of the twist-3 
gluon distribution function that could give indirect information about the orbital motion of the gluons
inside the proton.

The remainder of this paper is organized as follows: In section II, we introduce definitions of the twist-3 gluon distribution functions relevant to our study and show some relations 
among them. In section III, we introduce the frame we will work on and show our derivation 
of the twist-3 cross section formula in detail. 
In section IV, we show some numerical simulations with simple models for the normalized 
structure functions that are accessible at the EIC. Section V is devoted to a summary of our study.

%----------------------- Section 2: Definitions of twist-3 functions ----------------------

\section{Definitions of twist-3 gluon distribution functions}

In this section, we recall the definitions of the twist-3 gluon distribution functions relevant to our study.
Two types of the twist-3 functions, the kinematical functions and the dynamical functions,
in general contribute to the SSA.
The kinematical functions of the transversely polarized proton 
can be expressed by the first $k^2_T/M_N^2$-moment of the gluon Sivers function
\cite{Mulders:2000sh,Koike:2019zxc}.
We show their definitions below.
\beq
\Phi^{\alpha\beta\gamma}_{\partial}(x)&=&\int{d\lambda\over 2\pi}e^{i\lambda x}
\la pS_{\perp}|F^{\beta n}
(0)F^{\alpha n}(\lambda n)|pS_{\perp}\ra(i\overleftarrow{\partial}^{\gamma}_{\perp})
\nonumber\\
&\equiv&\lim_{\xi_{\perp}\to 0}\int{d\lambda\over 2\pi}e^{i\lambda x}\la pS_{\perp}|
\Bigl(F^{\beta n}(0)[0,\infty n]\Bigr)_ai{d\over d\xi_{\perp\gamma}}
\Bigl([\infty n+\xi_{\perp},\lambda n+\xi_{\perp}]F^{\alpha n}(\lambda n+\xi_{\perp})\Bigr)_a|pS_{\perp}\ra
\nonumber\\
&=&{M_N\over 2}g^{\alpha\beta}_{\perp}\epsilon^{pnS_{\perp}\gamma}G^{(1)}_T(x)
+i{M_N\over 2}\epsilon^{pn \alpha\beta}S_{\perp}^{\gamma}\Delta G^{(1)}_T(x)
+{M_N\over 8}\Bigl(\epsilon^{pnS_{\perp}\{\alpha}g^{\beta\}\gamma}_{\perp}
+\epsilon^{pn\gamma\{\alpha}S^{\beta\}}_{\perp}\Bigr)\Delta H^{(1)}_{T}(x)+\cdots,\hspace{2mm}
\label{kinematical}
\eeq
where $p, S_{\perp}$ and $M_N$ are the proton's momentum, spin and mass respectively. 
We used the simplified notation 
$\epsilon^{pnS_{\perp}\gamma}=\epsilon^{\mu\nu\rho\gamma}p_{\mu}n_{\nu}S_{\perp\rho}$. 
$[0,\lambda n]$ denotes the gauge-link operator,
\beq
[0,\lambda n]\equiv {\rm P}\exp\Bigl(ig\int_{\lambda}^0dt\, A^n(tn)\Big),
\eeq
that guarantees the gauge-invariance of the matrix element.
$n$ is a light-like vector that satisfies $p\cdot n=1, n^2=0$.  
A naively $T$-odd observable like the SSA receives the contributions from 
$G^{(1)}_T(x)$ and $\Delta H^{(1)}_{T}(x)$.

The dynamical gluon distribution functions are defined by a matrix element composed
of three gluon field strength tensors\cite{Ji:1992eu,Beppu:2010qn}.
The dynamical functions are categorized into two types, $C$-even function $N(x_1,x_2)$ and 
$C$-odd function $O(x_1,x_2)$, reflecting the fact that there are two structure constants $if^{abc}$
and $d^{abc}$ in SU($N_c$) group,
\beq
N^{\alpha\beta\gamma}(x_1,x_2)&=&i\int{d\lambda\over 2\pi}\int{d\mu\over 2\pi}
e^{i\lambda x_1}e^{i\mu (x_2-x_1)}
\la pS_{\perp}|if^{bca}F_{b}^{\beta n}(0)
gF_c^{\gamma n}(\mu n)F_a^{\alpha n}(\lambda n)|pS_{\perp}\ra
\nonumber\\
&=&2iM_N\Bigl[g_{\perp}^{\alpha\beta}\epsilon^{\gamma pnS_{\perp}}N(x_1,x_2)
-g_{\perp}^{\beta\gamma}\epsilon^{\alpha pnS_{\perp}}N(x_2,x_2-x_1)
-g_{\perp}^{\alpha\gamma}\epsilon^{\beta pnS_{\perp}}N(x_1,x_1-x_2)\Bigr]+\cdots,
\label{C-even}
\eeq
\beq
O^{\alpha\beta\gamma}(x_1,x_2)&=&i\int{d\lambda\over 2\pi}\int{d\mu\over 2\pi}
e^{i\lambda x_1}e^{i\mu (x_2-x_1)}
\la pS_{\perp}|d^{bca}F_{b}^{\beta n}(0)
gF_c^{\gamma n}(\mu n)F_a^{\alpha n}(\lambda n)|pS_{\perp}\ra
\nonumber\\
&=&2iM_N\Bigl[g_{\perp}^{\alpha\beta}\epsilon^{\gamma pnS_{\perp}}O(x_1,x_2)
+g_{\perp}^{\beta\gamma}\epsilon^{\alpha pnS_{\perp}}O(x_2,x_2-x_1)
+g_{\perp}^{\alpha\gamma}\epsilon^{\beta pnS_{\perp}}O(x_1,x_1-x_2)\Bigr]+\cdots,
\label{C-odd}
\eeq
where we omitted gauge-links for simplicity. These dynamical functions have the following symmetries.
\beq
O(x_1,x_2)&=&O(x_2,x_1),\hspace{5mm}O(x_1,x_2)=O(-x_1,-x_2),
\nonumber\\
N(x_1,x_2)&=&N(x_2,x_1),\hspace{5mm}N(x_1,x_2)=-N(-x_1,-x_2).
\label{symmetries}
\eeq
The kinematical functions and the $C$-even function $N(x_1,x_2)$ are not independent of each other.
The following relations were derived in \cite{Koike:2019zxc}. 
\beq
G_T^{(1)}(x)=-4\pi(N(x,x)-N(x,0)),\hspace{5mm}\Delta H_T^{(1)}(x)=8\pi N(x,0).
\label{relations}
\eeq
They show the relationships between the first $k^2_T/M_N^2$-moment of the gluon 
Sivers function and the $C$-even twist-3 gluon distribution function in analogy with the relationship between
the twist-3 Qiu-Sterman function and the quark Sivers function\cite{Boer:2003cm,Kang:2011hk}.

%----------------------- Section 3: Calculation of the SSA in $J/\psi$ production ----------------------

\section{Calculation of the SSA in $J/\psi$ production in SIDIS}

\subsection{Unpolarized cross section for $J/\psi$ production in SIDIS}

%%%%%%%%%%%%%%%%%%%%%%%%%%
\begin{figure}[h]
\begin{center}
  \includegraphics[height=6cm,width=12cm]{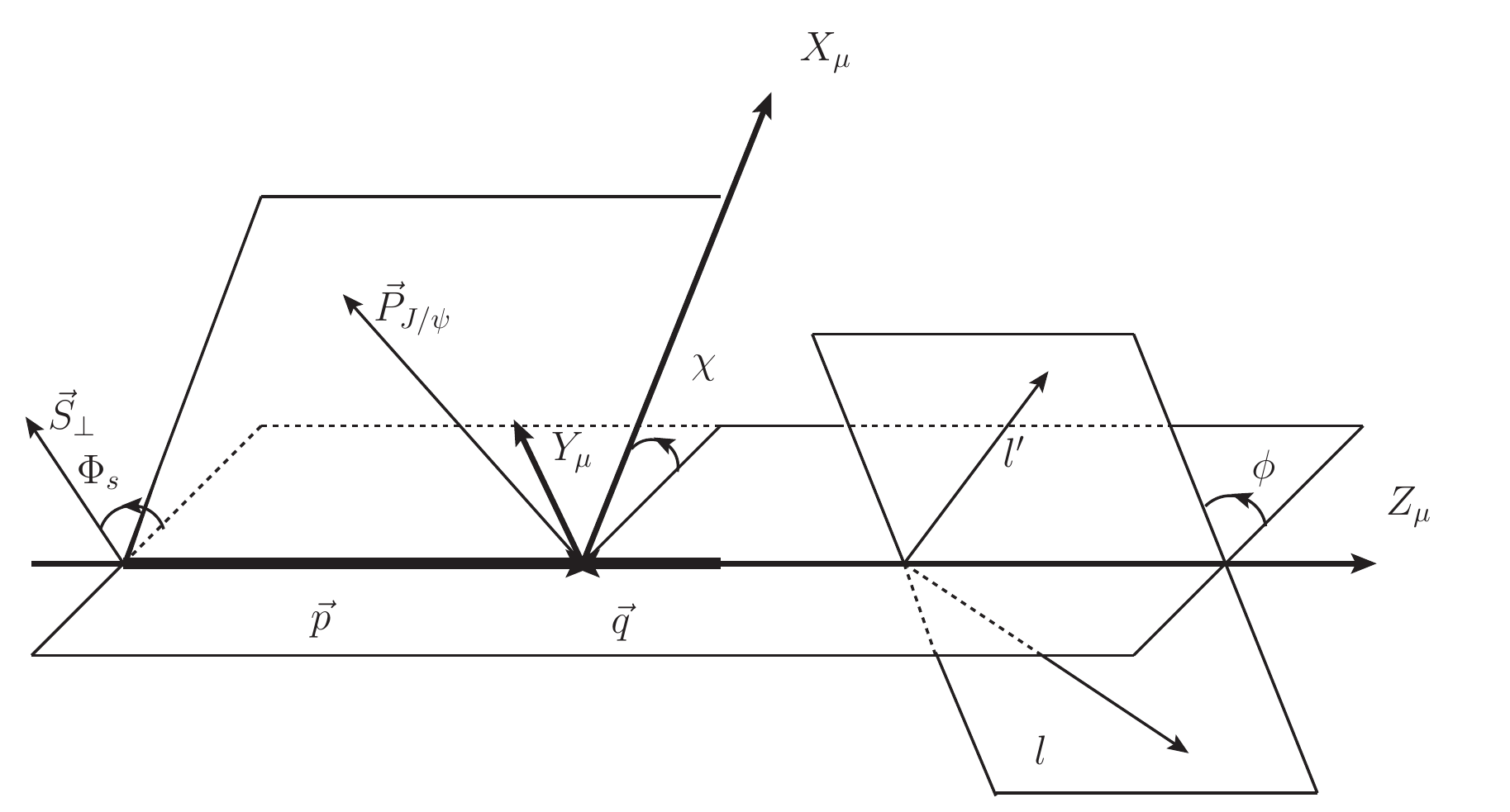}
\end{center}
 \caption{Schematic illustration of the scattering in the hadron frame.}
\label{frame}
\end{figure}
%%%%%%%%%%%%%%%%%%%%%%%%%%%

We calculate the SSA in $J/\psi$ production in SIDIS,
\beq
e(\ell)+p^{\uparrow}(p,S_{\perp})\to e(\ell')+J/\psi(P_{J/\psi})+X,
\eeq
in the hadron frame\cite{Meng:1991da}. 
It is convenient to use the following Lorentz invariant variables to express a cross section
formula in SIDIS.
\beq
S_{ep}&=&(p+\ell)^2,\hspace{5mm}
Q^2=-q^2=-(\ell-\ell')^2,\hspace{5mm}
x_{B}=\frac{Q^2}{2p\cdot q},\hspace{5mm}
z_f=\frac{p\cdot P_{J/\psi}}{p\cdot q}.
\eeq
All momenta and the spin vector of the polarized proton are given in this frame as
\beq
p&=&\Bigl({Q\over 2x_{B}},0,0,{Q\over 2x_{B}}\Bigr),\hspace{5mm}
q=\Bigl(0,0,0,-Q\Bigr),
\nonumber\\
P_{J/\psi}&=&{z_fQ\over 2}\Bigl(1+{P_T^2\over Q^2}+{m^2_{J/\psi}\over z_f^2Q^2},
{2P_T\over Q}\cos\chi
,{2P_T\over Q}\sin\chi,-1+{P_T^2\over Q^2}+{m^2_{J/\psi}\over z_f^2Q^2}\Bigr),
\nonumber\\
S_{\perp}&=&(0,\cos\Phi_S,\sin\Phi_S,0),
\eeq
where ${P_T}=|P_{J/\psi}^{\perp}|/z_f$ and $m_{J/\psi}$ is the mass of $J/\psi$. 
Using these variables,  the cross section formula is given by
\beq
\frac{d^{6}\sigma}{dx_{B}dQ^{2}dz_{f}dP_T^2d\phi d\chi}  =
\frac{\alpha ^{2}_{em}}{128\pi^4 S^{2}_{ep} x_{B}^2 Q^{2}}z_f
L_{\mu\nu}(\ell,\ell')W^{\mu\nu}(p,q,P_{J/\psi}), 
\eeq
where $\alpha_{em}=e^2/4\pi$ is the QED coupling constant, 
$\Phi_S$, $\chi$ and $\phi$ are the azimuthal angles of the proton's spin, the hadron plane and 
the lepton plane respectively as shown in FIG. \ref{frame}, the leptonic tensor is given by 
$L_{\mu\nu}(\ell,\ell')=2(\ell_\mu \ell'_\nu+\ell_\nu \ell'_{\mu})
-Q^2g^{\mu\nu}$.
The hadronic tensor $W_{\mu\nu}(p,q,P_{J/\psi})$ describes the scattering between the virtual photon and
the proton and the hadronization process of the charm quark pair into $J/\psi$.
We adopt non-relativistic QCD(NRQCD) framework\cite{Caswell:1985ui,Bodwin:1994jh} 
for the description of the hadronization mechanism 
of $J/\psi$. Within NRQCD, the $J/\psi$ production is illustrated as
\beq
e(\ell)+p(p)\to e(\ell')+\sum_nc\bar{c}[n](P_{J/\psi})+X,
\eeq 
where $n={}^3S_1^{[1]},{}^1S_0^{[8]},{}^3S_1^{[8]},\cdots$ 
denotes possible Fock states of the charm quark pair hadronizing into $J/\psi$. 
In this paper,  we focus on the color singlet contribution ${}^3S_1^{[1]}$ 
as the first attempt to calculate the twist-3 gluon distribution effect on the $J/\psi$ SSA.
The color-singlet hadronization gives the following structure in the spinor space.
\beq
{\cal N}\la{\cal O}^{J/\psi}({}^3S_1^{[1]})\ra
\slash{\epsilon}({\slash{P}_{J/\psi}}+m_{J/\psi}),\hspace{5mm}
\sum\epsilon^{\rho}\epsilon^{*\sigma}=-g^{\rho\sigma}
+{P_{J/\psi}^{\rho}P_{J/\psi}^{\sigma}\over m_{J/\psi}^2},
\eeq
where $\la{\cal O}^{J/\psi}({}^3S_1^{[1]})\ra$
is the long distance matrix element(LDME) that represents 
the hadronization effect of the charm quark pair with the quantum state ${}^3S_1^{[1]}$
into $J/\psi$.
We don't need the explicit form of the normalization factor ${\cal N}$ in our study
because it is completely canceled
at the level of the SSA that is given by the ratio of two cross sections.
For the unpolarized cross section, the hadronic tensor is given by
\beq
W^{\mu\nu}(p,q,P_{J/\psi})=\int^1_{0}{dx\over x}\,G(x)\,\la{\cal O}^{J/\psi}({}^3S_1^{[1]})\ra
\,w^{\mu\nu}(xp,q,P_{J/\psi}),
\eeq
where $G(x)$ is the unpolarized  gluon distribution function and $w^{\mu\nu}(xp,q,P_{J/\psi})$ represents the hard scattering between 
the virtual photon and a parton inside the proton. 
%%%%%%%%%%%%%%%%%%%%%%%%%%
\begin{figure}[h]
\begin{center}
  \includegraphics[height=12cm,width=15cm]{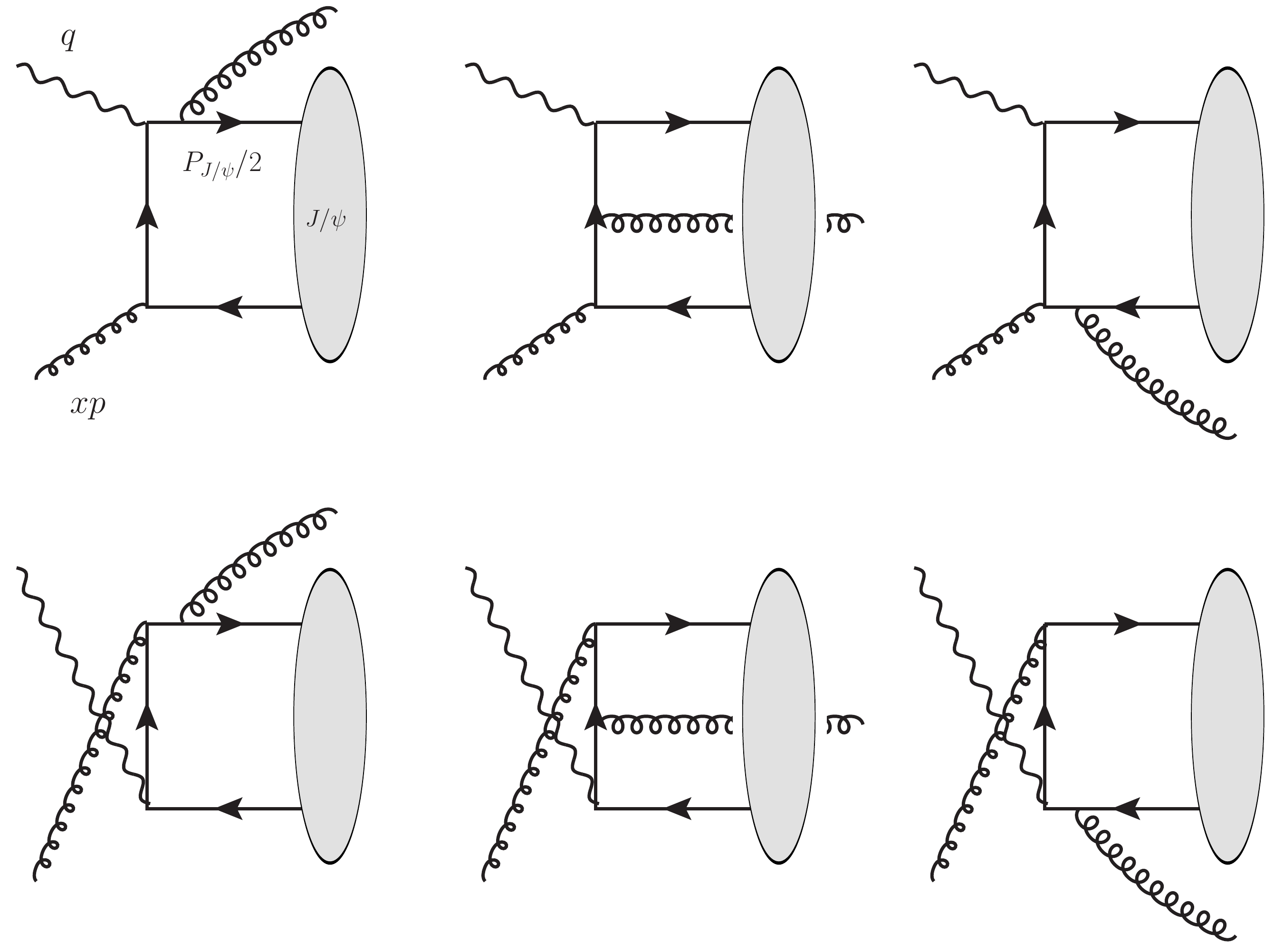}
\end{center}
 \caption{Diagrams that contribute to the unpolarized cross section. $w^{\mu\nu}(xp,q,P_{J/\psi})$ is given by
the squared amplitude of the sum of these six diagrams.}
\label{unpol_hard}
\end{figure}
%%%%%%%%%%%%%%%%%%%%%%%%%%%
We can calculate
$L^{\mu\nu}(\ell,\ell')w^{\mu\nu}(xp,q,P_{J/\psi})$ perturbatively by considering the diagrams in 
the leading-order(LO) with respect to the strong coupling constant
for the unpolarized cross section. The hadronic tensor $W_{\mu\nu}(p,q,P_{J/\psi})$ is
conventionally expanded in terms of 9 independent tensors ${\cal V}_i^{\mu\nu}$ ($i=1,2,\cdots 9$)
\cite{Meng:1991da} as
\beq
W^{\mu\nu}=\sum_{i=1}^9(W^{\rho\sigma}\widetilde{\cal V}_{i\rho\sigma}){\cal V}_i^{\mu\nu},
\eeq
where the inverse tensors $\widetilde{\cal V}_{i\rho\sigma}$ satisfy 
${\cal V}_i^{\mu\nu}\widetilde{\cal V}_{i'\mu\nu}=\delta_{ii'}$. Here we just show the 
explicit definitions of the symmetric
tensors that are relevant to our study.
\beq
&&{\cal V}_1^{\mu\nu}=X^\mu X^\nu+Y^\mu Y^\nu,
\hspace{5mm}
{\cal V}_2^{\mu\nu}=g^{\mu\nu}+Z^\mu Z^\nu,
\hspace{5mm}
{\cal V}_3^{\mu\nu}=T^\mu X^\nu+X^\mu T^\nu,
\nonumber\\
&&{\cal V}_4^{\mu\nu}=X^\mu X^\nu-Y^\mu Y^\nu,
\hspace{5mm}
{\cal V}_8^{\mu\nu}=T^\mu Y^\nu+Y^\mu T^\nu,
\hspace{5mm}
{\cal V}_9^{\mu\nu}=X^\mu Y^\nu+Y^\mu X^\nu,
\nonumber
\eeq
\beq
&&\widetilde{\cal V}_1^{\mu\nu}=
\frac{1}{2}(2T^\mu T^\nu+X^\mu X^\nu+Y^\mu Y^\nu),
\hspace{5mm}
\widetilde{\cal V}_2^{\mu\nu}=T^\mu T^\nu,
\hspace{5mm}
\widetilde{\cal V}_3^{\mu\nu}=-\frac{1}{2}(T^\mu X^\nu+X^\mu T^\nu),
\nonumber\\
&&\widetilde{\cal V}_4^{\mu\nu}=\frac{1}{2}(X^\mu X^\nu-Y^\mu Y^\nu),
\hspace{5mm}
\widetilde{\cal V}_8^{\mu\nu}=-\frac{1}{2}(T^\mu Y^\nu+Y^\mu T^\nu),
\hspace{5mm}
\widetilde{\cal V}_9^{\mu\nu}=\frac{1}{2}(X^\mu Y^\nu+Y^\mu X^\nu),
\nonumber
\eeq
where each vector is defined by
\beq
T^{\mu}=(1,0,0,0),\hspace{5mm}
X^{\mu}=(0,\cos\chi,\sin\chi,0),\hspace{5mm}
Y^{\mu}=(0,-\sin\chi,\cos\chi,0),\hspace{5mm}
Z^{\mu}=(0,0,0,1).
\eeq
Then we can calculate $L_{\mu\nu}W^{\mu\nu}$ as
\beq
L_{\mu\nu}W^{\mu\nu}=\sum_{i=1,\cdots, 4, 8, 9}[L_{\mu\nu}{\cal V}_i^{\mu\nu}][W_{\rho\sigma}\widetilde{\cal V}_{i}^{\rho\sigma}]=Q^2\sum_{i=1,\cdots, 4, 8, 9}{\cal A}_i(\phi-\chi)[W_{\rho\sigma}\widetilde{\cal V}_i^{\rho\sigma}],
\eeq
where the azimuthal dependences ${\cal A}_i(\varphi)$ are given by
\beq
&&{\cal A}_1(\varphi)={4\over y^2}(1-y+{y^2\over 2}),\hspace{5mm}
{\cal A}_2(\varphi)=-2,\hspace{5mm}
{\cal A}_3(\varphi)=-{4\over y^2}(2-y)\sqrt{1-y}\,{\rm cos}\varphi,
\nonumber\\
&&{\cal A}_4(\varphi)={4\over y^2}(1-y)\,{\rm cos}2\varphi,\hspace{5mm}
{\cal A}_8(\varphi)=-{4\over y^2}(2-y)\sqrt{1-y}\,{\rm sin}\varphi,\hspace{5mm}
{\cal A}_9(\varphi)={4\over y^2}(1-y)\,{\rm sin}2\varphi,
\label{azimuthal}
\eeq
where $\displaystyle y={Q^2\over x_BS_{ep}}$.
We can derive the unpolarized cross section formula by computing all the diagrams 
in FIG. \ref{unpol_hard} as
\beq
&&\frac{d^{6}\sigma}{dx_{B}dQ^{2}dz_{f}dP^2_Td\phi d\chi}
\nonumber\\
&=&
\frac{\alpha ^{2}_{em}\alpha^2_se^2_c
}{4\pi S^{2}_{ep} x_{B}^2Q^2}
\Bigl({\cal N}\la{\cal O}^{J/\psi}({}^3S_1^{[1]})\ra\Bigr)
\sum_{i=1,\cdots, 4, 8, 9}{\cal A}_i(\phi-\chi)
\int^1_{0}{dx\over x}\,G(x)\,
\hat{\sigma}_i\,\delta\Bigl[{P_T^2\over Q^2}
-\Bigl(1-{1\over \hat{x}}+{m_{J/\psi}^2\over z_fQ^2}\Bigr)\Bigl(1-{1\over z_f}\Bigr)\Bigr],\hspace{3mm}
\label{unpol}
\eeq
where $\hat{x}=x_B/x$, $\alpha_s$ is the strong coupling constant and 
$e_{c}$ is the electric charge of the charm quark.
We show all hard cross sections $\hat{\sigma}_i$ in Appendix A because they are lengthy.
The result in the $J/\psi$ rest frame is also available in \cite{DAlesio:2021yws}.

\subsection{Twist-3 polarized cross section for $J/\psi$ production in SIDIS}

We next calculate the twist-3 polarized cross section formula.
The twist-3 quark distribution effect could also contribute to the SSA and it was calculated in 
$pp$ collision\cite{Schafer:2013wca}. However, it is out of scope of the present study because it vanishes
in  the color-singlet case in SIDIS.
%%%%%%%%%%%%%%%%%%%%%%%%%%
\begin{figure}[h]
\begin{center}
  \includegraphics[height=4cm,width=4cm]{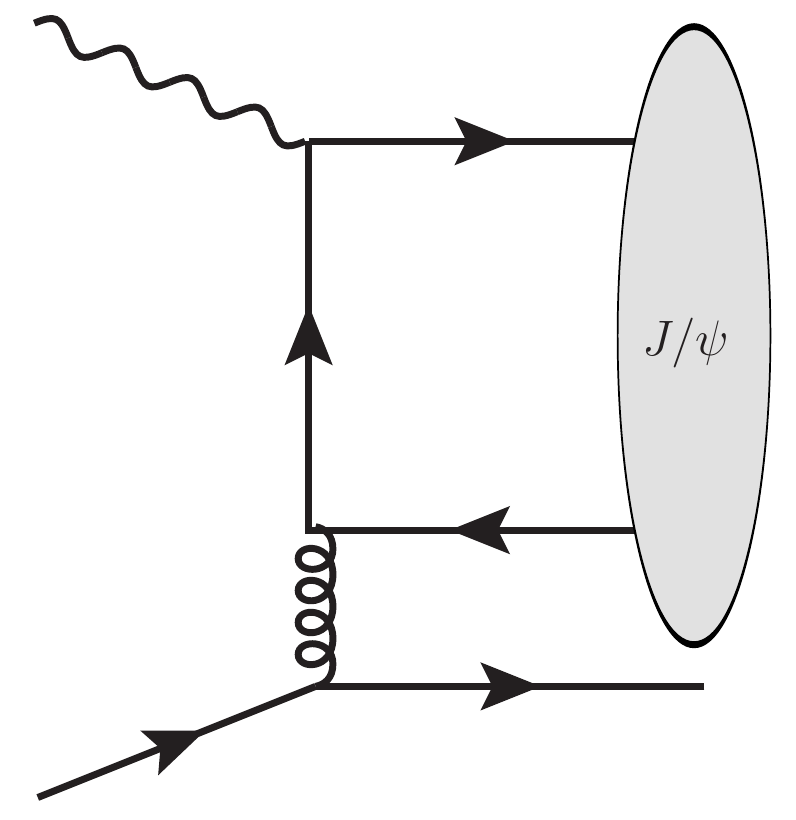}
\end{center}
 \caption*{A typical diagram given by the quark distribution effect. 
This is canceled in the color-singlet case in SIDIS.}
\end{figure}
%%%%%%%%%%%%%%%%%%%%%%%%%%%
The general formula for the twist-3 gluon 
contribution in SIDIS was derived in \cite{Yoshida:2022vnf} as
\beq
W^{\rm twist-3}_{\rho\sigma}(p,q,P_{J/\psi})&=&\la{\cal O}^{J/\psi}({}^3S_1^{[1]})\ra
\Bigl\{
\omega^{\mu}_{\ \alpha}\omega^{\nu}_{\ \beta}\omega^{\lambda}_{\ \gamma}
\int{dx\over x^2}\Phi^{\alpha\beta\gamma}_{\partial}(x)
{\partial\over \partial k^{\lambda}}S_{\mu\nu,\rho\sigma}(k)\Bigr|_{k=xp}
\nonumber\\
&&-{1\over 2}\omega^{\mu}_{\ \alpha}\omega^{\nu}_{\ \beta}\omega^{\lambda}_{\ \gamma}
\int dx_1\int dx_2 \Bigl[{-if^{abc}\over N_c(N_c^2-1)}N^{\alpha\beta\gamma}(x_1,x_2)
+{N_cd^{abc}\over (N_c^2-4)(N_c^2-1)}O^{\alpha\beta\gamma}(x_1,x_2)\Bigr]
\nonumber\\
&&\times {1\over x_1-i\epsilon}{1\over x_2+i\epsilon}
{1\over x_2-x_1-i\epsilon}S^{abc}_{\mu\nu\lambda,\rho\sigma}(x_1p,x_2p)\Bigl\},
\eeq
where $\omega^{\mu}_{\ \alpha}=g^{\mu}_{\ \alpha}-p^{\mu}n_{\alpha}$. 
$S_{\mu\nu,\rho\sigma}(k)$ is given by 
\beq
S_{\mu\nu,\rho\sigma}(k)=H_{\mu\nu,\rho\sigma}(k)2\pi\delta\Bigl[(k+q-P_{J/\psi})^2\Bigr],
\eeq
where the hard part $H_{\mu\nu,\rho\sigma}(k)$ is given by the diagrams in FIG. \ref{unpol_hard} 
by replacing the momentum $xp$ with $k$. $S^{abc}_{\mu\nu\lambda,\rho\sigma}(x_1p,x_2p)$
can be separated into two parts as
%%%%%%%%%%%%%%%%%%%%%%%%%%
\begin{figure}[h]
\begin{center}
  \includegraphics[height=8cm,width=12cm]{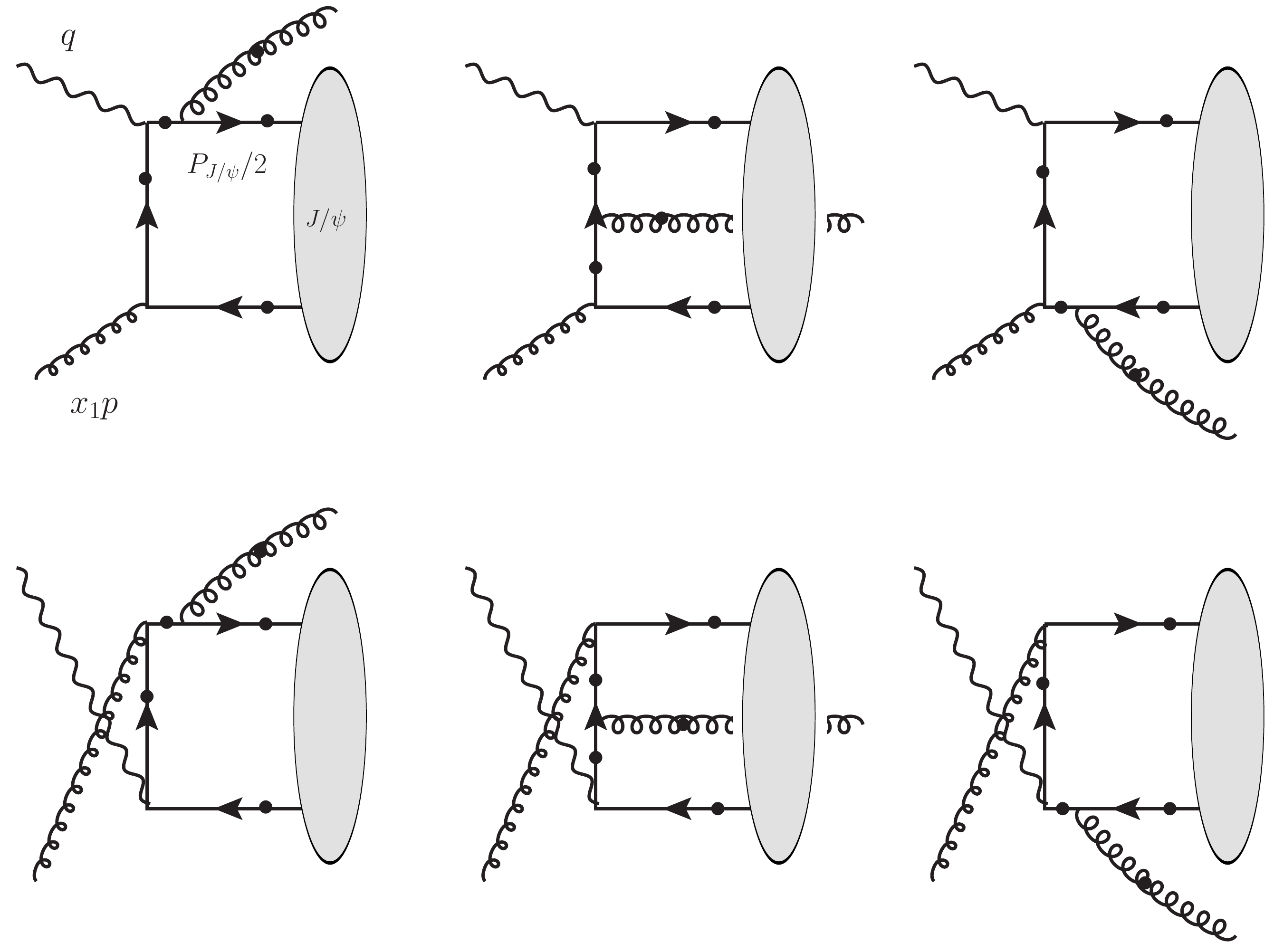}
\end{center}
\begin{center}
  \includegraphics[height=3cm,width=8cm]{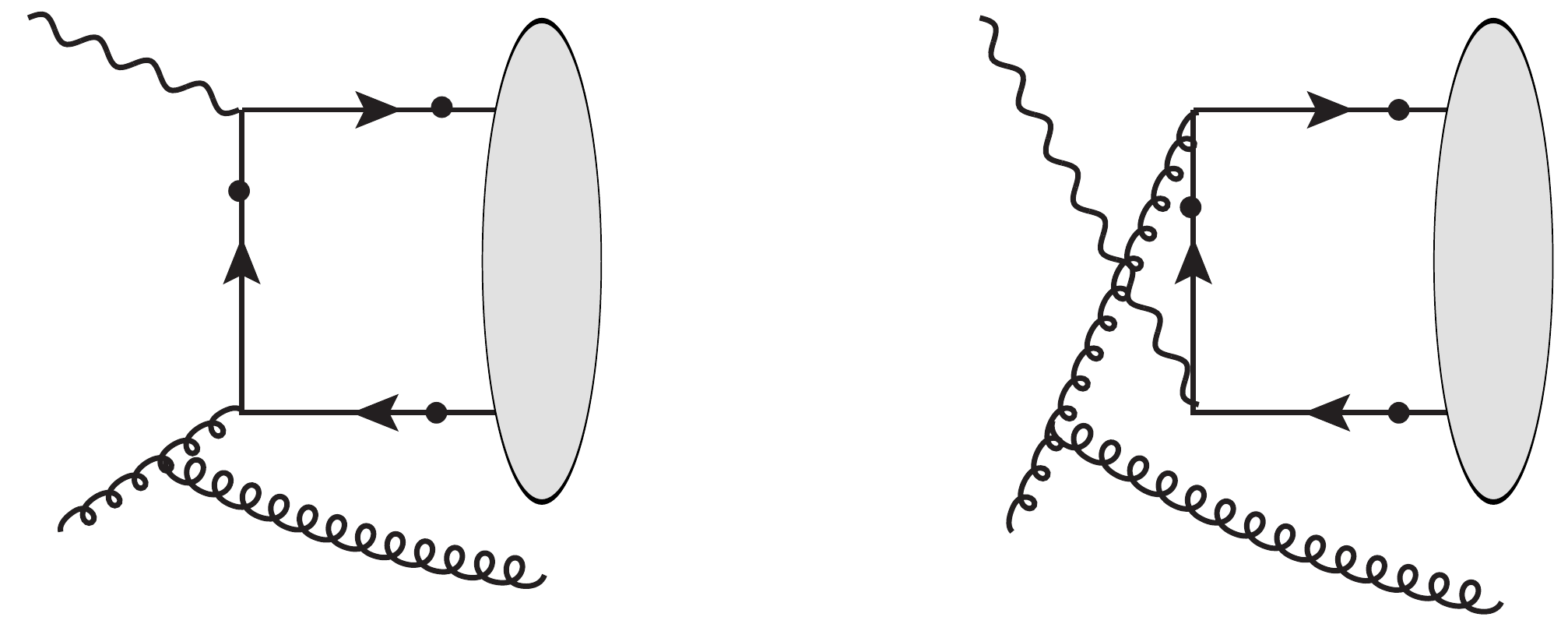}
\end{center}
 \caption{Diagrams that contribute to $H^{abc}_{L\mu\nu\lambda,\rho\sigma}(x_1p,x_2p)$. 
The external gluon line with the momentum $(x_2-x_1)p$ is connected to one of the black dots.
Thus there are 36 diagrams in this figure. We ignored some diagrams that obviously vanish.}
\label{polarized_hard}
\end{figure}
%%%%%%%%%%%%%%%%%%%%%%%%%%%
\beq
S^{abc}_{\mu\nu\lambda,\rho\sigma}(x_1p,x_2p)
=H^{abc}_{L\mu\nu\lambda,\rho\sigma}(x_1p,x_2p)2\pi\delta\Bigl[(x_2p+q-P_{J/\psi})^2\Bigr]
+H^{abc}_{R\mu\nu\lambda,\rho\sigma}(x_1p,x_2p)2\pi\delta\Bigl[(x_1p+q-P_{J/\psi})^2\Bigr].
\eeq
$H^{abc}_{L\mu\nu\lambda,\rho\sigma}(x_1p,x_2p)$ is given by the product of the diagrams 
in FIG. \ref{polarized_hard}
and the complex conjugate of the diagrams in FIG. \ref{unpol_hard}.
$H^{abc}_{R\mu\nu\lambda,\rho\sigma}(x_1p,x_2p)$ is the complex conjugate of 
$H^{abc}_{L\mu\nu\lambda,\rho\sigma}(x_1p,x_2p)$. 
From our direct calculation of the diagrams, we have observed that 
$H^{abc}_{L\mu\nu\lambda,\rho\sigma}(x_1p,x_2p)$ and 
$H^{abc}_{R\mu\nu\lambda,\rho\sigma}(x_1p,x_2p)$ have the following structures.
\beq
{1\over x_1-i\epsilon}
{1\over x_2-x_1-i\epsilon}H^{abc}_{L\mu\nu\lambda,\rho\sigma}(x_1p,x_2p)
&=&{1\over x_1-i\epsilon}H^{1\,abc}_{L\mu\nu\lambda,\rho\sigma}(x_2p)
+{x_2\over (x_1-i\epsilon)^2}H^{2\,abc}_{L\mu\nu\lambda,\rho\sigma}(x_2p)
\nonumber\\
&&+{1\over x_2-x_1-i\epsilon}H^{3\,abc}_{L\mu\nu\lambda,\rho\sigma}(x_2p)
+{x_2\over (x_2-x_1-i\epsilon)^2}H^{4\,abc}_{L\mu\nu\lambda,\rho\sigma}(x_2p)
\nonumber\\
&&+{1\over x_1-A x+i\epsilon}H^{5\,abc}_{L\mu\nu\lambda,\rho\sigma}(x_2p)
+{1\over x_1-(1-A)x-i\epsilon}H^{6\,abc}_{L\mu\nu\lambda,\rho\sigma}(x_2p),
\nonumber\\
{1\over x_2+i\epsilon}
{1\over x_2-x_1-i\epsilon}H^{abc}_{R\mu\nu\lambda,\rho\sigma}(x_1p,x_2p)
&=&{1\over x_2+i\epsilon}H^{1\,abc}_{R\mu\nu\lambda,\rho\sigma}(x_1p)
+{x_1\over (x_2+i\epsilon)^2}H^{2\,abc}_{R\mu\nu\lambda,\rho\sigma}(x_1p)
\nonumber\\
&&+{1\over x_2-x_1-i\epsilon}H^{3\,abc}_{R\mu\nu\lambda,\rho\sigma}(x_1p)
+{x_1\over (x_2-x_1-i\epsilon)^2}H^{4\,abc}_{R\mu\nu\lambda,\rho\sigma}(x_1p)
\nonumber\\
&&+{1\over x_2-A x-i\epsilon}H^{5\,abc}_{R\mu\nu\lambda,\rho\sigma}(x_1p)
+{1\over x_2-(1-A)x+i\epsilon}H^{6\,abc}_{R\mu\nu\lambda,\rho\sigma}(x_1p),\hspace{5mm}
\eeq
where $A={Q^2(1+\hat{x}-z_f)+m_{J/\psi}^2\hat{x}\over Q^2(2-z_f)}$.
Substituting (\ref{kinematical}), (\ref{C-even})
and (\ref{C-odd}) into $W^{\rm twist-3}_{\rho\sigma}(p,q,P_{J/\psi})$, we can derive the following form.
\beq
W^{\rm twist-3}_{\rho\sigma}(p,q,P_{J/\psi})&=&2\pi\la{\cal O}^{J/\psi}({}^3S_1^{[1]})\ra
\int{dx\over x^2}\delta\Bigl((xp+q-P_{J/\psi})^2\Bigr)
\nonumber\\
&&\Biggl\{\Bigl(x{d\over dx}G_T^{(1)}(x)-2{G_T^{(1)}(x)}\Bigr)
H^{G1}_{\rho\sigma}+{G_T^{(1)}(x)}H^{G2}_{\rho\sigma}
\nonumber\\
&&+\Bigl(x{d\over dx}\Delta H_T^{(1)}(x)-2{\Delta H_T^{(1)}(x)}\Bigr)
H^{H1}_{\rho\sigma}+{\Delta H_T^{(1)}(x)}H^{H2}_{\rho\sigma}
\nonumber\\
&&+\int dx'\sum_i\Bigl[\Bigl({1\over x-x'-i\epsilon}H^{Ni}_{1L\rho\sigma}
+{x\over (x-x'-i\epsilon)^2}H^{Ni}_{2L\rho\sigma}
+{1\over x'-A x+i\epsilon}H^{Ni}_{3L\rho\sigma}
\Bigr)N^i(x',x)
\nonumber\\
&&+\Bigl({1\over x-x'+i\epsilon}H^{Ni}_{1R\rho\sigma}
+{x\over (x-x'+i\epsilon)^2}H^{Ni}_{2R\rho\sigma}
+{1\over x'-A x-i\epsilon}H^{Ni}_{3R\rho\sigma}
\Bigr)N^i(x,x')
\nonumber\\
&&+\Bigl({1\over x-x'-i\epsilon}H^{Oi}_{1L\rho\sigma}
+{x\over (x-x'-i\epsilon)^2}H^{Oi}_{2L\rho\sigma}
+{1\over x'-A x+i\epsilon}H^{Oi}_{3L\rho\sigma}
\Bigr)O^i(x',x)
\nonumber\\
&&+\Bigl({1\over x-x'+i\epsilon}H^{Oi}_{1R\rho\sigma}
+{x\over (x-x'+i\epsilon)^2}H^{Oi}_{2R\rho\sigma}
+{1\over x'-A x-i\epsilon}H^{Oi}_{3R\rho\sigma}
\Bigr)O^i(x,x')
\Bigr]
\Biggr\},
\eeq
where we used the shorthand notations
\beq
N^{1,2,3}(x',x)=\{N(x',x),N(x,x-x'),N(x',x'-x)\},\hspace{5mm}O^{1,2,3}(x',x)=\{O(x',x),O(x,x-x'),O(x',x'-x)\}.
\eeq
We changed the integral variable as $x'\to x-x'$ 
for the denominators $1/(x'\pm i\epsilon)$ and 
$1/(x'-(1-A)x\pm i\epsilon)$ and used the symmetries (\ref{symmetries}).
One can derive the cross section formula by taking a contraction with tensors 
$\widetilde{\cal V}^{\rho\sigma}_i$. We have observed from our direct calculation that the contribution from
the $C$-odd function $O(x_1,x_2)$ is exactly canceled. This can be naturally understood from 
the fact that the charm-anticharm pair is charge neutral.
We can eliminate $x'$-integral by performing the following contour integrations. 
\beq
&&{1\over x'-x-i\epsilon}-{1\over x'-x+i\epsilon}=2\pi i\delta(x'-x),\hspace{7mm}
{1\over x'-Ax-i\epsilon}-{1\over x'-Ax+i\epsilon}=2\pi i\delta(x'-Ax),
\nonumber\\
&&\hspace{23mm}
{1\over (x'-x+i\epsilon)^2}-{1\over (x'-x-i\epsilon)^2}=2\pi i{\partial\over \partial x'}\delta(x'-x).
\label{derivative_term}
\eeq
The kinematical functions 
$G_T^{(1)}(x)$ and $\Delta H_T^{(1)}(x)$ can be eliminated by using the relations (\ref{relations}).
As a result, we can write down the cross section formula only in terms of the $C$-even function $N(x_1,x_2)$
as
\beq
&&\frac{d^{6}\Delta\sigma}{dx_{B}dQ^{2}dz_{f}dP_T^2d\phi d\chi}
\nonumber\\
&=&
\frac{\alpha ^{2}_{em}\alpha_s^2e_c^2(2\pi M_N)}{4\pi S^{2}_{ep} x_{B}^2Q^2}
\Bigl({\cal N}\la{\cal O}^{J/\psi}({}^3S_1^{[1]})\ra\Bigr)
\sum_{i=1,\cdots, 4, 8, 9}{\cal A}_i(\phi-\chi){\cal S}_i(\Phi_S-\chi)
\int {dx\over x^2}\delta\Bigl[{P_T^2\over Q^2}
-\Bigl(1-{1\over \hat{x}}+{m_{J/\psi}^2\over z_fQ^2}\Bigr)\Bigl(1-{1\over z_f}\Bigr)\Bigr]
\nonumber\\
&&\times \Bigl[N(x,x)\sigma^{N1}_i+N(x,0)\sigma^{N2}_i+N(x,A x)\sigma^{N3}_i
+N(x,(1-A)x)\sigma^{N4}_i+N(A x,-(1-A) x)\sigma^{N5}_i
\Bigr],
\label{polarized}
\eeq
where ${\cal S}_i(\Phi_S-\chi)=\sin(\Phi_S-\chi)(i=1,2,3,4),\ \cos(\Phi_S-\chi)(i=8,9)$. 
All hard cross sections are shown in Appendix A. It turns out that 
the derivative terms ${d\over dx}N(x,x)$ and ${d\over dx}N(x,0)$ given by (\ref{derivative_term})
are exactly canceled in the color singlet
channel, which is consistent with the statement made in \cite{Yuan:2008vn}.
However, we have observed that non-zero contributions from 
non-derivative terms $N(x,x)$ and $N(x,0)$ survive even in the color-singlet case. 
In addition, there are other contributions
$N(x,A x)$, $N(x,(1-A)x)$ and $N(Ax,Ax-x)$
that can be regarded as the hard-pole contribution in the sense of the conventional pole
calculation. A similar contribution was also observed in the case of the twist-3 quark distribution
\cite{Schafer:2013wca}.
If the color-singlet contribution is dominant in $J/\psi$ production, LDME is exactly canceled between
the unpolarized cross section (\ref{unpol}) and the polarized cross section (\ref{polarized}) in the ratio.
Thus we can conclude that the SSA in the $J/\psi$ production is an ideal observable to investigate 
the $C$-even twist-3 gluon distribution function $N(x_1,x_2)$ by eliminating other nonperturbative effects
like the $C$-odd type twist-3 effect and the hadronization effect of $J/\psi$.

%------------------  Section 4: Numerical calculation ------------

\section{Numerical calculation for the SSA in the $J/\psi$ production}

We perform numerical simulations of the $J/\psi$ SSA
for the kinematics accessible at the future EIC experiment. The polarized cross section (\ref{polarized})
can be expanded in terms of five structure functions ${\cal F}_i\ (i=1,2,\cdots 5)$ as
\beq
\frac{d^{6}\Delta\sigma}{dx_{B}dQ^{2}dz_{f}dP_T^2d\phi d\chi}
&=&\sin(\phi_h-\phi_S)({\cal F}_1+{\cal F}_2\cos\phi_h+{\cal F}_3\cos 2\phi_h)
\nonumber\\
&&+\cos(\phi_h-\phi_S)({\cal F}_4\sin\phi_h+{\cal F}_5\sin 2\phi_h),
\eeq
where the azimuthal dependences are defined by
\beq
\phi_h=\phi-\chi,\hspace{5mm}\phi_h-\phi_S=\Phi_S-\chi.
\eeq
The unpolarized cross section is given by
\beq
\frac{d^{6}\sigma}{dx_{B}dQ^{2}dz_{f}dP_T^2d\phi d\chi}
=\sigma_1^{\rm U}+\sigma_2^{\rm U}\cos\phi_h+\sigma_3^{\rm U}\cos 2\phi_h.
\eeq
We calculate five normalized structure functions \cite{Beppu:2012vi},
\beq
{{\cal F}_1\over \sigma_1^{\rm U}},\hspace{4mm}
{{\cal F}_2\over 2\sigma_1^{\rm U}},\hspace{4mm}
{{\cal F}_3\over 2\sigma_1^{\rm U}},\hspace{4mm}
{{\cal F}_4\over 2\sigma_1^{\rm U}},\hspace{4mm}
{{\cal F}_5\over 2\sigma_1^{\rm U}}.
\label{structure_functions}
\eeq
We here show the explicit form of ${\cal F}_1/ \sigma_1^{\rm U}$ that can be derived from 
(\ref{azimuthal}), (\ref{unpol}) and (\ref{polarized}).
\beq
{{\cal F}_1\over \sigma_1^{\rm U}}&=&{2\pi M_N\over [{4\over y^2}(1-y+{y^2\over 2})\hat{\sigma}_1
-2\hat{\sigma}_2]\bar{x}G(\bar{x})}\Bigl[{4\over y^2}(1-y+{y^2\over 2})
\Bigl(\sum_{i=1}^{5}N^i(\bar{x})\sigma^{Ni}_1\Bigr)
-2\Bigl(\sum_{i=1}^{5}N^i(\bar{x})\sigma^{Ni}_2\Bigr)\Bigl],
\eeq
where we defined
\beq
N^{1,2,3,4,5}(x)=\{N(x,x),\ N(x,0),\ N(x,Ax),\ N(x,(1-A)x),\ N(Ax,-(1-A)x)\},
\eeq
\beq
\bar{x}=x_B\Bigl(1+{m_{J/\psi}^2\over z_fQ^2}+{P_T^2\over Q^2}{z_f\over 1-z_{f}}\Bigr).
\eeq
%%%%%%%%%%%%%%%%%%%%%%%%%%
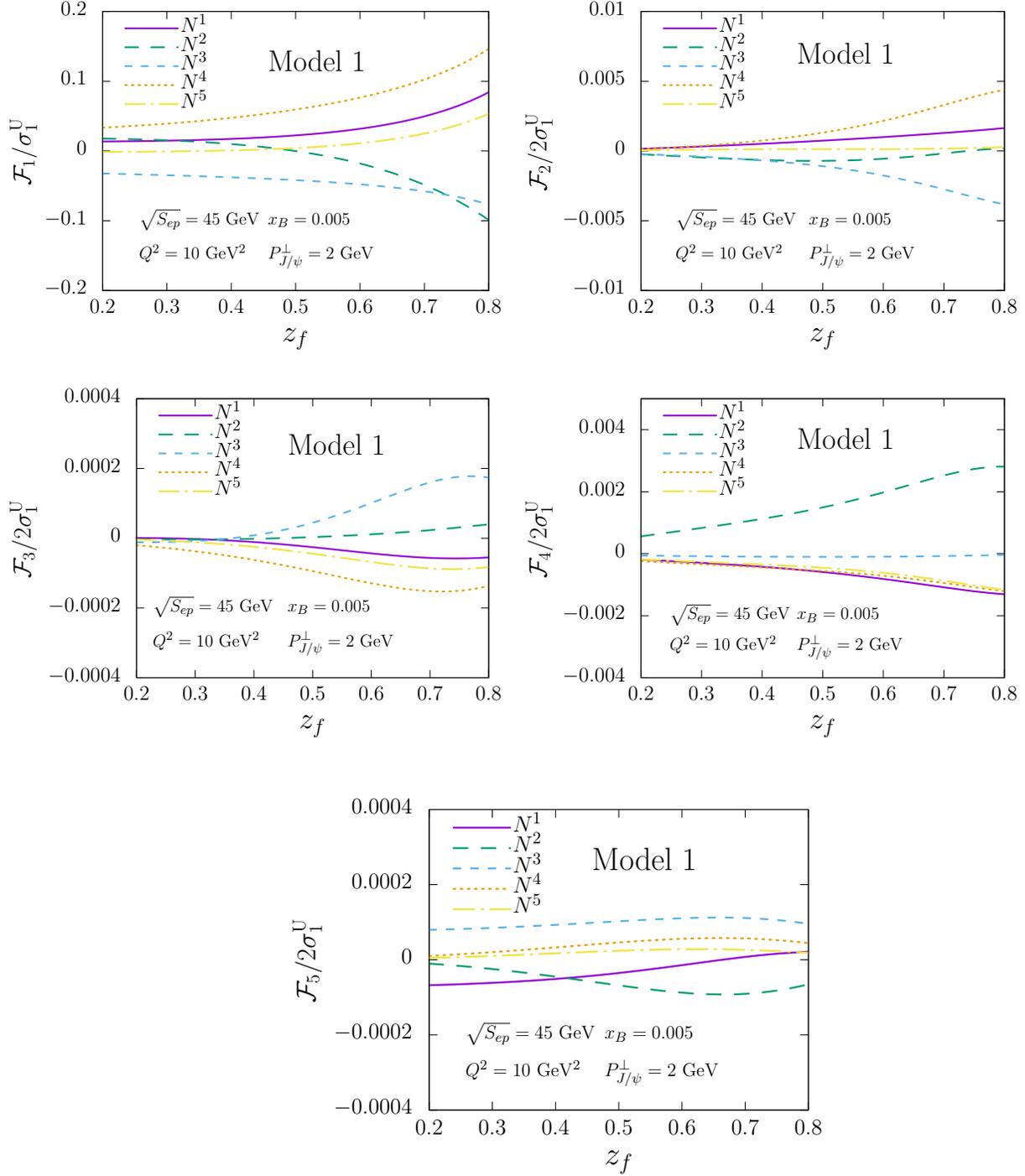
\begin{figure}[H]
\begin{minipage}{0.45\hsize}
  \input{F1_model1.tex}
\end{minipage}
\begin{minipage}{0.45\hsize}
  \input{F2_model1.tex}
\end{minipage}

\begin{minipage}{0.45\hsize}
  \input{F3_model1.tex}
\end{minipage}
\begin{minipage}{0.45\hsize}
  \input{F4_model1.tex}
\end{minipage}

\begin{center}
\input{F5_model1.tex}
\end{center}
\caption{Numerical calculations for the normalized structure functions in (\ref{structure_functions}).
$N^{1,2,3,4,5}$ respectively show the contributions from the five functions 
$N(x,x)$, $N(x,0)$, $N(x,A x)$, $N(x,(1-A)x)$, $N(A x,-(1-A)x)$
with the model 1 function (\ref{model1}).}
\end{figure}
%%%%%%%%%%%%%%%%%%%%%%%%%%%
\noindent
One can easily derive other normalized structure functions in the same way.  
The LDME is exactly canceled as stated above and, therefore, the nonperturbative
effect simply arises from ratios of twist-3 and twist-2 gluon distributions.
\newpage
%%%%%%%%%%%%%%%%%%%%%%%%%%
\begin{figure}[H]
\begin{minipage}{0.45\hsize}
  \input{F1_model2.tex}
\end{minipage}
\begin{minipage}{0.45\hsize}
  \input{F2_model2.tex}
\end{minipage}

\begin{minipage}{0.45\hsize}
  \input{F3_model2.tex}
\end{minipage}
\begin{minipage}{0.45\hsize}
  \input{F4_model2.tex}
\end{minipage}

\begin{center}
\input{F5_model2.tex}
\end{center}
\caption{Numerical calculations for the normalized structure functions
with the model 2 function (\ref{model2}).}
\end{figure}
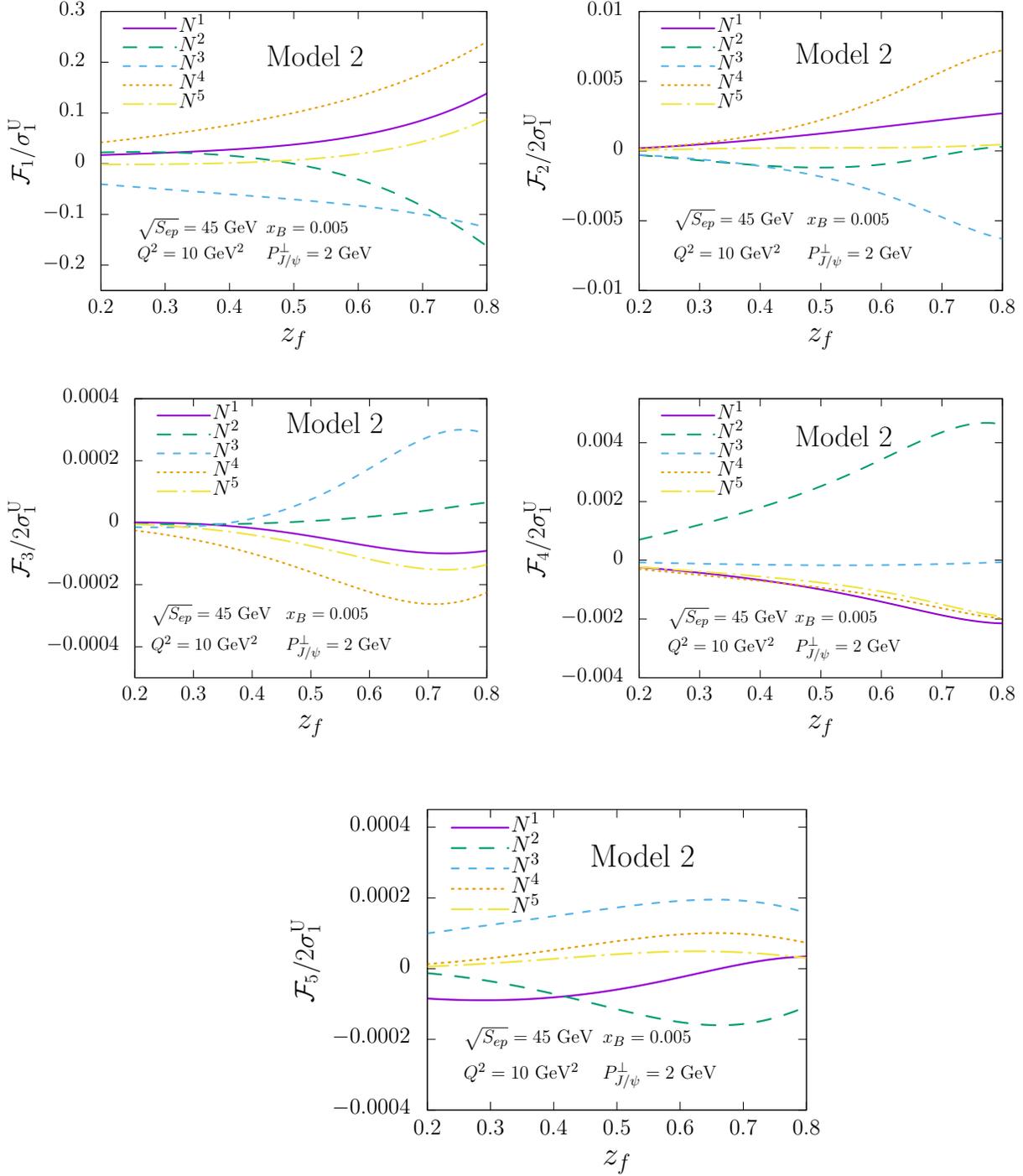
%%%%%%%%%%%%%%%%%%%%%%%%%%%
\noindent
The $C$-even function $N(x_1,x_2)$ has not been well constrained by experiment so far. 
We use the following simple models used in \cite{Koike:2011mb}.
\beq
&&{\rm model\ 1}:\hspace{3mm} 0.002xG(x),
\label{model1}
\\
&&{\rm model\ 2}:\hspace{3mm} 0.0005\sqrt{x}G(x).
\label{model2}
\eeq
Experimental investigations of the twist-3 gluon contributions were reported in the past couple of years
\cite{PHENIX:2021irw,PHENIX:2022znm}. The magnitudes of the above models are consistent with 
the upper bound of the data.
%We use CTEQ6L \cite{Pumplin:2002vw} for the gluon distribution function $G(x)$ and 
%set the factorization scale $\mu$ at $\mu^2=Q^2+m_{J/\psi}^2+P^2_{J\perp}$. 
Each structure function depends on five types of
$C$-even functions \{$N(x,x)$, $N(x,0)$, $N(x,A x)$, $N(x,(1-A)x)$,
$N(A x,-(1-A)x)$\}.
We separately plot the contributions from those five functions by substituting
one of the models into each function.
Models for $N(x,A x)$, $N(x,(1-A)x)$,
$N(A x,-(1-A)x)$ should take the $A$-dependence into account for realistic simulations.
However, we use the above $A$-independent models for these functions 
because our current knowledge about the functions is very limited. 
Our simulations are still beneficial because they show some differences in the qualitative behavior
among the hard cross sections.
We perform our simulations with typical EIC kinematic valuables
\cite{Accardi:2012qut}: $\sqrt{S_{ep}}=45$ GeV, $Q^2=10$ GeV$^2$, $x_{B}=0.005$, 
$P_{J/\psi}^{\perp}=2$ GeV. FIG. 4 and 5 respectively show our simulations with the model 1 and 2 
for the five structure functions. The contributions from the five types of functions show different 
qualitative behaviors in each normalized structure function. These simulations could help to clarify
which function is dominant in the $J/\psi$ SSA by comparing with the data from EIC. 
Our simulations show that the difference in the $x$-dependence 
between the model 1 and 2 does not cause 
a significant change in the qualitative behaviors. However, we find that the magnitudes of the contributions
are uniformly increased in the model 2 compared to the model 1. This reflects the fact that the model 2 
is more singular with respect to $x$ and, therefore, it is enhanced at the small value of $x_{B}$.
The observation of the enhancement could give us the information about the small-$x$ behavior of the
$C$-even function. The future EIC experiment plans to 
investigate the proton structure in a wide range of $x_{B}$, $10^{-3}\lesssim x_{B}\lesssim 0.5$.
The measurement of $J/\psi$ SSA at the EIC is a great opportunity to understand the small-$x$
behavior of the higher twist gluon distribution function.

%------------------  Section 5: Summary ----------------- 

\section{Summary}

We performed the first calculation for the twist-3 gluon contribution to the $J/\psi$ SSA
within the collinear framework.  We observed that one of the possible twist-3 contributions
, $C$-odd type twist-3 gluon contribution, is exactly canceled in the spin-dependent cross section formula.
The cancellation partially happens to also the $C$-even type function. The derivative terms 
${d\over dx}N(x,x)$ and ${d\over dx}N(x,0)$ are exactly canceled, while nonderivative
terms $N(x,x)$ and $N(x,0)$ survive in the cross section.
Once it was stated that the soft-gluon-pole type contribution is exactly canceled 
in the color singlet contribution in SIDIS\cite{Yuan:2008vn}. 
However, our result shows that this statement is not valid
in high-$P_{J/\psi}^{\perp}$ region where the relationship between the TMD and the collinear frameworks 
does not hold. In addition, we obtained the contributions from another type of the pole contributions
whose existence was pointed out in \cite{Schafer:2013wca} in the case of the twist-3 quark distribution.
We have completed the LO cross section formula for the $J/\psi$ SSA
in the color-singlet case.  Our result enables future investigations of the twist-3 gluon distribution function
at the EIC. We performed some numerical simulations for the structure functions in the polarized cross section
at the EIC kinematics.  The qualitative differences among the five types of the $C$-even functions could 
help to pin down the dominant contribution to the $J/\psi$ SSA.  Our simulations show the correlation 
between the magnitudes of the structure functions and the $x$-dependence of the twist-3 gluon
distribution function. Future investigations in a wide range of the Bjorken variable 
at the EIC will provide rich information about the little-known twist-3 gluon distribution function.

%-------------------- Appendix ---------------------

\appendix

\section{List of hard cross sections}

\noindent
(1) Hard cross sections of the unpolarized cross section in (\ref{unpol})

\beq
\hat{\sigma}_1&=&{1\over z_f^2[Q^2(-1+\hat{x})+m_{J/\psi}^2\hat{x}]^2[m_{J/\psi}^2\hat{x}+
Q^2(1+\hat{x}-z_f)]^2}
\nonumber\\
&&\times 64m_{J/\psi}^2\hat{x}^2\Bigl(m_{J/\psi}^6\hat{x}^2(1-z_f+z_f^2)
+m_{J/\psi}^2Q^4[1+(-2+4\hat{x}-7\hat{x}^2)z_f+(3-18\hat{x}+22\hat{x}^2)z_f^2
\nonumber\\
&&-2(1-8\hat{x}+9\hat{x}^2)z_f^3+(1-6\hat{x}+6\hat{x}^2)z_f^4]
+Q^6(-1+\hat{x})z_f[-z_f+\hat{x}(-1+2z_f)]+m_{J/\psi}^4Q^2\hat{x}[z_f(-3+3z_f-2z_f^2)
\nonumber\\
&&+\hat{x}(-5+17z_f-15z_f^2+6z_f^3)]\Bigr)
\eeq
\beq
\hat{\sigma}_2&=&{1\over z_f^2[Q^2(-1+\hat{x})+
m_{J/\psi}^2\hat{x}]^2[m_{J/\psi}^2\hat{x}+Q^2(1+\hat{x}-z_f)]^2}
\nonumber\\
&&\times 64m_{J/\psi}^2Q^2\hat{x}^2\Bigl(Q^4(-1+\hat{x})^2z_f^2+
2m_{J/\psi}^2Q^2(-1+\hat{x})\hat{x}z_f(-1+6z_f-6z_f^2+2z_f^3)
\nonumber\\
&&+m_{J/\psi}^4\hat{x}^2(-2+10z_f-11z_f^2+4z_f^3)
\Bigr)
\eeq
\beq
\hat{\sigma}_3&=&
{1\over z_f[Q^2(-1+\hat{x})+
m_{J/\psi}^2\hat{x}]^2[m_{J/\psi}^2\hat{x}+Q^2(1+\hat{x}-z_f)]^2}
\nonumber\\
&&\times 64m_{J/\psi}^2QP_T\hat{x}^3\Bigl(Q^4(-1+\hat{x})z_f+m_{J/\psi}^4\hat{x}(3-4z_f+2z_f^2)+
m_{J/\psi}^2Q^2[z_f(-3+4z_f-2z_f^2)
\nonumber\\
&&+\hat{x}(-1+9z_f-10z_f^2+4z_f^3)]\Bigr)
\eeq
\beq
\hat{\sigma}_4&=&-{64m_{J/\psi}^4\hat{x}^3(-1+z_f)[m_{J/\psi}^2\hat{x}+Q^2(-1+\hat{x})z_f][m_{J/\psi}^2+
Q^2(-1+4z_f-2z_f^2)]\over 
z_f^2[Q^2(-1+\hat{x})+m_{J/\psi}^2\hat{x}]^2[m_{J/\psi}^2\hat{x}+Q^2(1+\hat{x}-z_f)]^2}
\eeq
\beq
\hat{\sigma}_8&=&\hat{\sigma}_9=0
\eeq

\

\noindent
(2) Hard cross sections of $N(x,x)$ in (\ref{polarized})

\beq
\hat{\sigma}^{N1}_1&=&{-1\over (1-z_f)z_f^2Q^2[Q^2(-1+\hat{x})+m_{J/\psi}^2\hat{x}]^3[m_{J/\psi}^2\hat{x}+
Q^2(1+\hat{x}-z_f)]^3}
\nonumber\\
&&\times 128m_{J/\psi}^2P_T\hat{x}^3\Bigl(2m_{J/\psi}^{10}\hat{x}^4(5-6z_f+3z_f^2)+
4Q^{10}(-1+\hat{x})z_f[1-5z_f+3z_f^2-z_f^3
\nonumber\\
&&+\hat{x}^3(-1+2z_f)+
\hat{x}^2(z_f-4z_f^2)
+\hat{x}(-2+7z_f-3z_f^2+2z_f^3)]
\nonumber\\
&&+2m_{J/\psi}^8Q^2\hat{x}^3[-2z_f(8-9z_f+5z_f^2)+
\hat{x}(-9+58z_f-65z_f^2+26z_f^3)]
\nonumber\\
&&+m_{J/\psi}^6Q^4\hat{x}^2[18-34z_f+77z_f^2-63z_f^3+30z_f^4-
16\hat{x}z_f(-3+19z_f-20z_f^2+8z_f^3)
\nonumber\\
&&+2\hat{x}^2(-33+112z_f-65z_f^2-18z_f^3+24z_f^4)]
+m_{J/\psi}^4Q^6\hat{x}[z_f(-54+106z_f-123z_f^2+71z_f^3-24z_f^4)
\nonumber\\
&&-4\hat{x}^2z_f(-25+84z_f-27z_f^2-32z_f^3+24z_f^4)+
2\hat{x}^3(-19+24z_f+81z_f^2-114z_f^3+48z_f^4)
\nonumber\\
&&+2\hat{x}(-9+65z_f-123z_f^2+224z_f^3-177z_f^4+62z_f^5)]
+m_{J/\psi}^2Q^8[4-12z_f+47z_f^2-74z_f^3+67z_f^4-32z_f^5+8z_f^6
\nonumber\\
&&+4\hat{x}^4z_f(-13+41z_f-35z_f^2+12z_f^3)-
8\hat{x}^3z_f(-3+31z_f^2-32z_f^3+12z_f^4)
\nonumber\\
&&+\hat{x}z_f(28-220z_f+363z_f^2-359z_f^3+188z_f^4-48z_f^5)
\nonumber\\
&&+\hat{x}^2(12-84z_f+229z_f^2-157z_f^3+152z_f^4-116z_f^5+
48z_f^6)]\Bigr)
\nonumber\\
\eeq
\beq
\hat{\sigma}^{N1}_2&=&
{-1\over (1-z_f)z_f^2[Q^2(-1+\hat{x})+m_{J/\psi}^2\hat{x}]^3[m_{J/\psi}^2\hat{x}+Q^2(1+\hat{x}-z_f)]^3}
\nonumber\\
&&\times 512m_{J/\psi}^2P_T\hat{x}^3\Bigl(Q^8(-1+\hat{x})^2z_f^2(3+\hat{x}^2-2z_f-2\hat{x}z_f+
z_f^2)+m_{J/\psi}^8\hat{x}^4(-4+20z_f-23z_f^2+8z_f^3)
\nonumber\\
&&+2m_{J/\psi}^6Q^2\hat{x}^3[z_f(5-26z_f+29z_f^2-10z_f^3)+
2\hat{x}(-2+9z_f-6z_f^2-2z_f^3+2z_f^4)]
\nonumber\\
&&+m_{J/\psi}^4Q^4\hat{x}^2[-4+24z_f-50z_f^2+84z_f^3-67z_f^4+20z_f^5+
2\hat{x}z_f(6-29z_f+13z_f^2+12z_f^3-8z_f^4)
\nonumber\\
&&+2\hat{x}^2(-2+6z_f+11z_f^2-20z_f^3+8z_f^4)]
+2m_{J/\psi}^2Q^6\hat{x}z_f[3-21z_f+36z_f^2-35z_f^3+18z_f^4-4z_f^5
\nonumber\\
&&+2\hat{x}^3(-1+6z_f-6z_f^2+2z_f^3)+\hat{x}^2(1-4z_f-17z_f^2+22z_f^3-8z_f^4)+
\hat{x}(-2+13z_f-7z_f^2+9z_f^3-10z_f^4+4z_f^5)]\Bigr)
\nonumber\\
\eeq
\beq
\hat{\sigma}^{N1}_3&=&
{1\over z_f^3Q[Q^2(-1+\hat{x})+m_{J/\psi}^2\hat{x}]^3[m_{J/\psi}^2\hat{x}+
Q^2(1+\hat{x}-z_f)]^3}
\nonumber\\
&&\times 128m_{J/\psi}^2\hat{x}^2\Bigl(2m_{J/\psi}^{10}\hat{x}^5(13-21z_f+10z_f^2)+
Q^{10}(-1+\hat{x})^2z_f^2[4\hat{x}^3+3(-1+z_f)-8\hat{x}^2z_f+
\hat{x}(9-8z_f+4z_f^2)]
\nonumber\\
&&+m_{J/\psi}^8Q^2\hat{x}^4[z_f(-83+129z_f-62z_f^2)+
\hat{x}(46-78z_f^2+52z_f^3)]
\nonumber\\
&&+m_{J/\psi}^6Q^4\hat{x}^3[22-48z_f+148z_f^2-177z_f^3+80z_f^4-
\hat{x}z_f(101+53z_f-226z_f^2+136z_f^3)
\nonumber\\
&&+2\hat{x}^2(7+62z_f-79z_f^2+14z_f^3+16z_f^4)]
+m_{J/\psi}^2Q^8\hat{x}z_f[4+35z_f-92z_f^2+100z_f^3-58z_f^4+16z_f^5
\nonumber\\
&&+\hat{x}^4(-2+66z_f-76z_f^2+32z_f^3)+
\hat{x}^3(1-71z_f-50z_f^2+120z_f^3-64z_f^4)
\nonumber\\
&&+\hat{x}^2(-2+94z_f-15z_f^2-6z_f^3-28z_f^4+32z_f^5)
-\hat{x}(1+124z_f-233z_f^2+246z_f^3-150z_f^4+48z_f^5)]
\nonumber\\
&&+m_{J/\psi}^4Q^6\hat{x}^2[z_f(-63+143z_f-192z_f^2+148z_f^3-54z_f^4)
+2\hat{x}^3(-3+40z_f+z_f^2-50z_f^3+32z_f^4)
\nonumber\\
&&-\hat{x}^2z_f(17+245z_f-246z_f^2+16z_f^3+64z_f^4)+
\hat{x}(-10+110z_f-123z_f^2+248z_f^3-282z_f^4+132z_f^5)]\Bigr)
\eeq
\beq
\hat{\sigma}^{N1}_4&=&{1\over z_f^2Q^2[Q^2(-1+
\hat{x})+m_{J/\psi}^2\hat{x}]^3[m_{J/\psi}^2\hat{x}+Q^2(1+\hat{x}-z_f)]^3}
\nonumber\\
&&\times128m_{J/\psi}^4P_T\hat{x}^3\Bigl(2m_{J/\psi}^8\hat{x}^4(-5+3z_f)+
2m_{J/\psi}^6Q^2\hat{x}^3[-6(-2+z_f)z_f+\hat{x}(-7-11z_f+10z_f^2)]
\nonumber\\
&&+m_{J/\psi}^4Q^4\hat{x}^2[-6-13z_f^2+6z_f^3+12\hat{x}z_f(1+6z_f-4z_f^2)
+2\hat{x}^2(1-27z_f+6z_f^2+8z_f^3)]
\nonumber\\
&&+Q^8(-1+\hat{x})z_f[-3(-1+z_f)z_f+4\hat{x}^3(2-7z_f+4z_f^2)
-4\hat{x}^2(-1+7z_f-16z_f^2+8z_f^3)
\nonumber\\
&&+\hat{x}(12-57z_f+76z_f^2-52z_f^3+16z_f^4)]
+m_{J/\psi}^2Q^6\hat{x}[-z_f(-10+4z_f+z_f^2)-4\hat{x}^2z_f(4-21z_f+8z_f^3)
\nonumber\\
&&+2\hat{x}^3(3-9z_f-18z_f^2+16z_f^3)+2\hat{x}(5-28z_f+35z_f^2-47z_f^3+22z_f^4)]\Bigr)
\eeq
\beq
\hat{\sigma}^{N1}_8&=&{1\over z_f^3Q[Q^2(-1+\hat{x})+
m_{J/\psi}^2\hat{x}]^2[m_{J/\psi}^2\hat{x}+Q^2(1+\hat{x}-z_f)]^2}
\nonumber\\
&&\times 128m_{J/\psi}^2\hat{x}^2\Bigl(-Q^6(-1+\hat{x})z_f^2+2m_{J/\psi}^6\hat{x}^3(1-3z_f+2z_f^2)
\nonumber\\
&&+m_{J/\psi}^4Q^2\hat{x}^2(1-3z_f+2z_f^2)[-3z_f+2\hat{x}(1+z_f)]
+m_{J/\psi}^2Q^4\hat{x}z_f[z_f(-1-2z_f+2z_f^2)
\nonumber\\
&&+\hat{x}^2(2-6z_f+4z_f^2)+\hat{x}(-1+z_f+4z_f^2-4z_f^3)]\Bigr)
\eeq
\beq
\hat{\sigma}^{N1}_9&=&{128m_{J/\psi}^4P_T\hat{x}^3\over z_f^2Q^2[Q^2(-1+\hat{x})+
m_{J/\psi}^2\hat{x}]^2[m_{J/\psi}^2\hat{x}+Q^2(1+\hat{x}-z_f)]^2}
\Bigl(2m_{J/\psi}^4\hat{x}^2(-1+z_f)
\nonumber\\
&&+2m_{J/\psi}^2Q^2\hat{x}(-1+z_f)(\hat{x}-z_f+2\hat{x}z_f)+
Q^4z_f[4\hat{x}^2(-1+z_f)+z_f-4\hat{x}(-1+z_f)z_f]\Bigr)
\nonumber\\
\eeq

\

\noindent
(3) Hard cross sections of $N(x,0)$ in (\ref{polarized})

\beq
\hat{\sigma}^{N2}_1&=&{1\over (1-z_f)z_f^2Q^2[Q^2(-1+\hat{x})+m_{J/\psi}^2\hat{x}]^3[m_{J/\psi}^2\hat{x}+
Q^2(1+\hat{x}-z_f)]^3}
\nonumber\\
&&\times 128m_{J/\psi}^2P_T\hat{x}^3\Bigr(2m_{J/\psi}^{10}\hat{x}^4(9-10z_f+7z_f^2)+
4Q^{10}(-1+\hat{x})z_f[3-13z_f+13z_f^2-5z_f^3+\hat{x}^3(-3+6z_f)
\nonumber\\
&&+\hat{x}^2(-2+9z_f-12z_f^2)+\hat{x}z_f(3-5z_f+6z_f^2)]
+2m_{J/\psi}^8Q^2\hat{x}^3[-2z_f(13-14z_f+9z_f^2)
\nonumber\\
&&+\hat{x}(-21+118z_f-117z_f^2+50z_f^3)]
+m_{J/\psi}^6Q^4\hat{x}^2[42-90z_f+131z_f^2-89z_f^3+42z_f^4
\nonumber\\
&&-4\hat{x}z_f(-27+143z_f-142z_f^2+58z_f^3)
+6\hat{x}^2(-23+72z_f-31z_f^2-14z_f^3+16z_f^4)]
\nonumber\\
&&+m_{J/\psi}^4Q^6\hat{x}[z_f(-94+206z_f-209z_f^2+105z_f^3-32z_f^4)
-4\hat{x}^2z_f(-46+133z_f-11z_f^2-76z_f^3+48z_f^4)
\nonumber\\
&&+2\hat{x}^3(-39+32z_f+205z_f^2-234z_f^3+96z_f^4)
+2\hat{x}(-17+145z_f-345z_f^2+514z_f^3-361z_f^4+118z_f^5)]
\nonumber\\
&&+m_{J/\psi}^2Q^8[-4+12z_f+37z_f^2-106z_f^3+109z_f^4-52z_f^5
+12z_f^6+4\hat{x}^4z_f(-31+93z_f-71z_f^2+24z_f^3)
\nonumber\\
&&-4\hat{x}^3z_f(-7-27z_f+152z_f^2-134z_f^3+48z_f^4)
+\hat{x}z_f(96-572z_f+1041z_f^2-969z_f^3+460z_f^4-104z_f^5)
\nonumber\\
&&+\hat{x}^2(20-92z_f+211z_f^2-199z_f^3+316z_f^4-244z_f^5+
96z_f^6)]\Bigr)
\nonumber\\
\eeq
\beq
\hat{\sigma}^{N2}_2&=&{1\over (1-z_f)z_f^2[Q^2(-1+\hat{x})+m_{J/\psi}^2\hat{x}]^3[m_{J/\psi}^2\hat{x}+
Q^2(1+\hat{x}-z_f)]^3}
\nonumber\\
&&\times 512m_{J/\psi}^2P_T\hat{x}^3\Bigl(Q^8(-1+\hat{x})^2z_f^2(5+3\hat{x}^2+\hat{x}(4-6z_f)-
6z_f+3z_f^2)+m_{J/\psi}^8\hat{x}^4(-8+40z_f-45z_f^2+16z_f^3)
\nonumber\\
&&+2m_{J/\psi}^6Q^2\hat{x}^3[z_f(9-46z_f+51z_f^2-18z_f^3)
+\hat{x}(-8+36z_f-22z_f^2-8z_f^3+8z_f^4)]
\nonumber\\
&&+m_{J/\psi}^4Q^4\hat{x}^2[-8+48z_f-106z_f^2+160z_f^3-121z_f^4+36z_f^5
+2\hat{x}z_f(10-45z_f+11z_f^2+28z_f^3-16z_f^4)
\nonumber\\
&&+\hat{x}^2(-8+24z_f+50z_f^2-80z_f^3+32z_f^4)]
+2m_{J/\psi}^2Q^6\hat{x}z_f[7-45z_f+84z_f^2-79z_f^3+38z_f^4-8z_f^5
\nonumber\\
&&+\hat{x}^3(-4+26z_f-24z_f^2+8z_f^3)
+\hat{x}^2(1-43z_f^2+46z_f^3-16z_f^4)
+\hat{x}(-4+19z_f-17z_f^2+25z_f^3-22z_f^4+8z_f^5)]\Bigr)
\nonumber\\
\eeq
\beq
\hat{\sigma}^{N2}_3&=&{-1\over z_f^3Q(Q^2(-1+\hat{x})+m_{J/\psi}^2\hat{x})^3(m_{J/\psi}^2\hat{x}+
Q^2(1+\hat{x}-z_f))^3}
\nonumber\\
&&\times 128m_{J/\psi}^2\hat{x}^2\Bigr(2m_{J/\psi}^{10}\hat{x}^5(25-37z_f+18z_f^2)+
Q^{10}(-1+\hat{x})^2z_f^2[-9+12\hat{x}^3+\hat{x}^2(10-24z_f)+15z_f-6z_f^2
\nonumber\\
&&+\hat{x}(5-12z_f+12z_f^2)]+m_{J/\psi}^8Q^2\hat{x}^4[z_f(-149+217z_f-106z_f^2)+
2\hat{x}(43+16z_f-79z_f^2+50z_f^3)]
\nonumber\\
&&+m_{J/\psi}^6Q^4\hat{x}^3[54-160z_f+346z_f^2-339z_f^3+144z_f^4
-\hat{x}z_f(163+187z_f-454z_f^2+256z_f^3)
\nonumber\\
&&+\hat{x}^2(22+284z_f-294z_f^2+44z_f^3+64z_f^4)]
+m_{J/\psi}^2Q^8\hat{x}z_f[12+75z_f-268z_f^2+312z_f^3-166z_f^4+40z_f^5
\nonumber\\
&&+2\hat{x}^4(-1+77z_f-78z_f^2+32z_f^3)
+\hat{x}^3(7-137z_f-150z_f^2+256z_f^3-128z_f^4)
\nonumber\\
&&+\hat{x}(1-184z_f+487z_f^2-602z_f^3+354z_f^4-104z_f^5)
+\hat{x}^2(-18+92z_f+87z_f^2-30z_f^3-60z_f^4+64z_f^5)]
\nonumber\\
&&+m_{J/\psi}^4Q^6\hat{x}^2[z_f(-153+443z_f-562z_f^2+362z_f^3-114z_f^4)
+2\hat{x}^3(-7+88z_f+21z_f^2-106z_f^3+64z_f^4)
\nonumber\\
&&-\hat{x}^2z_f(7+527z_f-434z_f^2+128z_f^4)
+\hat{x}(-10+142z_f-263z_f^2+608z_f^3-602z_f^4+260z_f^5)]\Bigr)
\eeq
\beq
\hat{\sigma}^{N2}_4&=&{-1\over z_f^2Q^2[Q^2(-1+
\hat{x})+m_{J/\psi}^2\hat{x}]^3[m_{J/\psi}^2\hat{x}+Q^2(1+\hat{x}-z_f)]^3}
\nonumber\\
&&\times 128m_{J/\psi}^4P_T\hat{x}^3\Bigl(6m_{J/\psi}^8\hat{x}^4(-3+z_f)
+2m_{J/\psi}^6Q^2\hat{x}^3[2(11-3z_f)z_f+\hat{x}(-11-31z_f+18z_f^2)]
\nonumber\\
&&+m_{J/\psi}^4Q^4\hat{x}^2[-22+40z_f-67z_f^2+18z_f^3
+8\hat{x}z_f(2+21z_f-11z_f^2)+2\hat{x}^2(5-63z_f+6z_f^2+16z_f^3)]
\nonumber\\
&&+Q^8z_f[4\hat{x}^4(4-15z_f+8z_f^2)+3z_f(-11+23z_f-16z_f^2+4z_f^3)
-8\hat{x}^3(1-13z_f^2+8z_f^3)
\nonumber\\
&&+\hat{x}(-8+88z_f-153z_f^2+124z_f^3-40z_f^4)+\hat{x}^2(16-43z_f-36z_f^3+32z_f^4)]
\nonumber\\
&&+m_{J/\psi}^2Q^6\hat{x}[z_f(58-112z_f+89z_f^2-24z_f^3)
-4\hat{x}^2z_f(9-45z_f-4z_f^2+16z_f^3)
\nonumber\\
&&+2\hat{x}^3(7-21z_f-42z_f^2+32z_f^3)
+2\hat{x}(5-36z_f+57z_f^2-103z_f^3+46z_f^4)]\Bigr)
\eeq
\beq
\hat{\sigma}^{N2}_8&=&{-1\over z_f^3Q[Q^2(-1+\hat{x})+m_{J/\psi}^2\hat{x}]^3[m_{J/\psi}^2\hat{x}+
Q^2(1+\hat{x}-z_f)]^3}
\nonumber\\
&&128m_{J/\psi}^2\hat{x}^2\Bigl(-Q^{10}(-1+\hat{x})^2z_f^2(3+2\hat{x}^2+\hat{x}(5-4z_f)-
5z_f+2z_f^2)+2m_{J/\psi}^{10}\hat{x}^5(1-3z_f+2z_f^2)
\nonumber\\
&&+m_{J/\psi}^6Q^4\hat{x}^3[6-24z_f+50z_f^2-53z_f^3+24z_f^4
+\hat{x}z_f(-21+27z_f+2z_f^2-16z_f^3)
\nonumber\\
&&+6\hat{x}^2(1-2z_f-z_f^2+2z_f^3)]
+m_{J/\psi}^8Q^2\hat{x}^4[z_f(-11+23z_f-14z_f^2)
+2\hat{x}(3-8z_f+3z_f^2+2z_f^3)]
\nonumber\\
&&+m_{J/\psi}^2Q^8\hat{x}z_f[\hat{x}^4(2-6z_f+4z_f^2)
+\hat{x}^3(1-23z_f+30z_f^2-16z_f^3)
\nonumber\\
&&+z_f(17-56z_f+64z_f^2-34z_f^3+8z_f^4)
+\hat{x}^2(6-24z_f+53z_f^2-46z_f^3+20z_f^4)
\nonumber\\
&&-\hat{x}(9-36z_f+31z_f^2+2z_f^3-14z_f^4+8z_f^5)]
+m_{J/\psi}^4Q^6\hat{x}^2[2\hat{x}^3(1-7z_f^2+6z_f^3)
-\hat{x}^2z_f(9+17z_f-46z_f^2+32z_f^3)
\nonumber\\
&&+z_f(-19+73z_f-102z_f^2+70z_f^3-22z_f^4)+
\hat{x}(6-18z_f+27z_f^2-4z_f^3-22z_f^4+20z_f^5)]\Bigr)
\eeq
\beq
\hat{\sigma}^{N2}_9&=&{-1\over z_f^2Q^2[Q^2(-1+\hat{x})+
m_{J/\psi}^2\hat{x}]^3[m_{J/\psi}^2\hat{x}+Q^2(1+\hat{x}-z_f)]^3}
\nonumber\\
&&\times 128m_{J/\psi}^4P_T\hat{x}^3\Bigl(2m_{J/\psi}^8\hat{x}^4(-1+z_f)+
2m_{J/\psi}^6Q^2\hat{x}^3[-2(-2+z_f)z_f+\hat{x}(-3+z_f+2z_f^2)]
\nonumber\\
&&+m_{J/\psi}^4Q^4\hat{x}^2[-6+12z_f-17z_f^2+6z_f^3
+6\hat{x}^2(-1-z_f+2z_f^2)-4\hat{x}z_f(-3-4z_f+4z_f^2)]
\nonumber\\
&&+Q^8z_f[4\hat{x}^4(-1+z_f)-4\hat{x}^3(1-6z_f+4z_f^2)
+z_f(-11+23z_f-16z_f^2+4z_f^3)
\nonumber\\
&&+\hat{x}^2(-12+31z_f-44z_f^2+20z_f^3)
+\hat{x}(4-15z_f^2+20z_f^3-8z_f^4)]
\nonumber\\
&&+m_{J/\psi}^2Q^6\hat{x}[4\hat{x}^2(11-8z_f)z_f^2+2\hat{x}^3(-1-5z_f+6z_f^2)
+z_f(14-32z_f+27z_f^2-8z_f^3)
\nonumber\\
&&+2\hat{x}(-3+7z_f^2-19z_f^3+10z_f^4)]\Bigr)
\eeq

\

\noindent
(4) Hard cross sections of $N(x,A x)$ in (\ref{polarized})

\beq
\sigma^{N3}_1&=&
{-1\over (1-z_f)z_f^2[Q^2(-1+\hat{x})+\hat{x}m_{J/\psi}^2]^3[Q^2(1+\hat{x}-z_f)+\hat{x}m_{J/\psi}^2]^3}
\nonumber\\
&&\times 128Q^2P_T\hat{x}^3m_{J/\psi}^2\Bigl(4Q^6(-1+\hat{x})(-2+z_f)z_f[-z_f+
\hat{x}(-1+2z_f)]
\nonumber\\
&&+m_{J/\psi}^2Q^4[-8+25z_f-50z_f^2+53z_f^3-32z_f^4+8z_f^5
+\hat{x}^2(-4+103z_f-365z_f^2+446z_f^3-240z_f^4+48z_f^5)
\nonumber\\
&&-\hat{x}(4+48z_f-275z_f^2+383z_f^3-224z_f^4+48z_f^5)]
\nonumber\\
&&+Q^2m_{J/\psi}^4\hat{x}[-2+50z_f-83z_f^2+59z_f^3-16z_f^4+
\hat{x}(70-300z_f+396z_f^2-226z_f^3+48z_f^4)]
\nonumber\\
&&+m_{J/\psi}^6\hat{x}^2(-22+37z_f-27z_f^2+8z_f^3)\Bigr)
\eeq
\beq
\sigma^{N3}_2&=&
{-1\over (1-z_f)z_f^2[Q^2(-1+\hat{x})+\hat{x}m_{J/\psi}^2]^3[Q^2(1+\hat{x}-z_f)+\hat{x}m_{J/\psi}^2]^3}
\nonumber\\
&&512Q^4P_T\hat{x}^3(-2+z_f)m_{J/\psi}^2\Bigl(Q^4(-1+\hat{x})^2z_f^2+
2m_{J/\psi}^2Q^2(-1+\hat{x})\hat{x}z_f(-2+11z_f-12z_f^2+4z_f^3)
\nonumber\\
&&+m_{J/\psi}^4\hat{x}^2(-4+20z_f-23z_f^2+8z_f^3)\Bigr)
\eeq
\beq
\sigma^{N3}_3&=&{1\over z_f^3[Q^2(-1+\hat{x})+\hat{x}m_{J/\psi}^2]^3[Q^2(1+\hat{x}-z_f)+
\hat{x}m_{J/\psi}^2]^3}
\nonumber\\
&&\times 256Q^3\hat{x}^3m_{J/\psi}^2\Bigl(2Q^6(-1+\hat{x})^2(-2+z_f)z_f^2+
m_{J/\psi}^2Q^4z_f[-1-18z_f+41z_f^2-32z_f^3+8z_f^4
\nonumber\\
&&-2\hat{x}(1-41z_f+76z_f^2-52z_f^3+12z_f^4)+
\hat{x}^2(3-64z_f+111z_f^2-72z_f^3+16z_f^4)]
\nonumber\\
&&+m_{J/\psi}^4Q^2\hat{x}[1+43z_f-91z_f^2+67z_f^3-16z_f^4+
\hat{x}(7-91z_f+161z_f^2-107z_f^3+24z_f^4)]
\nonumber\\
&&+m_{J/\psi}^6\hat{x}^2(-25+50z_f-35z_f^2+8z_f^3)\Bigr)
\eeq
\beq
\sigma^{N3}_4&=&{1\over z_f^2[Q^2(-1+\hat{x})+
\hat{x}m_{J/\psi}^2]^3[Q^2(1+\hat{x}-z_f)+\hat{x}m_{J/\psi}^2]^3}
\nonumber\\
&&\times 128Q^2P_T\hat{x}^3m_{J/\psi}^4\Bigl(Q^4z_f[3-3z_f+
\hat{x}(12-67z_f+64z_f^2-16z_f^3)+
\hat{x}^2(-15+70z_f-64z_f^2+16z_f^3)]
\nonumber\\
&&+m_{J/\psi}^2Q^2\hat{x}[-2-16z_f+11z_f^2+
2\hat{x}(-7+43z_f-37z_f^2+8z_f^3)]+
m_{J/\psi}^4\hat{x}^2(18-11z_f)\Bigr)
\eeq
\beq
\sigma^{N3}_8&=&-{256Q^3\hat{x}^3(1-z_f)^2m_{J/\psi}^4[Q^2(-1+\hat{x})z_f+\hat{x}m_{J/\psi}^2]
\over z_f^3[Q^2(-1+\hat{x})+\hat{x}m_{J/\psi}^2]^2[Q^2(1+\hat{x}-z_f)+\hat{x}m_{J/\psi}^2]^3}
\eeq
\beq
\sigma^{N3}_9&=&-{128Q^2P_T\hat{x}^3m_{J/\psi}^4\Bigl(Q^2z_f[1-z_f+\hat{x}(-3+2z_f)]+
m_{J/\psi}^2\hat{x}(-2+z_f)\Bigr)
\over z_f^2[Q^2(-1+\hat{x})+\hat{x}m_{J/\psi}^2]^2[Q^2(1+\hat{x}-z_f)+\hat{x}m_{J/\psi}^2]^3}
\eeq

\

\noindent
(5) Hard cross sections of $N(x,(1-A)x)$ in (\ref{polarized})

\beq
\sigma^{N4}_1&=&{1\over (1-
z_f)z_f^2[Q^2(-1+\hat{x})+\hat{x}m_{J/\psi}^2]^3[Q^2(1+\hat{x}-z_f)+\hat{x}m_{J/\psi}^2]^3}
\nonumber\\
&&\times 128Q^2P_T\hat{x}^3m_{J/\psi}^2\Bigr(4Q^6(-1+\hat{x})z_f[-1+6z_f-3z_f^2+
\hat{x}(3-9z_f+4z_f^2)]+
m_{J/\psi}^2Q^4[-8+15z_f-34z_f^2+37z_f^3
\nonumber\\
&&-24z_f^4+6z_f^5+\hat{x}(4-80z_f+361z_f^2-449z_f^3+248z_f^4-52z_f^5)
\nonumber\\
&&+\hat{x}^2(-4+101z_f-379z_f^2+446z_f^3-236z_f^4+48z_f^5)]
\nonumber\\
&&+m_{J/\psi}^4Q^2\hat{x}[2+46z_f-69z_f^2+49z_f^3-12z_f^4+
\hat{x}(70-310z_f+400z_f^2-236z_f^3+52z_f^4)]
\nonumber\\
&&+m_{J/\psi}^6\hat{x}^2(-22+33z_f-25z_f^2+6z_f^3)\Bigr)
\eeq
\beq
\sigma^{N4}_2&=&{1\over (1-
z_f)z_f^2[Q^2(-1+\hat{x})+\hat{x}m_{J/\psi}^2]^3[Q^2(1+\hat{x}-z_f)+\hat{x}m_{J/\psi}^2]^3}
\nonumber\\
&&\times 1024Q^4P_T\hat{x}^3(-2+z_f)m_{J/\psi}^2\Bigl(Q^4(-1+\hat{x})^2z_f^2+
2m_{J/\psi}^2Q^2(-1+\hat{x})\hat{x}z_f(-1+6z_f-6z_f^2+2z_f^3)
\nonumber\\
&&+m_{J/\psi}^4\hat{x}^2(-2+10z_f-11z_f^2+4z_f^3)\Bigr)
\eeq
\beq
\sigma^{N4}_3&=&{-1\over z_f^3[Q^2(-1+\hat{x})+
\hat{x}m_{J/\psi}^2]^3[Q^2(1+\hat{x}-z_f)+\hat{x}m_{J/\psi}^2]^3}
\nonumber\\
&&\times 128Q^3\hat{x}^2m_{J/\psi}^2\Bigl(Q^6(-1+\hat{x})^2z_f^2[3-3z_f+\hat{x}(-13+8z_f)]
+m_{J/\psi}^2Q^4\hat{x}z_f[-2-55z_f+112z_f^2-80z_f^3+20z_f^4
\nonumber\\
&&-4\hat{x}z_f(-46+83z_f-55z_f^2+13z_f^3)+
\hat{x}^2(2-129z_f+220z_f^2-140z_f^3+32z_f^4)]
\nonumber\\
&&+m_{J/\psi}^4Q^2\hat{x}^2[-2+110z_f-215z_f^2+157z_f^3-40z_f^4+
\hat{x}(14-184z_f+319z_f^2-216z_f^3+52z_f^4)]
\nonumber\\
&&+m_{J/\psi}^6\hat{x}^3(-50+102z_f-77z_f^2+20z_f^3)\Bigr)
\eeq
\beq
\sigma^{N4}_4&=&{-1\over z_f^2(Q^2(-1+\hat{x})+
\hat{x}m_{J/\psi}^2)^3(Q^2(1+\hat{x}-z_f)+\hat{x}m_{J/\psi}^2)^3}
\nonumber\\
&&\times 128Q^2P_T\hat{x}^3m_{J/\psi}^4\Bigl(Q^4z_f[\hat{x}(12-65z_f+68z_f^2-20z_f^3)
\nonumber\\
&&+3(-1+5z_f-6z_f^2+2z_f^3)
+\hat{x}^2(-17+66z_f-60z_f^2+16z_f^3)]
\nonumber\\
&&+m_{J/\psi}^2Q^2\hat{x}[2-32z_f+37z_f^2-12z_f^3+
2\hat{x}(-7+38z_f-36z_f^2+10z_f^3)]
+m_{J/\psi}^4\hat{x}^2(18-19z_f+6z_f^2)\Bigr)\hspace{3mm}
\eeq
\beq
\sigma^{N4}_8&=&{-1\over 
z_f^3(Q^2(-1+\hat{x})+\hat{x}m_{J/\psi}^2)^3(Q^2(1+\hat{x}-z_f)+\hat{x}m_{J/\psi}^2)^2}
\nonumber\\
&&\times 128Q^3\hat{x}^2m_{J/\psi}^2\Bigl(Q^4(-1+\hat{x})^2z_f^2+
2m_{J/\psi}^2Q^2(-1+\hat{x})\hat{x}z_f(-1+6z_f-6z_f^2+2z_f^3)
\nonumber\\
&&+m_{J/\psi}^4\hat{x}^2(-2+10z_f-11z_f^2+4z_f^3)\Bigr)
\eeq
\beq
\sigma^{N4}_9&=&-{128Q^2P_T\hat{x}^3m_{J/\psi}^4\Bigr(Q^2z_f[-1+4z_f-2z_f^2+
\hat{x}(5-10z_f+4z_f^2)]+m_{J/\psi}^2\hat{x}(2-5z_f+2z_f^2)\Bigr)\over 
z_f^2[Q^2(-1+\hat{x})+\hat{x}m_{J/\psi}^2]^3[Q^2(1+\hat{x}-z_f)+\hat{x}m_{J/\psi}^2]^2}
\eeq

\

\noindent
(6) Hard cross sections of $N(A x,-(1-A)x)$ in (\ref{polarized})

\beq
\sigma^{N5}_1&=&{1\over (1-
z_f)z_f^2[Q^2(-1+\hat{x})+\hat{x}m_{J/\psi}^2]^3[Q^2(1+\hat{x}-z_f)+\hat{x}m_{J/\psi}^2]^3}
\nonumber\\
&&\times 256Q^2P_T\hat{x}^3m_{J/\psi}^2\Bigl(2Q^6(-1+\hat{x})^2z_f(1-4z_f+2z_f^2)
+m_{J/\psi}^2Q^4(z_f^2(-10+17z_f-12z_f^2+3z_f^3)
\nonumber\\
&&-2\hat{x}z_f(16-79z_f+108z_f^2-62z_f^3+13z_f^4)
+2\hat{x}^2(-2+23z_f-84z_f^2+107z_f^3-59z_f^4+12z_f^5))
\nonumber\\
&&+m_{J/\psi}^4Q^2\hat{x}(2z_f(12-20z_f+13z_f^2-3z_f^3)
+\hat{x}(30-149z_f+210z_f^2-123z_f^3+26z_f^4))
\nonumber\\
&&+m_{J/\psi}^6\hat{x}^2(-14+23z_f-14z_f^2+3z_f^3)\Bigr)
\eeq
\beq
\sigma^{N5}_2&=&{1\over (1-
z_f)z_f^2[Q^2(-1+\hat{x})+\hat{x}m_{J/\psi}^2]^3[Q^2(1+\hat{x}-z_f)+\hat{x}m_{J/\psi}^2]^3}
\nonumber\\
&&\times 512Q^4P_T\hat{x}^3(-2+z_f)m_{J/\psi}^2\Bigl(Q^4(-1+\hat{x})^2z_f^2
+2m_{J/\psi}^2Q^2(-1+\hat{x})\hat{x}z_f(-2+11z_f-12z_f^2+4z_f^3)
\nonumber\\
&&+m_{J/\psi}^4\hat{x}^2(-4+20z_f-23z_f^2+8z_f^3)\Bigr)
\eeq
\beq
\sigma^{N5}_3&=&{-1\over z_f^3[Q^2(-1+\hat{x})+
\hat{x}m_{J/\psi}^2]^3[Q^2(1+\hat{x}-z_f)+\hat{x}m_{J/\psi}^2]^3}
\nonumber\\
&&\times 128Q^3\hat{x}^2m_{J/\psi}^2\Bigl(Q^6(-1+\hat{x})^2z_f^2[3-3z_f+\hat{x}(-5+4z_f)]
+m_{J/\psi}^2Q^4\hat{x}z_f[-4-43z_f+106z_f^2-80z_f^3+20z_f^4
\nonumber\\
&&-4\hat{x}(1-39z_f+79z_f^2-55z_f^3+13z_f^4)
+\hat{x}^2(8-113z_f+210z_f^2-140z_f^3+32z_f^4)]
\nonumber\\
&&+m_{J/\psi}^4Q^2\hat{x}^2[z_f(100-221z_f+163z_f^2-40z_f^3)
+\hat{x}(12-166z_f+321z_f^2-222z_f^3+52z_f^4)]
\nonumber\\
&&+m_{J/\psi}^6\hat{x}^3(-52+114z_f-83z_f^2+20z_f^3)\Bigr)
\eeq
\beq
\sigma^{N5}_4&=&{-1\over z_f^2(Q^2(-1+\hat{x})+\hat{x}m_{J/\psi}^2)^3(Q^2(1+\hat{x}-z_f)
+\hat{x}m_{J/\psi}^2)^3}
\nonumber\\
&&\times 256Q^2P_T\hat{x}^3(-2+
z_f)m_{J/\psi}^4\Bigl(Q^4z_f[3(-1+z_f)z_f+2\hat{x}^2(2-7z_f+4z_f^2)
-2\hat{x}(1-7z_f+5z_f^2)]
\nonumber\\
&&+m_{J/\psi}^2Q^2\hat{x}[2(4-3z_f)z_f+\hat{x}(3-17z_f+10z_f^2)]
+m_{J/\psi}^4\hat{x}^2(-5+3z_f)\Bigr)
\eeq
\beq
\sigma^{N5}_8&=&{-1\over z_f^3(Q^2(-1+\hat{x})+
\hat{x}m_{J/\psi}^2)^3(Q^2(1+\hat{x}-z_f)+\hat{x}m_{J/\psi}^2)^3}
\nonumber\\
&&\times 128Q^3\hat{x}^2m_{J/\psi}^2\Bigl(Q^6(-1+\hat{x})^2(1+\hat{x}-z_f)z_f^2
+m_{J/\psi}^2Q^4\hat{x}z_f[z_f(-9+22z_f-16z_f^2+4z_f^3)
\nonumber\\
&&+\hat{x}^2(-4+17z_f-14z_f^2+4z_f^3)
-4\hat{x}(-1+2z_f+2z_f^2-3z_f^3+z_f^4)]
\nonumber\\
&&+m_{J/\psi}^4Q^2\hat{x}^2[z_f(12-35z_f+29z_f^2-8z_f^3)+
\hat{x}(-4+10z_f+3z_f^2-10z_f^3+4z_f^4)]
\nonumber\\
&&+m_{J/\psi}^6\hat{x}^3(-4+14z_f-13z_f^2+4z_f^3)\Bigr)
\eeq
\beq
\sigma^{N5}_9&=&
{256Q^2P_T\hat{x}^3(-2+z_f)(1-
z_f)m_{J/\psi}^4\Bigl(Q^4z_f(2\hat{x}^2+z_f-2\hat{x}z_f)
+m_{J/\psi}^2Q^2\hat{x}(\hat{x}-2z_f+2\hat{x}z_f)+m_{J/\psi}^4\hat{x}^2\Bigr)\over 
z_f^2[Q^2(-1+\hat{x})+\hat{x}m_{J/\psi}^2]^3[Q^2(1+\hat{x}-z_f)+\hat{x}m_{J/\psi}^2]^3}
\hspace{7mm}
\eeq

%--------------------- Acknowledgements ----------

\section*{Acknowledgements}

This work is supported by the National Natural Science Foundation of China under Grants No.\ 12022512 and No.\ 12035007, by the Guangdong Major Project of Basic and Applied Basic Research No.\ 2020B030103000, No. 2022A1515010683 and No. 2020A1515010794 and research startup funding at South China
Normal University.

\end{document}

%% file: F1_model1.tex
\scalebox{0.65}[0.65]{
\begin{tikzpicture}[gnuplot]
%% generated with GNUPLOT 5.4p5 (Lua 5.4; terminal rev. Jun 2020, script rev. 115)
%% 6/20/2023 8:52:55 AM
\tikzset{every node/.append style={font={\fontsize{15.0pt}{18.0pt}\selectfont}}}
\path (0.000,0.000) rectangle (12.500,8.750);
\gpcolor{color=gp lt color border}
\gpsetlinetype{gp lt border}
\gpsetdashtype{gp dt solid}
\gpsetlinewidth{1.00}
\draw[gp path] (2.256,1.478)--(2.436,1.478);
\draw[gp path] (11.671,1.478)--(11.491,1.478);
\node[gp node right,font={\fontsize{15.0pt}{18.0pt}\selectfont}] at (1.980,1.478) {$-0.2$};
\draw[gp path] (2.256,3.180)--(2.436,3.180);
\draw[gp path] (11.671,3.180)--(11.491,3.180);
\node[gp node right,font={\fontsize{15.0pt}{18.0pt}\selectfont}] at (1.980,3.180) {$-0.1$};
\draw[gp path] (2.256,4.883)--(2.436,4.883);
\draw[gp path] (11.671,4.883)--(11.491,4.883);
\node[gp node right,font={\fontsize{15.0pt}{18.0pt}\selectfont}] at (1.980,4.883) {$0$};
\draw[gp path] (2.256,6.585)--(2.436,6.585);
\draw[gp path] (11.671,6.585)--(11.491,6.585);
\node[gp node right,font={\fontsize{15.0pt}{18.0pt}\selectfont}] at (1.980,6.585) {$0.1$};
\draw[gp path] (2.256,8.287)--(2.436,8.287);
\draw[gp path] (11.671,8.287)--(11.491,8.287);
\node[gp node right,font={\fontsize{15.0pt}{18.0pt}\selectfont}] at (1.980,8.287) {$0.2$};
\draw[gp path] (2.256,1.478)--(2.256,1.658);
\draw[gp path] (2.256,8.287)--(2.256,8.107);
\node[gp node center,font={\fontsize{15.0pt}{18.0pt}\selectfont}] at (2.256,1.016) {$0.2$};
\draw[gp path] (3.825,1.478)--(3.825,1.658);
\draw[gp path] (3.825,8.287)--(3.825,8.107);
\node[gp node center,font={\fontsize{15.0pt}{18.0pt}\selectfont}] at (3.825,1.016) {$0.3$};
\draw[gp path] (5.394,1.478)--(5.394,1.658);
\draw[gp path] (5.394,8.287)--(5.394,8.107);
\node[gp node center,font={\fontsize{15.0pt}{18.0pt}\selectfont}] at (5.394,1.016) {$0.4$};
\draw[gp path] (6.963,1.478)--(6.963,1.658);
\draw[gp path] (6.963,8.287)--(6.963,8.107);
\node[gp node center,font={\fontsize{15.0pt}{18.0pt}\selectfont}] at (6.963,1.016) {$0.5$};
\draw[gp path] (8.533,1.478)--(8.533,1.658);
\draw[gp path] (8.533,8.287)--(8.533,8.107);
\node[gp node center,font={\fontsize{15.0pt}{18.0pt}\selectfont}] at (8.533,1.016) {$0.6$};
\draw[gp path] (10.102,1.478)--(10.102,1.658);
\draw[gp path] (10.102,8.287)--(10.102,8.107);
\node[gp node center,font={\fontsize{15.0pt}{18.0pt}\selectfont}] at (10.102,1.016) {$0.7$};
\draw[gp path] (11.671,1.478)--(11.671,1.658);
\draw[gp path] (11.671,8.287)--(11.671,8.107);
\node[gp node center,font={\fontsize{15.0pt}{18.0pt}\selectfont}] at (11.671,1.016) {$0.8$};
\draw[gp path] (2.256,8.287)--(2.256,1.478)--(11.671,1.478)--(11.671,8.287)--cycle;
\node[gp node left,font={\fontsize{13.0pt}{15.6pt}\selectfont}] at (3.041,3.180) {$\sqrt{S_{ep}}=45$ GeV};
\node[gp node left,font={\fontsize{13.0pt}{15.6pt}\selectfont}] at (3.041,2.329) {$Q^2=10$ GeV${}^2$};
\node[gp node left,font={\fontsize{13.0pt}{15.6pt}\selectfont}] at (6.179,3.180) {$x_{B}=0.005$};
\node[gp node left,font={\fontsize{13.0pt}{15.6pt}\selectfont}] at (6.179,2.329) {$P_{J/\psi}^{\perp}=2$ GeV};
\node[gp node left,font={\fontsize{20.0pt}{24.0pt}\selectfont}] at (6.179,6.925) {Model 1};
\node[gp node center,rotate=-270,font={\fontsize{18.0pt}{21.6pt}\selectfont}] at (0.438,4.882) {${\cal F}_1/\sigma^{{\rm U}}_1$};
\node[gp node center,font={\fontsize{22.0pt}{26.4pt}\selectfont}] at (6.963,0.323) {$z_f$};
\node[gp node right,font={\fontsize{15.0pt}{18.0pt}\selectfont}] at (4.920,7.876) {$N^1$};
\gpcolor{rgb color={0.580,0.000,0.827}}
\gpsetdashtype{gp dt 1}
\gpsetlinewidth{3.00}
\draw[gp path] (2.808,7.876)--(4.092,7.876);
\draw[gp path] (2.256,5.114)--(2.413,5.115)--(2.570,5.116)--(2.727,5.118)--(2.884,5.119)%
  --(3.041,5.121)--(3.198,5.123)--(3.354,5.125)--(3.511,5.128)--(3.668,5.131)--(3.825,5.133)%
  --(3.982,5.137)--(4.139,5.140)--(4.296,5.143)--(4.453,5.147)--(4.610,5.152)--(4.767,5.156)%
  --(4.924,5.161)--(5.081,5.166)--(5.237,5.171)--(5.394,5.177)--(5.551,5.184)--(5.708,5.190)%
  --(5.865,5.197)--(6.022,5.205)--(6.179,5.213)--(6.336,5.222)--(6.493,5.231)--(6.650,5.241)%
  --(6.807,5.252)--(6.963,5.263)--(7.120,5.275)--(7.277,5.287)--(7.434,5.301)--(7.591,5.315)%
  --(7.748,5.331)--(7.905,5.347)--(8.062,5.365)--(8.219,5.383)--(8.376,5.403)--(8.533,5.424)%
  --(8.690,5.446)--(8.847,5.470)--(9.003,5.496)--(9.160,5.523)--(9.317,5.552)--(9.474,5.582)%
  --(9.631,5.615)--(9.788,5.650)--(9.945,5.687)--(10.102,5.727)--(10.259,5.770)--(10.416,5.815)%
  --(10.573,5.863)--(10.730,5.915)--(10.886,5.970)--(11.043,6.030)--(11.200,6.093)--(11.357,6.162)%
  --(11.514,6.236)--(11.671,6.316);
\gpcolor{color=gp lt color border}
\node[gp node right,font={\fontsize{15.0pt}{18.0pt}\selectfont}] at (4.920,7.414) {$N^2$};
\gpcolor{rgb color={0.000,0.620,0.451}}
\gpsetdashtype{gp dt 2}
\draw[gp path] (2.808,7.414)--(4.092,7.414);
\draw[gp path] (2.256,5.184)--(2.413,5.182)--(2.570,5.179)--(2.727,5.177)--(2.884,5.173)%
  --(3.041,5.170)--(3.198,5.165)--(3.354,5.161)--(3.511,5.156)--(3.668,5.150)--(3.825,5.144)%
  --(3.982,5.137)--(4.139,5.130)--(4.296,5.122)--(4.453,5.114)--(4.610,5.105)--(4.767,5.096)%
  --(4.924,5.086)--(5.081,5.075)--(5.237,5.064)--(5.394,5.051)--(5.551,5.038)--(5.708,5.025)%
  --(5.865,5.010)--(6.022,4.995)--(6.179,4.978)--(6.336,4.961)--(6.493,4.942)--(6.650,4.923)%
  --(6.807,4.902)--(6.963,4.880)--(7.120,4.857)--(7.277,4.832)--(7.434,4.806)--(7.591,4.779)%
  --(7.748,4.749)--(7.905,4.719)--(8.062,4.686)--(8.219,4.652)--(8.376,4.615)--(8.533,4.577)%
  --(8.690,4.536)--(8.847,4.493)--(9.003,4.448)--(9.160,4.400)--(9.317,4.350)--(9.474,4.297)%
  --(9.631,4.241)--(9.788,4.182)--(9.945,4.121)--(10.102,4.056)--(10.259,3.987)--(10.416,3.916)%
  --(10.573,3.841)--(10.730,3.762)--(10.886,3.679)--(11.043,3.592)--(11.200,3.501)--(11.357,3.405)%
  --(11.514,3.305)--(11.671,3.199);
\gpcolor{color=gp lt color border}
\node[gp node right,font={\fontsize{15.0pt}{18.0pt}\selectfont}] at (4.920,6.952) {$N^3$};
\gpcolor{rgb color={0.337,0.706,0.914}}
\gpsetdashtype{gp dt 3}
\draw[gp path] (2.808,6.952)--(4.092,6.952);
\draw[gp path] (2.256,4.333)--(2.413,4.329)--(2.570,4.325)--(2.727,4.322)--(2.884,4.318)%
  --(3.041,4.314)--(3.198,4.309)--(3.354,4.305)--(3.511,4.301)--(3.668,4.296)--(3.825,4.292)%
  --(3.982,4.287)--(4.139,4.282)--(4.296,4.278)--(4.453,4.273)--(4.610,4.267)--(4.767,4.262)%
  --(4.924,4.257)--(5.081,4.251)--(5.237,4.246)--(5.394,4.240)--(5.551,4.234)--(5.708,4.228)%
  --(5.865,4.221)--(6.022,4.215)--(6.179,4.208)--(6.336,4.201)--(6.493,4.194)--(6.650,4.186)%
  --(6.807,4.179)--(6.963,4.171)--(7.120,4.162)--(7.277,4.154)--(7.434,4.145)--(7.591,4.135)%
  --(7.748,4.125)--(7.905,4.115)--(8.062,4.104)--(8.219,4.093)--(8.376,4.081)--(8.533,4.069)%
  --(8.690,4.055)--(8.847,4.042)--(9.003,4.027)--(9.160,4.012)--(9.317,3.996)--(9.474,3.978)%
  --(9.631,3.960)--(9.788,3.941)--(9.945,3.921)--(10.102,3.899)--(10.259,3.876)--(10.416,3.851)%
  --(10.573,3.825)--(10.730,3.797)--(10.886,3.767)--(11.043,3.735)--(11.200,3.700)--(11.357,3.663)%
  --(11.514,3.623)--(11.671,3.579);
\gpcolor{color=gp lt color border}
\node[gp node right,font={\fontsize{15.0pt}{18.0pt}\selectfont}] at (4.920,6.490) {$N^4$};
\gpcolor{rgb color={0.902,0.624,0.000}}
\gpsetdashtype{gp dt 4}
\draw[gp path] (2.808,6.490)--(4.092,6.490);
\draw[gp path] (2.256,5.452)--(2.413,5.460)--(2.570,5.468)--(2.727,5.477)--(2.884,5.486)%
  --(3.041,5.496)--(3.198,5.506)--(3.354,5.516)--(3.511,5.527)--(3.668,5.538)--(3.825,5.549)%
  --(3.982,5.561)--(4.139,5.574)--(4.296,5.587)--(4.453,5.600)--(4.610,5.614)--(4.767,5.629)%
  --(4.924,5.644)--(5.081,5.659)--(5.237,5.675)--(5.394,5.692)--(5.551,5.709)--(5.708,5.727)%
  --(5.865,5.745)--(6.022,5.765)--(6.179,5.784)--(6.336,5.805)--(6.493,5.826)--(6.650,5.848)%
  --(6.807,5.871)--(6.963,5.894)--(7.120,5.919)--(7.277,5.944)--(7.434,5.970)--(7.591,5.998)%
  --(7.748,6.026)--(7.905,6.055)--(8.062,6.086)--(8.219,6.118)--(8.376,6.151)--(8.533,6.185)%
  --(8.690,6.221)--(8.847,6.259)--(9.003,6.298)--(9.160,6.339)--(9.317,6.382)--(9.474,6.426)%
  --(9.631,6.473)--(9.788,6.522)--(9.945,6.574)--(10.102,6.628)--(10.259,6.685)--(10.416,6.745)%
  --(10.573,6.809)--(10.730,6.876)--(10.886,6.946)--(11.043,7.022)--(11.200,7.101)--(11.357,7.187)%
  --(11.514,7.277)--(11.671,7.375);
\gpcolor{color=gp lt color border}
\node[gp node right,font={\fontsize{15.0pt}{18.0pt}\selectfont}] at (4.920,6.028) {$N^5$};
\gpcolor{rgb color={0.941,0.894,0.259}}
\gpsetdashtype{gp dt 5}
\draw[gp path] (2.808,6.028)--(4.092,6.028);
\draw[gp path] (2.256,4.861)--(2.413,4.861)--(2.570,4.862)--(2.727,4.863)--(2.884,4.863)%
  --(3.041,4.864)--(3.198,4.865)--(3.354,4.866)--(3.511,4.868)--(3.668,4.869)--(3.825,4.871)%
  --(3.982,4.872)--(4.139,4.874)--(4.296,4.876)--(4.453,4.879)--(4.610,4.881)--(4.767,4.884)%
  --(4.924,4.887)--(5.081,4.890)--(5.237,4.893)--(5.394,4.897)--(5.551,4.901)--(5.708,4.905)%
  --(5.865,4.910)--(6.022,4.915)--(6.179,4.920)--(6.336,4.926)--(6.493,4.932)--(6.650,4.938)%
  --(6.807,4.945)--(6.963,4.953)--(7.120,4.961)--(7.277,4.970)--(7.434,4.980)--(7.591,4.990)%
  --(7.748,5.001)--(7.905,5.013)--(8.062,5.026)--(8.219,5.039)--(8.376,5.054)--(8.533,5.070)%
  --(8.690,5.087)--(8.847,5.105)--(9.003,5.125)--(9.160,5.146)--(9.317,5.169)--(9.474,5.193)%
  --(9.631,5.219)--(9.788,5.247)--(9.945,5.277)--(10.102,5.310)--(10.259,5.344)--(10.416,5.381)%
  --(10.573,5.421)--(10.730,5.463)--(10.886,5.508)--(11.043,5.557)--(11.200,5.609)--(11.357,5.664)%
  --(11.514,5.723)--(11.671,5.787);
\gpcolor{color=gp lt color border}
\gpsetdashtype{gp dt solid}
\gpsetlinewidth{1.00}
\draw[gp path] (2.256,8.287)--(2.256,1.478)--(11.671,1.478)--(11.671,8.287)--cycle;
%% coordinates of the plot area
\gpdefrectangularnode{gp plot 1}{\pgfpoint{2.256cm}{1.478cm}}{\pgfpoint{11.671cm}{8.287cm}}
\end{tikzpicture}
}
%% gnuplot variables

%% file: F2_model1.tex
\scalebox{0.65}[0.65]{
\begin{tikzpicture}[gnuplot]
%% generated with GNUPLOT 5.4p5 (Lua 5.4; terminal rev. Jun 2020, script rev. 115)
%% 6/20/2023 9:18:54 AM
\tikzset{every node/.append style={font={\fontsize{15.0pt}{18.0pt}\selectfont}}}
\path (0.000,0.000) rectangle (12.500,8.750);
\gpcolor{color=gp lt color border}
\gpsetlinetype{gp lt border}
\gpsetdashtype{gp dt solid}
\gpsetlinewidth{1.00}
\draw[gp path] (2.808,1.478)--(2.988,1.478);
\draw[gp path] (11.671,1.478)--(11.491,1.478);
\node[gp node right] at (2.532,1.478) {$-0.01$};
\draw[gp path] (2.808,3.180)--(2.988,3.180);
\draw[gp path] (11.671,3.180)--(11.491,3.180);
\node[gp node right] at (2.532,3.180) {$-0.005$};
\draw[gp path] (2.808,4.883)--(2.988,4.883);
\draw[gp path] (11.671,4.883)--(11.491,4.883);
\node[gp node right] at (2.532,4.883) {$0$};
\draw[gp path] (2.808,6.585)--(2.988,6.585);
\draw[gp path] (11.671,6.585)--(11.491,6.585);
\node[gp node right] at (2.532,6.585) {$0.005$};
\draw[gp path] (2.808,8.287)--(2.988,8.287);
\draw[gp path] (11.671,8.287)--(11.491,8.287);
\node[gp node right] at (2.532,8.287) {$0.01$};
\draw[gp path] (2.808,1.478)--(2.808,1.658);
\draw[gp path] (2.808,8.287)--(2.808,8.107);
\node[gp node center] at (2.808,1.016) {$0.2$};
\draw[gp path] (4.285,1.478)--(4.285,1.658);
\draw[gp path] (4.285,8.287)--(4.285,8.107);
\node[gp node center] at (4.285,1.016) {$0.3$};
\draw[gp path] (5.762,1.478)--(5.762,1.658);
\draw[gp path] (5.762,8.287)--(5.762,8.107);
\node[gp node center] at (5.762,1.016) {$0.4$};
\draw[gp path] (7.239,1.478)--(7.239,1.658);
\draw[gp path] (7.239,8.287)--(7.239,8.107);
\node[gp node center] at (7.239,1.016) {$0.5$};
\draw[gp path] (8.717,1.478)--(8.717,1.658);
\draw[gp path] (8.717,8.287)--(8.717,8.107);
\node[gp node center] at (8.717,1.016) {$0.6$};
\draw[gp path] (10.194,1.478)--(10.194,1.658);
\draw[gp path] (10.194,8.287)--(10.194,8.107);
\node[gp node center] at (10.194,1.016) {$0.7$};
\draw[gp path] (11.671,1.478)--(11.671,1.658);
\draw[gp path] (11.671,8.287)--(11.671,8.107);
\node[gp node center] at (11.671,1.016) {$0.8$};
\draw[gp path] (2.808,8.287)--(2.808,1.478)--(11.671,1.478)--(11.671,8.287)--cycle;
\node[gp node left,font={\fontsize{13.0pt}{15.6pt}\selectfont}] at (3.547,3.180) {$\sqrt{S_{ep}}=45$ GeV};
\node[gp node left,font={\fontsize{13.0pt}{15.6pt}\selectfont}] at (3.547,2.329) {$Q^2=10$ GeV${}^2$};
\node[gp node left,font={\fontsize{13.0pt}{15.6pt}\selectfont}] at (6.796,3.180) {$x_{B}=0.005$};
\node[gp node left,font={\fontsize{13.0pt}{15.6pt}\selectfont}] at (6.796,2.329) {$P_{J/\psi}^{\perp}=2$ GeV};
\node[gp node left,font={\fontsize{20.0pt}{24.0pt}\selectfont}] at (6.501,7.095) {Model 1};
\node[gp node center,rotate=-270,font={\fontsize{18.0pt}{21.6pt}\selectfont}] at (0.438,4.882) {${\cal F}_2/2\sigma^{\rm U}_1$};
\node[gp node center,font={\fontsize{22.0pt}{26.4pt}\selectfont}] at (7.239,0.323) {$z_f$};
\node[gp node right,font={\fontsize{15.0pt}{18.0pt}\selectfont}] at (5.472,7.876) {$N^1$};
\gpcolor{rgb color={0.580,0.000,0.827}}
\gpsetdashtype{gp dt 1}
\gpsetlinewidth{3.00}
\draw[gp path] (3.360,7.876)--(4.644,7.876);
\draw[gp path] (2.808,4.936)--(2.956,4.941)--(3.103,4.947)--(3.251,4.952)--(3.399,4.957)%
  --(3.547,4.963)--(3.694,4.968)--(3.842,4.974)--(3.990,4.980)--(4.137,4.986)--(4.285,4.992)%
  --(4.433,4.998)--(4.581,5.005)--(4.728,5.011)--(4.876,5.017)--(5.024,5.024)--(5.171,5.031)%
  --(5.319,5.037)--(5.467,5.044)--(5.615,5.051)--(5.762,5.058)--(5.910,5.065)--(6.058,5.072)%
  --(6.205,5.080)--(6.353,5.087)--(6.501,5.095)--(6.649,5.102)--(6.796,5.110)--(6.944,5.118)%
  --(7.092,5.126)--(7.239,5.134)--(7.387,5.142)--(7.535,5.151)--(7.683,5.159)--(7.830,5.168)%
  --(7.978,5.176)--(8.126,5.185)--(8.274,5.194)--(8.421,5.203)--(8.569,5.212)--(8.717,5.221)%
  --(8.864,5.231)--(9.012,5.240)--(9.160,5.250)--(9.308,5.260)--(9.455,5.270)--(9.603,5.280)%
  --(9.751,5.290)--(9.898,5.300)--(10.046,5.311)--(10.194,5.322)--(10.342,5.332)--(10.489,5.343)%
  --(10.637,5.355)--(10.785,5.366)--(10.932,5.378)--(11.080,5.390)--(11.228,5.402)--(11.376,5.415)%
  --(11.523,5.428)--(11.671,5.442);
\gpcolor{color=gp lt color border}
\node[gp node right,font={\fontsize{15.0pt}{18.0pt}\selectfont}] at (5.472,7.414) {$N^2$};
\gpcolor{rgb color={0.000,0.620,0.451}}
\gpsetdashtype{gp dt 2}
\draw[gp path] (3.360,7.414)--(4.644,7.414);
\draw[gp path] (2.808,4.802)--(2.956,4.795)--(3.103,4.787)--(3.251,4.779)--(3.399,4.771)%
  --(3.547,4.763)--(3.694,4.756)--(3.842,4.748)--(3.990,4.740)--(4.137,4.733)--(4.285,4.725)%
  --(4.433,4.718)--(4.581,4.711)--(4.728,4.704)--(4.876,4.697)--(5.024,4.690)--(5.171,4.684)%
  --(5.319,4.678)--(5.467,4.673)--(5.615,4.667)--(5.762,4.663)--(5.910,4.658)--(6.058,4.654)%
  --(6.205,4.651)--(6.353,4.648)--(6.501,4.645)--(6.649,4.643)--(6.796,4.642)--(6.944,4.642)%
  --(7.092,4.642)--(7.239,4.642)--(7.387,4.644)--(7.535,4.646)--(7.683,4.649)--(7.830,4.653)%
  --(7.978,4.657)--(8.126,4.663)--(8.274,4.669)--(8.421,4.676)--(8.569,4.684)--(8.717,4.693)%
  --(8.864,4.702)--(9.012,4.713)--(9.160,4.724)--(9.308,4.736)--(9.455,4.749)--(9.603,4.763)%
  --(9.751,4.777)--(9.898,4.792)--(10.046,4.807)--(10.194,4.822)--(10.342,4.838)--(10.489,4.854)%
  --(10.637,4.869)--(10.785,4.884)--(10.932,4.898)--(11.080,4.911)--(11.228,4.923)--(11.376,4.933)%
  --(11.523,4.941)--(11.671,4.946);
\gpcolor{color=gp lt color border}
\node[gp node right,font={\fontsize{15.0pt}{18.0pt}\selectfont}] at (5.472,6.952) {$N^3$};
\gpcolor{rgb color={0.337,0.706,0.914}}
\gpsetdashtype{gp dt 3}
\draw[gp path] (3.360,6.952)--(4.644,6.952);
\draw[gp path] (2.808,4.808)--(2.956,4.802)--(3.103,4.796)--(3.251,4.790)--(3.399,4.784)%
  --(3.547,4.778)--(3.694,4.771)--(3.842,4.764)--(3.990,4.757)--(4.137,4.750)--(4.285,4.742)%
  --(4.433,4.735)--(4.581,4.727)--(4.728,4.719)--(4.876,4.710)--(5.024,4.701)--(5.171,4.692)%
  --(5.319,4.682)--(5.467,4.672)--(5.615,4.662)--(5.762,4.651)--(5.910,4.640)--(6.058,4.628)%
  --(6.205,4.616)--(6.353,4.603)--(6.501,4.589)--(6.649,4.575)--(6.796,4.560)--(6.944,4.544)%
  --(7.092,4.528)--(7.239,4.511)--(7.387,4.493)--(7.535,4.474)--(7.683,4.454)--(7.830,4.433)%
  --(7.978,4.411)--(8.126,4.388)--(8.274,4.364)--(8.421,4.339)--(8.569,4.312)--(8.717,4.285)%
  --(8.864,4.256)--(9.012,4.226)--(9.160,4.195)--(9.308,4.163)--(9.455,4.129)--(9.603,4.095)%
  --(9.751,4.059)--(9.898,4.023)--(10.046,3.985)--(10.194,3.947)--(10.342,3.909)--(10.489,3.870)%
  --(10.637,3.831)--(10.785,3.792)--(10.932,3.754)--(11.080,3.716)--(11.228,3.680)--(11.376,3.644)%
  --(11.523,3.611)--(11.671,3.580);
\gpcolor{color=gp lt color border}
\node[gp node right,font={\fontsize{15.0pt}{18.0pt}\selectfont}] at (5.472,6.490) {$N^4$};
\gpcolor{rgb color={0.902,0.624,0.000}}
\gpsetdashtype{gp dt 4}
\draw[gp path] (3.360,6.490)--(4.644,6.490);
\draw[gp path] (2.808,4.930)--(2.956,4.936)--(3.103,4.943)--(3.251,4.950)--(3.399,4.957)%
  --(3.547,4.965)--(3.694,4.973)--(3.842,4.982)--(3.990,4.991)--(4.137,5.000)--(4.285,5.010)%
  --(4.433,5.020)--(4.581,5.031)--(4.728,5.042)--(4.876,5.054)--(5.024,5.066)--(5.171,5.079)%
  --(5.319,5.093)--(5.467,5.107)--(5.615,5.121)--(5.762,5.137)--(5.910,5.152)--(6.058,5.169)%
  --(6.205,5.186)--(6.353,5.204)--(6.501,5.223)--(6.649,5.243)--(6.796,5.263)--(6.944,5.285)%
  --(7.092,5.307)--(7.239,5.330)--(7.387,5.354)--(7.535,5.379)--(7.683,5.405)--(7.830,5.432)%
  --(7.978,5.461)--(8.126,5.490)--(8.274,5.520)--(8.421,5.552)--(8.569,5.584)--(8.717,5.618)%
  --(8.864,5.653)--(9.012,5.688)--(9.160,5.725)--(9.308,5.763)--(9.455,5.801)--(9.603,5.840)%
  --(9.751,5.880)--(9.898,5.921)--(10.046,5.962)--(10.194,6.003)--(10.342,6.045)--(10.489,6.086)%
  --(10.637,6.127)--(10.785,6.167)--(10.932,6.206)--(11.080,6.244)--(11.228,6.281)--(11.376,6.316)%
  --(11.523,6.348)--(11.671,6.378);
\gpcolor{color=gp lt color border}
\node[gp node right,font={\fontsize{15.0pt}{18.0pt}\selectfont}] at (5.472,6.028) {$N^5$};
\gpcolor{rgb color={0.941,0.894,0.259}}
\gpsetdashtype{gp dt 5}
\draw[gp path] (3.360,6.028)--(4.644,6.028);
\draw[gp path] (2.808,4.901)--(2.956,4.903)--(3.103,4.905)--(3.251,4.906)--(3.399,4.908)%
  --(3.547,4.910)--(3.694,4.911)--(3.842,4.913)--(3.990,4.914)--(4.137,4.916)--(4.285,4.917)%
  --(4.433,4.918)--(4.581,4.919)--(4.728,4.921)--(4.876,4.922)--(5.024,4.923)--(5.171,4.924)%
  --(5.319,4.924)--(5.467,4.925)--(5.615,4.926)--(5.762,4.926)--(5.910,4.927)--(6.058,4.927)%
  --(6.205,4.928)--(6.353,4.928)--(6.501,4.928)--(6.649,4.928)--(6.796,4.928)--(6.944,4.928)%
  --(7.092,4.928)--(7.239,4.928)--(7.387,4.928)--(7.535,4.928)--(7.683,4.928)--(7.830,4.928)%
  --(7.978,4.928)--(8.126,4.928)--(8.274,4.928)--(8.421,4.928)--(8.569,4.928)--(8.717,4.928)%
  --(8.864,4.929)--(9.012,4.929)--(9.160,4.930)--(9.308,4.931)--(9.455,4.932)--(9.603,4.934)%
  --(9.751,4.935)--(9.898,4.937)--(10.046,4.940)--(10.194,4.942)--(10.342,4.945)--(10.489,4.947)%
  --(10.637,4.951)--(10.785,4.954)--(10.932,4.957)--(11.080,4.961)--(11.228,4.965)--(11.376,4.969)%
  --(11.523,4.973)--(11.671,4.977);
\gpcolor{color=gp lt color border}
\gpsetdashtype{gp dt solid}
\gpsetlinewidth{1.00}
\draw[gp path] (2.808,8.287)--(2.808,1.478)--(11.671,1.478)--(11.671,8.287)--cycle;
%% coordinates of the plot area
\gpdefrectangularnode{gp plot 1}{\pgfpoint{2.808cm}{1.478cm}}{\pgfpoint{11.671cm}{8.287cm}}
\end{tikzpicture}
}
%% gnuplot variables

%% file: F3_model1.tex
\scalebox{0.65}[0.65]{
\begin{tikzpicture}[gnuplot]
%% generated with GNUPLOT 5.4p5 (Lua 5.4; terminal rev. Jun 2020, script rev. 115)
%% 6/20/2023 10:10:52 AM
\tikzset{every node/.append style={font={\fontsize{15.0pt}{18.0pt}\selectfont}}}
\path (0.000,0.000) rectangle (12.500,8.750);
\gpcolor{color=gp lt color border}
\gpsetlinetype{gp lt border}
\gpsetdashtype{gp dt solid}
\gpsetlinewidth{1.00}
\draw[gp path] (3.084,1.478)--(3.264,1.478);
\draw[gp path] (11.671,1.478)--(11.491,1.478);
\node[gp node right] at (2.808,1.478) {$-0.0004$};
\draw[gp path] (3.084,3.180)--(3.264,3.180);
\draw[gp path] (11.671,3.180)--(11.491,3.180);
\node[gp node right] at (2.808,3.180) {$-0.0002$};
\draw[gp path] (3.084,4.883)--(3.264,4.883);
\draw[gp path] (11.671,4.883)--(11.491,4.883);
\node[gp node right] at (2.808,4.883) {$0$};
\draw[gp path] (3.084,6.585)--(3.264,6.585);
\draw[gp path] (11.671,6.585)--(11.491,6.585);
\node[gp node right] at (2.808,6.585) {$0.0002$};
\draw[gp path] (3.084,8.287)--(3.264,8.287);
\draw[gp path] (11.671,8.287)--(11.491,8.287);
\node[gp node right] at (2.808,8.287) {$0.0004$};
\draw[gp path] (3.084,1.478)--(3.084,1.658);
\draw[gp path] (3.084,8.287)--(3.084,8.107);
\node[gp node center] at (3.084,1.016) {$0.2$};
\draw[gp path] (4.515,1.478)--(4.515,1.658);
\draw[gp path] (4.515,8.287)--(4.515,8.107);
\node[gp node center] at (4.515,1.016) {$0.3$};
\draw[gp path] (5.946,1.478)--(5.946,1.658);
\draw[gp path] (5.946,8.287)--(5.946,8.107);
\node[gp node center] at (5.946,1.016) {$0.4$};
\draw[gp path] (7.377,1.478)--(7.377,1.658);
\draw[gp path] (7.377,8.287)--(7.377,8.107);
\node[gp node center] at (7.377,1.016) {$0.5$};
\draw[gp path] (8.809,1.478)--(8.809,1.658);
\draw[gp path] (8.809,8.287)--(8.809,8.107);
\node[gp node center] at (8.809,1.016) {$0.6$};
\draw[gp path] (10.240,1.478)--(10.240,1.658);
\draw[gp path] (10.240,8.287)--(10.240,8.107);
\node[gp node center] at (10.240,1.016) {$0.7$};
\draw[gp path] (11.671,1.478)--(11.671,1.658);
\draw[gp path] (11.671,8.287)--(11.671,8.107);
\node[gp node center] at (11.671,1.016) {$0.8$};
\draw[gp path] (3.084,8.287)--(3.084,1.478)--(11.671,1.478)--(11.671,8.287)--cycle;
\node[gp node left,font={\fontsize{13.0pt}{15.6pt}\selectfont}] at (3.370,3.180) {$\sqrt{S_{ep}}=45$ GeV};
\node[gp node left,font={\fontsize{13.0pt}{15.6pt}\selectfont}] at (3.370,2.329) {$Q^2=10$ GeV${}^2$};
\node[gp node left,font={\fontsize{13.0pt}{15.6pt}\selectfont}] at (6.662,3.180) {$x_{B}=0.005$};
\node[gp node left,font={\fontsize{13.0pt}{15.6pt}\selectfont}] at (6.662,2.329) {$P_{J/\psi}^{\perp}=2$ GeV};
\node[gp node left,font={\fontsize{20.0pt}{24.0pt}\selectfont}] at (6.662,7.010) {Model 1};
\node[gp node center,rotate=-270,font={\fontsize{18.0pt}{21.6pt}\selectfont}] at (0.438,4.882) {${\cal F}_3/2\sigma^{\rm U}_1$};
\node[gp node center,font={\fontsize{22.0pt}{26.4pt}\selectfont}] at (7.377,0.323) {$z_f$};
\node[gp node right,font={\fontsize{15.0pt}{18.0pt}\selectfont}] at (5.748,7.876) {$N^1$};
\gpcolor{rgb color={0.580,0.000,0.827}}
\gpsetdashtype{gp dt 1}
\gpsetlinewidth{3.00}
\draw[gp path] (3.636,7.876)--(4.920,7.876);
\draw[gp path] (3.084,4.890)--(3.227,4.889)--(3.370,4.887)--(3.513,4.886)--(3.656,4.884)%
  --(3.800,4.881)--(3.943,4.878)--(4.086,4.875)--(4.229,4.872)--(4.372,4.867)--(4.515,4.863)%
  --(4.658,4.858)--(4.801,4.852)--(4.945,4.846)--(5.088,4.840)--(5.231,4.833)--(5.374,4.825)%
  --(5.517,4.817)--(5.660,4.808)--(5.803,4.799)--(5.946,4.789)--(6.089,4.779)--(6.233,4.768)%
  --(6.376,4.756)--(6.519,4.745)--(6.662,4.732)--(6.805,4.719)--(6.948,4.706)--(7.091,4.692)%
  --(7.234,4.678)--(7.377,4.664)--(7.521,4.649)--(7.664,4.634)--(7.807,4.619)--(7.950,4.603)%
  --(8.093,4.588)--(8.236,4.573)--(8.379,4.557)--(8.522,4.542)--(8.666,4.527)--(8.809,4.512)%
  --(8.952,4.498)--(9.095,4.484)--(9.238,4.471)--(9.381,4.458)--(9.524,4.447)--(9.667,4.436)%
  --(9.810,4.426)--(9.954,4.417)--(10.097,4.409)--(10.240,4.402)--(10.383,4.397)--(10.526,4.393)%
  --(10.669,4.390)--(10.812,4.389)--(10.955,4.389)--(11.099,4.391)--(11.242,4.394)--(11.385,4.399)%
  --(11.528,4.406)--(11.671,4.414);
\gpcolor{color=gp lt color border}
\node[gp node right,font={\fontsize{15.0pt}{18.0pt}\selectfont}] at (5.748,7.414) {$N^2$};
\gpcolor{rgb color={0.000,0.620,0.451}}
\gpsetdashtype{gp dt 2}
\draw[gp path] (3.636,7.414)--(4.920,7.414);
\draw[gp path] (3.084,4.856)--(3.227,4.855)--(3.370,4.854)--(3.513,4.853)--(3.656,4.852)%
  --(3.800,4.852)--(3.943,4.851)--(4.086,4.851)--(4.229,4.851)--(4.372,4.851)--(4.515,4.852)%
  --(4.658,4.852)--(4.801,4.853)--(4.945,4.854)--(5.088,4.855)--(5.231,4.857)--(5.374,4.859)%
  --(5.517,4.861)--(5.660,4.863)--(5.803,4.865)--(5.946,4.868)--(6.089,4.871)--(6.233,4.874)%
  --(6.376,4.878)--(6.519,4.882)--(6.662,4.886)--(6.805,4.890)--(6.948,4.894)--(7.091,4.899)%
  --(7.234,4.904)--(7.377,4.910)--(7.521,4.915)--(7.664,4.921)--(7.807,4.927)--(7.950,4.933)%
  --(8.093,4.940)--(8.236,4.947)--(8.379,4.954)--(8.522,4.962)--(8.666,4.969)--(8.809,4.978)%
  --(8.952,4.986)--(9.095,4.995)--(9.238,5.004)--(9.381,5.014)--(9.524,5.023)--(9.667,5.034)%
  --(9.810,5.045)--(9.954,5.056)--(10.097,5.067)--(10.240,5.079)--(10.383,5.092)--(10.526,5.105)%
  --(10.669,5.119)--(10.812,5.132)--(10.955,5.147)--(11.099,5.161)--(11.242,5.176)--(11.385,5.191)%
  --(11.528,5.206)--(11.671,5.221);
\gpcolor{color=gp lt color border}
\node[gp node right,font={\fontsize{15.0pt}{18.0pt}\selectfont}] at (5.748,6.952) {$N^3$};
\gpcolor{rgb color={0.337,0.706,0.914}}
\gpsetdashtype{gp dt 3}
\draw[gp path] (3.636,6.952)--(4.920,6.952);
\draw[gp path] (3.084,4.786)--(3.227,4.785)--(3.370,4.785)--(3.513,4.785)--(3.656,4.787)%
  --(3.800,4.789)--(3.943,4.791)--(4.086,4.795)--(4.229,4.800)--(4.372,4.806)--(4.515,4.812)%
  --(4.658,4.820)--(4.801,4.829)--(4.945,4.840)--(5.088,4.851)--(5.231,4.865)--(5.374,4.879)%
  --(5.517,4.895)--(5.660,4.912)--(5.803,4.932)--(5.946,4.952)--(6.089,4.975)--(6.233,4.999)%
  --(6.376,5.025)--(6.519,5.053)--(6.662,5.083)--(6.805,5.114)--(6.948,5.148)--(7.091,5.183)%
  --(7.234,5.221)--(7.377,5.260)--(7.521,5.301)--(7.664,5.344)--(7.807,5.389)--(7.950,5.435)%
  --(8.093,5.484)--(8.236,5.533)--(8.379,5.584)--(8.522,5.636)--(8.666,5.689)--(8.809,5.742)%
  --(8.952,5.796)--(9.095,5.851)--(9.238,5.905)--(9.381,5.958)--(9.524,6.011)--(9.667,6.062)%
  --(9.810,6.112)--(9.954,6.159)--(10.097,6.204)--(10.240,6.245)--(10.383,6.283)--(10.526,6.316)%
  --(10.669,6.344)--(10.812,6.367)--(10.955,6.383)--(11.099,6.394)--(11.242,6.397)--(11.385,6.393)%
  --(11.528,6.381)--(11.671,6.361);
\gpcolor{color=gp lt color border}
\node[gp node right,font={\fontsize{15.0pt}{18.0pt}\selectfont}] at (5.748,6.490) {$N^4$};
\gpcolor{rgb color={0.902,0.624,0.000}}
\gpsetdashtype{gp dt 4}
\draw[gp path] (3.636,6.490)--(4.920,6.490);
\draw[gp path] (3.084,4.712)--(3.227,4.699)--(3.370,4.686)--(3.513,4.673)--(3.656,4.659)%
  --(3.800,4.644)--(3.943,4.629)--(4.086,4.614)--(4.229,4.597)--(4.372,4.580)--(4.515,4.563)%
  --(4.658,4.545)--(4.801,4.526)--(4.945,4.507)--(5.088,4.487)--(5.231,4.466)--(5.374,4.444)%
  --(5.517,4.422)--(5.660,4.400)--(5.803,4.376)--(5.946,4.352)--(6.089,4.327)--(6.233,4.302)%
  --(6.376,4.276)--(6.519,4.249)--(6.662,4.222)--(6.805,4.195)--(6.948,4.166)--(7.091,4.138)%
  --(7.234,4.108)--(7.377,4.079)--(7.521,4.049)--(7.664,4.019)--(7.807,3.989)--(7.950,3.959)%
  --(8.093,3.929)--(8.236,3.899)--(8.379,3.869)--(8.522,3.840)--(8.666,3.811)--(8.809,3.784)%
  --(8.952,3.757)--(9.095,3.731)--(9.238,3.706)--(9.381,3.683)--(9.524,3.662)--(9.667,3.643)%
  --(9.810,3.626)--(9.954,3.611)--(10.097,3.600)--(10.240,3.591)--(10.383,3.585)--(10.526,3.583)%
  --(10.669,3.585)--(10.812,3.591)--(10.955,3.601)--(11.099,3.615)--(11.242,3.634)--(11.385,3.658)%
  --(11.528,3.686)--(11.671,3.719);
\gpcolor{color=gp lt color border}
\node[gp node right,font={\fontsize{15.0pt}{18.0pt}\selectfont}] at (5.748,6.028) {$N^5$};
\gpcolor{rgb color={0.941,0.894,0.259}}
\gpsetdashtype{gp dt 5}
\draw[gp path] (3.636,6.028)--(4.920,6.028);
\draw[gp path] (3.084,4.849)--(3.227,4.844)--(3.370,4.839)--(3.513,4.834)--(3.656,4.828)%
  --(3.800,4.822)--(3.943,4.816)--(4.086,4.809)--(4.229,4.801)--(4.372,4.793)--(4.515,4.785)%
  --(4.658,4.776)--(4.801,4.767)--(4.945,4.757)--(5.088,4.746)--(5.231,4.735)--(5.374,4.724)%
  --(5.517,4.712)--(5.660,4.699)--(5.803,4.686)--(5.946,4.672)--(6.089,4.658)--(6.233,4.643)%
  --(6.376,4.628)--(6.519,4.612)--(6.662,4.595)--(6.805,4.578)--(6.948,4.560)--(7.091,4.542)%
  --(7.234,4.524)--(7.377,4.505)--(7.521,4.486)--(7.664,4.466)--(7.807,4.446)--(7.950,4.426)%
  --(8.093,4.405)--(8.236,4.385)--(8.379,4.364)--(8.522,4.343)--(8.666,4.323)--(8.809,4.303)%
  --(8.952,4.283)--(9.095,4.264)--(9.238,4.245)--(9.381,4.227)--(9.524,4.210)--(9.667,4.194)%
  --(9.810,4.180)--(9.954,4.166)--(10.097,4.155)--(10.240,4.145)--(10.383,4.137)--(10.526,4.131)%
  --(10.669,4.128)--(10.812,4.127)--(10.955,4.129)--(11.099,4.133)--(11.242,4.141)--(11.385,4.152)%
  --(11.528,4.165)--(11.671,4.182);
\gpcolor{color=gp lt color border}
\gpsetdashtype{gp dt solid}
\gpsetlinewidth{1.00}
\draw[gp path] (3.084,8.287)--(3.084,1.478)--(11.671,1.478)--(11.671,8.287)--cycle;
%% coordinates of the plot area
\gpdefrectangularnode{gp plot 1}{\pgfpoint{3.084cm}{1.478cm}}{\pgfpoint{11.671cm}{8.287cm}}
\end{tikzpicture}
}	
%% gnuplot variables

%% file: F4_model1.tex
\scalebox{0.65}[0.65]{
\begin{tikzpicture}[gnuplot]
%% generated with GNUPLOT 5.4p5 (Lua 5.4; terminal rev. Jun 2020, script rev. 115)
%% 6/20/2023 10:24:03 AM
\tikzset{every node/.append style={font={\fontsize{15.0pt}{18.0pt}\selectfont}}}
\path (0.000,0.000) rectangle (12.500,8.750);
\gpcolor{color=gp lt color border}
\gpsetlinetype{gp lt border}
\gpsetdashtype{gp dt solid}
\gpsetlinewidth{1.00}
\draw[gp path] (2.808,1.478)--(2.988,1.478);
\draw[gp path] (11.671,1.478)--(11.491,1.478);
\node[gp node right] at (2.532,1.478) {$-0.004$};
\draw[gp path] (2.808,2.991)--(2.988,2.991);
\draw[gp path] (11.671,2.991)--(11.491,2.991);
\node[gp node right] at (2.532,2.991) {$-0.002$};
\draw[gp path] (2.808,4.504)--(2.988,4.504);
\draw[gp path] (11.671,4.504)--(11.491,4.504);
\node[gp node right] at (2.532,4.504) {$0$};
\draw[gp path] (2.808,6.017)--(2.988,6.017);
\draw[gp path] (11.671,6.017)--(11.491,6.017);
\node[gp node right] at (2.532,6.017) {$0.002$};
\draw[gp path] (2.808,7.530)--(2.988,7.530);
\draw[gp path] (11.671,7.530)--(11.491,7.530);
\node[gp node right] at (2.532,7.530) {$0.004$};
\draw[gp path] (2.808,1.478)--(2.808,1.658);
\draw[gp path] (2.808,8.287)--(2.808,8.107);
\node[gp node center] at (2.808,1.016) {$0.2$};
\draw[gp path] (4.285,1.478)--(4.285,1.658);
\draw[gp path] (4.285,8.287)--(4.285,8.107);
\node[gp node center] at (4.285,1.016) {$0.3$};
\draw[gp path] (5.762,1.478)--(5.762,1.658);
\draw[gp path] (5.762,8.287)--(5.762,8.107);
\node[gp node center] at (5.762,1.016) {$0.4$};
\draw[gp path] (7.239,1.478)--(7.239,1.658);
\draw[gp path] (7.239,8.287)--(7.239,8.107);
\node[gp node center] at (7.239,1.016) {$0.5$};
\draw[gp path] (8.717,1.478)--(8.717,1.658);
\draw[gp path] (8.717,8.287)--(8.717,8.107);
\node[gp node center] at (8.717,1.016) {$0.6$};
\draw[gp path] (10.194,1.478)--(10.194,1.658);
\draw[gp path] (10.194,8.287)--(10.194,8.107);
\node[gp node center] at (10.194,1.016) {$0.7$};
\draw[gp path] (11.671,1.478)--(11.671,1.658);
\draw[gp path] (11.671,8.287)--(11.671,8.107);
\node[gp node center] at (11.671,1.016) {$0.8$};
\draw[gp path] (2.808,8.287)--(2.808,1.478)--(11.671,1.478)--(11.671,8.287)--cycle;
\node[gp node left,font={\fontsize{13.0pt}{15.6pt}\selectfont}] at (3.399,2.991) {$\sqrt{S_{ep}}=45$ GeV};
\node[gp node left,font={\fontsize{13.0pt}{15.6pt}\selectfont}] at (3.399,2.235) {$Q^2=10$ GeV${}^2$};
\node[gp node left,font={\fontsize{13.0pt}{15.6pt}\selectfont}] at (6.501,2.991) {$x_{B}=0.005$};
\node[gp node left,font={\fontsize{13.0pt}{15.6pt}\selectfont}] at (6.501,2.235) {$P_{J/\psi}^{\perp}=2$ GeV};
\node[gp node left,font={\fontsize{20.0pt}{24.0pt}\selectfont}] at (6.501,7.152) {Model 1};
\node[gp node center,rotate=-270,font={\fontsize{18.0pt}{21.6pt}\selectfont}] at (0.438,4.882) {${\cal F}_4/2\sigma^{{\rm U}}_1$};
\node[gp node center,font={\fontsize{22.0pt}{26.4pt}\selectfont}] at (7.239,0.323) {$z_f$};
\node[gp node right,font={\fontsize{15.0pt}{18.0pt}\selectfont}] at (5.472,7.876) {$N^1$};
\gpcolor{rgb color={0.580,0.000,0.827}}
\gpsetdashtype{gp dt 1}
\gpsetlinewidth{3.00}
\draw[gp path] (3.360,7.876)--(4.644,7.876);
\draw[gp path] (2.808,4.362)--(2.956,4.354)--(3.103,4.347)--(3.251,4.339)--(3.399,4.331)%
  --(3.547,4.323)--(3.694,4.315)--(3.842,4.307)--(3.990,4.299)--(4.137,4.291)--(4.285,4.282)%
  --(4.433,4.273)--(4.581,4.264)--(4.728,4.255)--(4.876,4.246)--(5.024,4.237)--(5.171,4.227)%
  --(5.319,4.217)--(5.467,4.206)--(5.615,4.196)--(5.762,4.185)--(5.910,4.174)--(6.058,4.162)%
  --(6.205,4.151)--(6.353,4.138)--(6.501,4.126)--(6.649,4.113)--(6.796,4.100)--(6.944,4.086)%
  --(7.092,4.072)--(7.239,4.057)--(7.387,4.042)--(7.535,4.027)--(7.683,4.011)--(7.830,3.995)%
  --(7.978,3.978)--(8.126,3.961)--(8.274,3.943)--(8.421,3.925)--(8.569,3.907)--(8.717,3.888)%
  --(8.864,3.868)--(9.012,3.849)--(9.160,3.829)--(9.308,3.809)--(9.455,3.788)--(9.603,3.767)%
  --(9.751,3.746)--(9.898,3.726)--(10.046,3.705)--(10.194,3.684)--(10.342,3.664)--(10.489,3.644)%
  --(10.637,3.624)--(10.785,3.605)--(10.932,3.587)--(11.080,3.570)--(11.228,3.554)--(11.376,3.539)%
  --(11.523,3.526)--(11.671,3.515);
\gpcolor{color=gp lt color border}
\node[gp node right,font={\fontsize{15.0pt}{18.0pt}\selectfont}] at (5.472,7.414) {$N^2$};
\gpcolor{rgb color={0.000,0.620,0.451}}
\gpsetdashtype{gp dt 2}
\draw[gp path] (3.360,7.414)--(4.644,7.414);
\draw[gp path] (2.808,4.928)--(2.956,4.948)--(3.103,4.968)--(3.251,4.989)--(3.399,5.009)%
  --(3.547,5.030)--(3.694,5.050)--(3.842,5.071)--(3.990,5.091)--(4.137,5.112)--(4.285,5.133)%
  --(4.433,5.154)--(4.581,5.175)--(4.728,5.196)--(4.876,5.218)--(5.024,5.240)--(5.171,5.262)%
  --(5.319,5.285)--(5.467,5.308)--(5.615,5.332)--(5.762,5.356)--(5.910,5.381)--(6.058,5.406)%
  --(6.205,5.432)--(6.353,5.458)--(6.501,5.486)--(6.649,5.514)--(6.796,5.543)--(6.944,5.572)%
  --(7.092,5.603)--(7.239,5.635)--(7.387,5.667)--(7.535,5.700)--(7.683,5.735)--(7.830,5.770)%
  --(7.978,5.806)--(8.126,5.844)--(8.274,5.882)--(8.421,5.921)--(8.569,5.960)--(8.717,6.001)%
  --(8.864,6.042)--(9.012,6.084)--(9.160,6.126)--(9.308,6.168)--(9.455,6.210)--(9.603,6.252)%
  --(9.751,6.294)--(9.898,6.335)--(10.046,6.374)--(10.194,6.413)--(10.342,6.450)--(10.489,6.484)%
  --(10.637,6.517)--(10.785,6.546)--(10.932,6.571)--(11.080,6.593)--(11.228,6.610)--(11.376,6.623)%
  --(11.523,6.629)--(11.671,6.630);
\gpcolor{color=gp lt color border}
\node[gp node right,font={\fontsize{15.0pt}{18.0pt}\selectfont}] at (5.472,6.952) {$N^3$};
\gpcolor{rgb color={0.337,0.706,0.914}}
\gpsetdashtype{gp dt 3}
\draw[gp path] (3.360,6.952)--(4.644,6.952);
\draw[gp path] (2.808,4.459)--(2.956,4.457)--(3.103,4.455)--(3.251,4.454)--(3.399,4.452)%
  --(3.547,4.450)--(3.694,4.449)--(3.842,4.447)--(3.990,4.446)--(4.137,4.445)--(4.285,4.443)%
  --(4.433,4.442)--(4.581,4.441)--(4.728,4.439)--(4.876,4.438)--(5.024,4.437)--(5.171,4.436)%
  --(5.319,4.435)--(5.467,4.434)--(5.615,4.433)--(5.762,4.432)--(5.910,4.432)--(6.058,4.431)%
  --(6.205,4.430)--(6.353,4.430)--(6.501,4.429)--(6.649,4.429)--(6.796,4.428)--(6.944,4.428)%
  --(7.092,4.428)--(7.239,4.428)--(7.387,4.428)--(7.535,4.428)--(7.683,4.428)--(7.830,4.428)%
  --(7.978,4.428)--(8.126,4.429)--(8.274,4.429)--(8.421,4.430)--(8.569,4.431)--(8.717,4.432)%
  --(8.864,4.433)--(9.012,4.434)--(9.160,4.435)--(9.308,4.437)--(9.455,4.438)--(9.603,4.440)%
  --(9.751,4.442)--(9.898,4.444)--(10.046,4.446)--(10.194,4.448)--(10.342,4.450)--(10.489,4.452)%
  --(10.637,4.455)--(10.785,4.457)--(10.932,4.460)--(11.080,4.462)--(11.228,4.465)--(11.376,4.468)%
  --(11.523,4.470)--(11.671,4.473);
\gpcolor{color=gp lt color border}
\node[gp node right,font={\fontsize{15.0pt}{18.0pt}\selectfont}] at (5.472,6.490) {$N^4$};
\gpcolor{rgb color={0.902,0.624,0.000}}
\gpsetdashtype{gp dt 4}
\draw[gp path] (3.360,6.490)--(4.644,6.490);
\draw[gp path] (2.808,4.322)--(2.956,4.314)--(3.103,4.306)--(3.251,4.298)--(3.399,4.290)%
  --(3.547,4.282)--(3.694,4.274)--(3.842,4.267)--(3.990,4.259)--(4.137,4.251)--(4.285,4.244)%
  --(4.433,4.236)--(4.581,4.229)--(4.728,4.221)--(4.876,4.214)--(5.024,4.207)--(5.171,4.199)%
  --(5.319,4.192)--(5.467,4.184)--(5.615,4.177)--(5.762,4.169)--(5.910,4.161)--(6.058,4.154)%
  --(6.205,4.146)--(6.353,4.138)--(6.501,4.129)--(6.649,4.121)--(6.796,4.112)--(6.944,4.103)%
  --(7.092,4.094)--(7.239,4.085)--(7.387,4.075)--(7.535,4.065)--(7.683,4.054)--(7.830,4.043)%
  --(7.978,4.032)--(8.126,4.020)--(8.274,4.007)--(8.421,3.994)--(8.569,3.981)--(8.717,3.966)%
  --(8.864,3.951)--(9.012,3.936)--(9.160,3.920)--(9.308,3.903)--(9.455,3.885)--(9.603,3.867)%
  --(9.751,3.849)--(9.898,3.829)--(10.046,3.810)--(10.194,3.789)--(10.342,3.769)--(10.489,3.748)%
  --(10.637,3.727)--(10.785,3.706)--(10.932,3.685)--(11.080,3.665)--(11.228,3.645)--(11.376,3.626)%
  --(11.523,3.608)--(11.671,3.591);
\gpcolor{color=gp lt color border}
\node[gp node right,font={\fontsize{15.0pt}{18.0pt}\selectfont}] at (5.472,6.028) {$N^5$};
\gpcolor{rgb color={0.941,0.894,0.259}}
\gpsetdashtype{gp dt 5}
\draw[gp path] (3.360,6.028)--(4.644,6.028);
\draw[gp path] (2.808,4.368)--(2.956,4.361)--(3.103,4.355)--(3.251,4.348)--(3.399,4.342)%
  --(3.547,4.336)--(3.694,4.330)--(3.842,4.323)--(3.990,4.317)--(4.137,4.311)--(4.285,4.305)%
  --(4.433,4.299)--(4.581,4.292)--(4.728,4.286)--(4.876,4.280)--(5.024,4.274)--(5.171,4.267)%
  --(5.319,4.261)--(5.467,4.254)--(5.615,4.248)--(5.762,4.241)--(5.910,4.234)--(6.058,4.227)%
  --(6.205,4.220)--(6.353,4.212)--(6.501,4.204)--(6.649,4.197)--(6.796,4.188)--(6.944,4.180)%
  --(7.092,4.171)--(7.239,4.161)--(7.387,4.152)--(7.535,4.141)--(7.683,4.131)--(7.830,4.119)%
  --(7.978,4.108)--(8.126,4.095)--(8.274,4.082)--(8.421,4.068)--(8.569,4.054)--(8.717,4.039)%
  --(8.864,4.023)--(9.012,4.006)--(9.160,3.989)--(9.308,3.970)--(9.455,3.951)--(9.603,3.932)%
  --(9.751,3.911)--(9.898,3.890)--(10.046,3.868)--(10.194,3.846)--(10.342,3.823)--(10.489,3.800)%
  --(10.637,3.777)--(10.785,3.753)--(10.932,3.730)--(11.080,3.707)--(11.228,3.684)--(11.376,3.662)%
  --(11.523,3.641)--(11.671,3.622);
\gpcolor{color=gp lt color border}
\gpsetdashtype{gp dt solid}
\gpsetlinewidth{1.00}
\draw[gp path] (2.808,8.287)--(2.808,1.478)--(11.671,1.478)--(11.671,8.287)--cycle;
%% coordinates of the plot area
\gpdefrectangularnode{gp plot 1}{\pgfpoint{2.808cm}{1.478cm}}{\pgfpoint{11.671cm}{8.287cm}}
\end{tikzpicture}
}
%% gnuplot variables

%% file: F5_model1.tex
\scalebox{0.70}[0.70]{
\begin{tikzpicture}[gnuplot]
%% generated with GNUPLOT 5.4p5 (Lua 5.4; terminal rev. Jun 2020, script rev. 115)
%% 6/20/2023 10:54:46 AM
\tikzset{every node/.append style={font={\fontsize{15.0pt}{18.0pt}\selectfont}}}
\path (0.000,0.000) rectangle (12.500,8.750);
\gpcolor{color=gp lt color border}
\gpsetlinetype{gp lt border}
\gpsetdashtype{gp dt solid}
\gpsetlinewidth{1.00}
\draw[gp path] (3.084,1.478)--(3.264,1.478);
\draw[gp path] (11.671,1.478)--(11.491,1.478);
\node[gp node right] at (2.808,1.478) {$-0.0004$};
\draw[gp path] (3.084,3.180)--(3.264,3.180);
\draw[gp path] (11.671,3.180)--(11.491,3.180);
\node[gp node right] at (2.808,3.180) {$-0.0002$};
\draw[gp path] (3.084,4.883)--(3.264,4.883);
\draw[gp path] (11.671,4.883)--(11.491,4.883);
\node[gp node right] at (2.808,4.883) {$0$};
\draw[gp path] (3.084,6.585)--(3.264,6.585);
\draw[gp path] (11.671,6.585)--(11.491,6.585);
\node[gp node right] at (2.808,6.585) {$0.0002$};
\draw[gp path] (3.084,8.287)--(3.264,8.287);
\draw[gp path] (11.671,8.287)--(11.491,8.287);
\node[gp node right] at (2.808,8.287) {$0.0004$};
\draw[gp path] (3.084,1.478)--(3.084,1.658);
\draw[gp path] (3.084,8.287)--(3.084,8.107);
\node[gp node center] at (3.084,1.016) {$0.2$};
\draw[gp path] (4.515,1.478)--(4.515,1.658);
\draw[gp path] (4.515,8.287)--(4.515,8.107);
\node[gp node center] at (4.515,1.016) {$0.3$};
\draw[gp path] (5.946,1.478)--(5.946,1.658);
\draw[gp path] (5.946,8.287)--(5.946,8.107);
\node[gp node center] at (5.946,1.016) {$0.4$};
\draw[gp path] (7.377,1.478)--(7.377,1.658);
\draw[gp path] (7.377,8.287)--(7.377,8.107);
\node[gp node center] at (7.377,1.016) {$0.5$};
\draw[gp path] (8.809,1.478)--(8.809,1.658);
\draw[gp path] (8.809,8.287)--(8.809,8.107);
\node[gp node center] at (8.809,1.016) {$0.6$};
\draw[gp path] (10.240,1.478)--(10.240,1.658);
\draw[gp path] (10.240,8.287)--(10.240,8.107);
\node[gp node center] at (10.240,1.016) {$0.7$};
\draw[gp path] (11.671,1.478)--(11.671,1.658);
\draw[gp path] (11.671,8.287)--(11.671,8.107);
\node[gp node center] at (11.671,1.016) {$0.8$};
\draw[gp path] (3.084,8.287)--(3.084,1.478)--(11.671,1.478)--(11.671,8.287)--cycle;
\node[gp node left,font={\fontsize{13.0pt}{15.6pt}\selectfont}] at (3.800,3.180) {$\sqrt{S_{ep}}=45$ GeV};
\node[gp node left,font={\fontsize{13.0pt}{15.6pt}\selectfont}] at (3.800,2.329) {$Q^2=10$ GeV${}^2$};
\node[gp node left,font={\fontsize{13.0pt}{15.6pt}\selectfont}] at (6.948,3.180) {$x_{B}=0.005$};
\node[gp node left,font={\fontsize{13.0pt}{15.6pt}\selectfont}] at (6.948,2.329) {$P_{J/\psi}^{\perp}=2$ GeV};
\node[gp node left,font={\fontsize{20.0pt}{24.0pt}\selectfont}] at (6.662,7.010) {Model 1};
\node[gp node center,rotate=-270,font={\fontsize{18.0pt}{21.6pt}\selectfont}] at (0.438,4.882) {${\cal F}_5/2\sigma^{\rm U}_1$};
\node[gp node center,font={\fontsize{22.0pt}{26.4pt}\selectfont}] at (7.377,0.323) {$z_f$};
\node[gp node right,font={\fontsize{15.0pt}{18.0pt}\selectfont}] at (5.748,7.876) {$N^1$};
\gpcolor{rgb color={0.580,0.000,0.827}}
\gpsetdashtype{gp dt 1}
\gpsetlinewidth{3.00}
\draw[gp path] (3.636,7.876)--(4.920,7.876);
\draw[gp path] (3.084,4.306)--(3.227,4.311)--(3.370,4.315)--(3.513,4.319)--(3.656,4.324)%
  --(3.800,4.329)--(3.943,4.335)--(4.086,4.341)--(4.229,4.347)--(4.372,4.353)--(4.515,4.360)%
  --(4.658,4.367)--(4.801,4.374)--(4.945,4.382)--(5.088,4.390)--(5.231,4.399)--(5.374,4.408)%
  --(5.517,4.417)--(5.660,4.427)--(5.803,4.438)--(5.946,4.448)--(6.089,4.460)--(6.233,4.472)%
  --(6.376,4.484)--(6.519,4.496)--(6.662,4.510)--(6.805,4.523)--(6.948,4.538)--(7.091,4.552)%
  --(7.234,4.567)--(7.377,4.583)--(7.521,4.599)--(7.664,4.616)--(7.807,4.633)--(7.950,4.650)%
  --(8.093,4.668)--(8.236,4.686)--(8.379,4.704)--(8.522,4.723)--(8.666,4.742)--(8.809,4.761)%
  --(8.952,4.780)--(9.095,4.800)--(9.238,4.819)--(9.381,4.838)--(9.524,4.858)--(9.667,4.877)%
  --(9.810,4.895)--(9.954,4.913)--(10.097,4.931)--(10.240,4.948)--(10.383,4.964)--(10.526,4.980)%
  --(10.669,4.994)--(10.812,5.008)--(10.955,5.020)--(11.099,5.031)--(11.242,5.041)--(11.385,5.049)%
  --(11.528,5.055)--(11.671,5.060);
\gpcolor{color=gp lt color border}
\node[gp node right,font={\fontsize{15.0pt}{18.0pt}\selectfont}] at (5.748,7.414) {$N^2$};
\gpcolor{rgb color={0.000,0.620,0.451}}
\gpsetdashtype{gp dt 2}
\draw[gp path] (3.636,7.414)--(4.920,7.414);
\draw[gp path] (3.084,4.795)--(3.227,4.786)--(3.370,4.775)--(3.513,4.765)--(3.656,4.753)%
  --(3.800,4.741)--(3.943,4.729)--(4.086,4.716)--(4.229,4.702)--(4.372,4.688)--(4.515,4.673)%
  --(4.658,4.658)--(4.801,4.642)--(4.945,4.626)--(5.088,4.609)--(5.231,4.592)--(5.374,4.575)%
  --(5.517,4.557)--(5.660,4.538)--(5.803,4.520)--(5.946,4.501)--(6.089,4.481)--(6.233,4.462)%
  --(6.376,4.442)--(6.519,4.422)--(6.662,4.402)--(6.805,4.382)--(6.948,4.362)--(7.091,4.342)%
  --(7.234,4.322)--(7.377,4.303)--(7.521,4.283)--(7.664,4.264)--(7.807,4.246)--(7.950,4.228)%
  --(8.093,4.211)--(8.236,4.194)--(8.379,4.179)--(8.522,4.164)--(8.666,4.151)--(8.809,4.138)%
  --(8.952,4.127)--(9.095,4.118)--(9.238,4.110)--(9.381,4.104)--(9.524,4.100)--(9.667,4.098)%
  --(9.810,4.097)--(9.954,4.100)--(10.097,4.104)--(10.240,4.111)--(10.383,4.121)--(10.526,4.133)%
  --(10.669,4.148)--(10.812,4.165)--(10.955,4.185)--(11.099,4.208)--(11.242,4.234)--(11.385,4.262)%
  --(11.528,4.293)--(11.671,4.325);
\gpcolor{color=gp lt color border}
\node[gp node right,font={\fontsize{15.0pt}{18.0pt}\selectfont}] at (5.748,6.952) {$N^3$};
\gpcolor{rgb color={0.337,0.706,0.914}}
\gpsetdashtype{gp dt 3}
\draw[gp path] (3.636,6.952)--(4.920,6.952);
\draw[gp path] (3.084,5.562)--(3.227,5.565)--(3.370,5.569)--(3.513,5.573)--(3.656,5.577)%
  --(3.800,5.581)--(3.943,5.585)--(4.086,5.590)--(4.229,5.595)--(4.372,5.601)--(4.515,5.606)%
  --(4.658,5.612)--(4.801,5.618)--(4.945,5.625)--(5.088,5.631)--(5.231,5.638)--(5.374,5.645)%
  --(5.517,5.652)--(5.660,5.659)--(5.803,5.667)--(5.946,5.674)--(6.089,5.682)--(6.233,5.690)%
  --(6.376,5.698)--(6.519,5.706)--(6.662,5.714)--(6.805,5.723)--(6.948,5.731)--(7.091,5.739)%
  --(7.234,5.747)--(7.377,5.755)--(7.521,5.763)--(7.664,5.771)--(7.807,5.779)--(7.950,5.787)%
  --(8.093,5.794)--(8.236,5.801)--(8.379,5.807)--(8.522,5.813)--(8.666,5.819)--(8.809,5.824)%
  --(8.952,5.829)--(9.095,5.832)--(9.238,5.835)--(9.381,5.838)--(9.524,5.839)--(9.667,5.839)%
  --(9.810,5.839)--(9.954,5.837)--(10.097,5.834)--(10.240,5.829)--(10.383,5.823)--(10.526,5.816)%
  --(10.669,5.807)--(10.812,5.797)--(10.955,5.784)--(11.099,5.770)--(11.242,5.755)--(11.385,5.737)%
  --(11.528,5.718)--(11.671,5.696);
\gpcolor{color=gp lt color border}
\node[gp node right,font={\fontsize{15.0pt}{18.0pt}\selectfont}] at (5.748,6.490) {$N^4$};
\gpcolor{rgb color={0.902,0.624,0.000}}
\gpsetdashtype{gp dt 4}
\draw[gp path] (3.636,6.490)--(4.920,6.490);
\draw[gp path] (3.084,4.970)--(3.227,4.978)--(3.370,4.985)--(3.513,4.993)--(3.656,5.001)%
  --(3.800,5.010)--(3.943,5.019)--(4.086,5.028)--(4.229,5.037)--(4.372,5.046)--(4.515,5.056)%
  --(4.658,5.066)--(4.801,5.076)--(4.945,5.087)--(5.088,5.097)--(5.231,5.108)--(5.374,5.119)%
  --(5.517,5.130)--(5.660,5.141)--(5.803,5.152)--(5.946,5.163)--(6.089,5.175)--(6.233,5.186)%
  --(6.376,5.197)--(6.519,5.209)--(6.662,5.220)--(6.805,5.231)--(6.948,5.242)--(7.091,5.253)%
  --(7.234,5.264)--(7.377,5.275)--(7.521,5.285)--(7.664,5.295)--(7.807,5.305)--(7.950,5.314)%
  --(8.093,5.323)--(8.236,5.331)--(8.379,5.339)--(8.522,5.346)--(8.666,5.353)--(8.809,5.358)%
  --(8.952,5.363)--(9.095,5.368)--(9.238,5.371)--(9.381,5.374)--(9.524,5.375)--(9.667,5.376)%
  --(9.810,5.375)--(9.954,5.374)--(10.097,5.371)--(10.240,5.367)--(10.383,5.362)--(10.526,5.355)%
  --(10.669,5.348)--(10.812,5.339)--(10.955,5.329)--(11.099,5.317)--(11.242,5.305)--(11.385,5.291)%
  --(11.528,5.276)--(11.671,5.260);
\gpcolor{color=gp lt color border}
\node[gp node right,font={\fontsize{15.0pt}{18.0pt}\selectfont}] at (5.748,6.028) {$N^5$};
\gpcolor{rgb color={0.941,0.894,0.259}}
\gpsetdashtype{gp dt 5}
\draw[gp path] (3.636,6.028)--(4.920,6.028);
\draw[gp path] (3.084,4.922)--(3.227,4.926)--(3.370,4.930)--(3.513,4.935)--(3.656,4.939)%
  --(3.800,4.944)--(3.943,4.949)--(4.086,4.954)--(4.229,4.959)--(4.372,4.965)--(4.515,4.970)%
  --(4.658,4.976)--(4.801,4.982)--(4.945,4.987)--(5.088,4.993)--(5.231,4.999)--(5.374,5.005)%
  --(5.517,5.011)--(5.660,5.018)--(5.803,5.024)--(5.946,5.030)--(6.089,5.036)--(6.233,5.042)%
  --(6.376,5.048)--(6.519,5.054)--(6.662,5.060)--(6.805,5.066)--(6.948,5.072)--(7.091,5.077)%
  --(7.234,5.083)--(7.377,5.088)--(7.521,5.093)--(7.664,5.097)--(7.807,5.102)--(7.950,5.106)%
  --(8.093,5.109)--(8.236,5.113)--(8.379,5.116)--(8.522,5.118)--(8.666,5.120)--(8.809,5.122)%
  --(8.952,5.123)--(9.095,5.123)--(9.238,5.123)--(9.381,5.123)--(9.524,5.122)--(9.667,5.120)%
  --(9.810,5.118)--(9.954,5.115)--(10.097,5.111)--(10.240,5.107)--(10.383,5.103)--(10.526,5.098)%
  --(10.669,5.092)--(10.812,5.086)--(10.955,5.079)--(11.099,5.072)--(11.242,5.064)--(11.385,5.056)%
  --(11.528,5.047)--(11.671,5.039);
\gpcolor{color=gp lt color border}
\gpsetdashtype{gp dt solid}
\gpsetlinewidth{1.00}
\draw[gp path] (3.084,8.287)--(3.084,1.478)--(11.671,1.478)--(11.671,8.287)--cycle;
%% coordinates of the plot area
\gpdefrectangularnode{gp plot 1}{\pgfpoint{3.084cm}{1.478cm}}{\pgfpoint{11.671cm}{8.287cm}}
\end{tikzpicture}
}
%% gnuplot variables

%% file: F1_model2.tex
\scalebox{0.65}[0.65]{
\begin{tikzpicture}[gnuplot]
%% generated with GNUPLOT 5.4p5 (Lua 5.4; terminal rev. Jun 2020, script rev. 115)
%% 6/20/2023 11:21:11 AM
\tikzset{every node/.append style={font={\fontsize{15.0pt}{18.0pt}\selectfont}}}
\path (0.000,0.000) rectangle (12.500,8.750);
\gpcolor{color=gp lt color border}
\gpsetlinetype{gp lt border}
\gpsetdashtype{gp dt solid}
\gpsetlinewidth{1.00}
\draw[gp path] (2.256,2.097)--(2.436,2.097);
\draw[gp path] (11.671,2.097)--(11.491,2.097);
\node[gp node right] at (1.980,2.097) {$-0.2$};
\draw[gp path] (2.256,3.335)--(2.436,3.335);
\draw[gp path] (11.671,3.335)--(11.491,3.335);
\node[gp node right] at (1.980,3.335) {$-0.1$};
\draw[gp path] (2.256,4.573)--(2.436,4.573);
\draw[gp path] (11.671,4.573)--(11.491,4.573);
\node[gp node right] at (1.980,4.573) {$0$};
\draw[gp path] (2.256,5.811)--(2.436,5.811);
\draw[gp path] (11.671,5.811)--(11.491,5.811);
\node[gp node right] at (1.980,5.811) {$0.1$};
\draw[gp path] (2.256,7.049)--(2.436,7.049);
\draw[gp path] (11.671,7.049)--(11.491,7.049);
\node[gp node right] at (1.980,7.049) {$0.2$};
\draw[gp path] (2.256,8.287)--(2.436,8.287);
\draw[gp path] (11.671,8.287)--(11.491,8.287);
\node[gp node right] at (1.980,8.287) {$0.3$};
\draw[gp path] (2.256,1.478)--(2.256,1.658);
\draw[gp path] (2.256,8.287)--(2.256,8.107);
\node[gp node center] at (2.256,1.016) {$0.2$};
\draw[gp path] (3.825,1.478)--(3.825,1.658);
\draw[gp path] (3.825,8.287)--(3.825,8.107);
\node[gp node center] at (3.825,1.016) {$0.3$};
\draw[gp path] (5.394,1.478)--(5.394,1.658);
\draw[gp path] (5.394,8.287)--(5.394,8.107);
\node[gp node center] at (5.394,1.016) {$0.4$};
\draw[gp path] (6.963,1.478)--(6.963,1.658);
\draw[gp path] (6.963,8.287)--(6.963,8.107);
\node[gp node center] at (6.963,1.016) {$0.5$};
\draw[gp path] (8.533,1.478)--(8.533,1.658);
\draw[gp path] (8.533,8.287)--(8.533,8.107);
\node[gp node center] at (8.533,1.016) {$0.6$};
\draw[gp path] (10.102,1.478)--(10.102,1.658);
\draw[gp path] (10.102,8.287)--(10.102,8.107);
\node[gp node center] at (10.102,1.016) {$0.7$};
\draw[gp path] (11.671,1.478)--(11.671,1.658);
\draw[gp path] (11.671,8.287)--(11.671,8.107);
\node[gp node center] at (11.671,1.016) {$0.8$};
\draw[gp path] (2.256,8.287)--(2.256,1.478)--(11.671,1.478)--(11.671,8.287)--cycle;
\node[gp node left,font={\fontsize{13.0pt}{15.6pt}\selectfont}] at (3.041,2.964) {$\sqrt{S_{ep}}=45$ GeV};
\node[gp node left,font={\fontsize{13.0pt}{15.6pt}\selectfont}] at (3.041,2.345) {$Q^2=10$ GeV${}^2$};
\node[gp node left,font={\fontsize{13.0pt}{15.6pt}\selectfont}] at (6.179,2.964) {$x_{B}=0.005$};
\node[gp node left,font={\fontsize{13.0pt}{15.6pt}\selectfont}] at (6.179,2.345) {$P_{J/\psi}^{\perp}=2$ GeV};
\node[gp node left,font={\fontsize{20.0pt}{24.0pt}\selectfont}] at (6.179,7.049) {Model 2};
\node[gp node center,rotate=-270,font={\fontsize{18.0pt}{21.6pt}\selectfont}] at (0.438,4.882) {${\cal F}_1/\sigma^{\rm U}_1$};
\node[gp node center,font={\fontsize{22.0pt}{26.4pt}\selectfont}] at (6.963,0.323) {$z_f$};
\node[gp node right,font={\fontsize{15.0pt}{18.0pt}\selectfont}] at (4.920,7.876) {$N^1$};
\gpcolor{rgb color={0.580,0.000,0.827}}
\gpsetdashtype{gp dt 1}
\gpsetlinewidth{3.00}
\draw[gp path] (2.808,7.876)--(4.092,7.876);
\draw[gp path] (2.256,4.783)--(2.413,4.788)--(2.570,4.794)--(2.727,4.799)--(2.884,4.804)%
  --(3.041,4.810)--(3.198,4.815)--(3.354,4.821)--(3.511,4.826)--(3.668,4.832)--(3.825,4.839)%
  --(3.982,4.845)--(4.139,4.851)--(4.296,4.858)--(4.453,4.865)--(4.610,4.873)--(4.767,4.880)%
  --(4.924,4.888)--(5.081,4.897)--(5.237,4.906)--(5.394,4.915)--(5.551,4.925)--(5.708,4.935)%
  --(5.865,4.946)--(6.022,4.957)--(6.179,4.969)--(6.336,4.981)--(6.493,4.995)--(6.650,5.009)%
  --(6.807,5.023)--(6.963,5.039)--(7.120,5.056)--(7.277,5.073)--(7.434,5.092)--(7.591,5.111)%
  --(7.748,5.132)--(7.905,5.154)--(8.062,5.177)--(8.219,5.202)--(8.376,5.228)--(8.533,5.255)%
  --(8.690,5.284)--(8.847,5.315)--(9.003,5.347)--(9.160,5.382)--(9.317,5.418)--(9.474,5.457)%
  --(9.631,5.498)--(9.788,5.541)--(9.945,5.586)--(10.102,5.634)--(10.259,5.685)--(10.416,5.738)%
  --(10.573,5.794)--(10.730,5.854)--(10.886,5.917)--(11.043,5.983)--(11.200,6.053)--(11.357,6.127)%
  --(11.514,6.205)--(11.671,6.287);
\gpcolor{color=gp lt color border}
\node[gp node right,font={\fontsize{15.0pt}{18.0pt}\selectfont}] at (4.920,7.414) {$N^2$};
\gpcolor{rgb color={0.000,0.620,0.451}}
\gpsetdashtype{gp dt 2}
\draw[gp path] (2.808,7.414)--(4.092,7.414);
\draw[gp path] (2.256,4.847)--(2.413,4.851)--(2.570,4.853)--(2.727,4.855)--(2.884,4.857)%
  --(3.041,4.857)--(3.198,4.857)--(3.354,4.857)--(3.511,4.855)--(3.668,4.853)--(3.825,4.849)%
  --(3.982,4.846)--(4.139,4.841)--(4.296,4.835)--(4.453,4.829)--(4.610,4.821)--(4.767,4.813)%
  --(4.924,4.803)--(5.081,4.793)--(5.237,4.781)--(5.394,4.769)--(5.551,4.755)--(5.708,4.740)%
  --(5.865,4.724)--(6.022,4.706)--(6.179,4.687)--(6.336,4.667)--(6.493,4.645)--(6.650,4.622)%
  --(6.807,4.597)--(6.963,4.570)--(7.120,4.541)--(7.277,4.511)--(7.434,4.478)--(7.591,4.444)%
  --(7.748,4.407)--(7.905,4.368)--(8.062,4.327)--(8.219,4.283)--(8.376,4.237)--(8.533,4.188)%
  --(8.690,4.136)--(8.847,4.082)--(9.003,4.024)--(9.160,3.964)--(9.317,3.900)--(9.474,3.834)%
  --(9.631,3.764)--(9.788,3.691)--(9.945,3.614)--(10.102,3.534)--(10.259,3.451)--(10.416,3.365)%
  --(10.573,3.275)--(10.730,3.182)--(10.886,3.086)--(11.043,2.987)--(11.200,2.885)--(11.357,2.779)%
  --(11.514,2.671)--(11.671,2.560);
\gpcolor{color=gp lt color border}
\node[gp node right,font={\fontsize{15.0pt}{18.0pt}\selectfont}] at (4.920,6.952) {$N^3$};
\gpcolor{rgb color={0.337,0.706,0.914}}
\gpsetdashtype{gp dt 3}
\draw[gp path] (2.808,6.952)--(4.092,6.952);
\draw[gp path] (2.256,4.073)--(2.413,4.060)--(2.570,4.047)--(2.727,4.034)--(2.884,4.022)%
  --(3.041,4.009)--(3.198,3.997)--(3.354,3.984)--(3.511,3.972)--(3.668,3.960)--(3.825,3.948)%
  --(3.982,3.936)--(4.139,3.924)--(4.296,3.912)--(4.453,3.900)--(4.610,3.888)--(4.767,3.876)%
  --(4.924,3.864)--(5.081,3.852)--(5.237,3.840)--(5.394,3.828)--(5.551,3.816)--(5.708,3.803)%
  --(5.865,3.791)--(6.022,3.779)--(6.179,3.766)--(6.336,3.753)--(6.493,3.740)--(6.650,3.727)%
  --(6.807,3.714)--(6.963,3.700)--(7.120,3.687)--(7.277,3.673)--(7.434,3.658)--(7.591,3.644)%
  --(7.748,3.629)--(7.905,3.613)--(8.062,3.598)--(8.219,3.582)--(8.376,3.565)--(8.533,3.548)%
  --(8.690,3.530)--(8.847,3.512)--(9.003,3.493)--(9.160,3.473)--(9.317,3.452)--(9.474,3.431)%
  --(9.631,3.409)--(9.788,3.386)--(9.945,3.362)--(10.102,3.337)--(10.259,3.311)--(10.416,3.284)%
  --(10.573,3.256)--(10.730,3.226)--(10.886,3.195)--(11.043,3.162)--(11.200,3.128)--(11.357,3.092)%
  --(11.514,3.054)--(11.671,3.014);
\gpcolor{color=gp lt color border}
\node[gp node right,font={\fontsize{15.0pt}{18.0pt}\selectfont}] at (4.920,6.490) {$N^4$};
\gpcolor{rgb color={0.902,0.624,0.000}}
\gpsetdashtype{gp dt 4}
\draw[gp path] (2.808,6.490)--(4.092,6.490);
\draw[gp path] (2.256,5.091)--(2.413,5.108)--(2.570,5.126)--(2.727,5.144)--(2.884,5.162)%
  --(3.041,5.181)--(3.198,5.199)--(3.354,5.219)--(3.511,5.238)--(3.668,5.258)--(3.825,5.279)%
  --(3.982,5.300)--(4.139,5.321)--(4.296,5.343)--(4.453,5.365)--(4.610,5.388)--(4.767,5.412)%
  --(4.924,5.436)--(5.081,5.460)--(5.237,5.486)--(5.394,5.512)--(5.551,5.538)--(5.708,5.566)%
  --(5.865,5.594)--(6.022,5.623)--(6.179,5.652)--(6.336,5.683)--(6.493,5.714)--(6.650,5.746)%
  --(6.807,5.779)--(6.963,5.813)--(7.120,5.848)--(7.277,5.884)--(7.434,5.921)--(7.591,5.959)%
  --(7.748,5.999)--(7.905,6.039)--(8.062,6.081)--(8.219,6.124)--(8.376,6.168)--(8.533,6.214)%
  --(8.690,6.261)--(8.847,6.310)--(9.003,6.361)--(9.160,6.413)--(9.317,6.467)--(9.474,6.523)%
  --(9.631,6.580)--(9.788,6.640)--(9.945,6.702)--(10.102,6.766)--(10.259,6.832)--(10.416,6.901)%
  --(10.573,6.972)--(10.730,7.046)--(10.886,7.122)--(11.043,7.202)--(11.200,7.285)--(11.357,7.371)%
  --(11.514,7.460)--(11.671,7.553);
\gpcolor{color=gp lt color border}
\node[gp node right,font={\fontsize{15.0pt}{18.0pt}\selectfont}] at (4.920,6.028) {$N^5$};
\gpcolor{rgb color={0.941,0.894,0.259}}
\gpsetdashtype{gp dt 5}
\draw[gp path] (2.808,6.028)--(4.092,6.028);
\draw[gp path] (2.256,4.554)--(2.413,4.553)--(2.570,4.554)--(2.727,4.554)--(2.884,4.554)%
  --(3.041,4.555)--(3.198,4.555)--(3.354,4.556)--(3.511,4.558)--(3.668,4.559)--(3.825,4.560)%
  --(3.982,4.562)--(4.139,4.564)--(4.296,4.566)--(4.453,4.569)--(4.610,4.571)--(4.767,4.574)%
  --(4.924,4.578)--(5.081,4.581)--(5.237,4.585)--(5.394,4.590)--(5.551,4.594)--(5.708,4.599)%
  --(5.865,4.605)--(6.022,4.611)--(6.179,4.618)--(6.336,4.625)--(6.493,4.633)--(6.650,4.641)%
  --(6.807,4.650)--(6.963,4.660)--(7.120,4.670)--(7.277,4.681)--(7.434,4.694)--(7.591,4.707)%
  --(7.748,4.721)--(7.905,4.736)--(8.062,4.752)--(8.219,4.770)--(8.376,4.789)--(8.533,4.809)%
  --(8.690,4.831)--(8.847,4.854)--(9.003,4.879)--(9.160,4.906)--(9.317,4.935)--(9.474,4.965)%
  --(9.631,4.998)--(9.788,5.033)--(9.945,5.070)--(10.102,5.110)--(10.259,5.152)--(10.416,5.196)%
  --(10.573,5.243)--(10.730,5.293)--(10.886,5.346)--(11.043,5.402)--(11.200,5.460)--(11.357,5.522)%
  --(11.514,5.586)--(11.671,5.654);
\gpcolor{color=gp lt color border}
\gpsetdashtype{gp dt solid}
\gpsetlinewidth{1.00}
\draw[gp path] (2.256,8.287)--(2.256,1.478)--(11.671,1.478)--(11.671,8.287)--cycle;
%% coordinates of the plot area
\gpdefrectangularnode{gp plot 1}{\pgfpoint{2.256cm}{1.478cm}}{\pgfpoint{11.671cm}{8.287cm}}
\end{tikzpicture}
}
%% gnuplot variables

%% file: F2_model2.tex
\scalebox{0.65}[0.65]{
\begin{tikzpicture}[gnuplot]
%% generated with GNUPLOT 5.4p5 (Lua 5.4; terminal rev. Jun 2020, script rev. 115)
%% 6/20/2023 12:18:03 PM
\tikzset{every node/.append style={font={\fontsize{15.0pt}{18.0pt}\selectfont}}}
\path (0.000,0.000) rectangle (12.500,8.750);
\gpcolor{color=gp lt color border}
\gpsetlinetype{gp lt border}
\gpsetdashtype{gp dt solid}
\gpsetlinewidth{1.00}
\draw[gp path] (2.808,1.478)--(2.988,1.478);
\draw[gp path] (11.671,1.478)--(11.491,1.478);
\node[gp node right] at (2.532,1.478) {$-0.01$};
\draw[gp path] (2.808,3.180)--(2.988,3.180);
\draw[gp path] (11.671,3.180)--(11.491,3.180);
\node[gp node right] at (2.532,3.180) {$-0.005$};
\draw[gp path] (2.808,4.883)--(2.988,4.883);
\draw[gp path] (11.671,4.883)--(11.491,4.883);
\node[gp node right] at (2.532,4.883) {$0$};
\draw[gp path] (2.808,6.585)--(2.988,6.585);
\draw[gp path] (11.671,6.585)--(11.491,6.585);
\node[gp node right] at (2.532,6.585) {$0.005$};
\draw[gp path] (2.808,8.287)--(2.988,8.287);
\draw[gp path] (11.671,8.287)--(11.491,8.287);
\node[gp node right] at (2.532,8.287) {$0.01$};
\draw[gp path] (2.808,1.478)--(2.808,1.658);
\draw[gp path] (2.808,8.287)--(2.808,8.107);
\node[gp node center] at (2.808,1.016) {$0.2$};
\draw[gp path] (4.285,1.478)--(4.285,1.658);
\draw[gp path] (4.285,8.287)--(4.285,8.107);
\node[gp node center] at (4.285,1.016) {$0.3$};
\draw[gp path] (5.762,1.478)--(5.762,1.658);
\draw[gp path] (5.762,8.287)--(5.762,8.107);
\node[gp node center] at (5.762,1.016) {$0.4$};
\draw[gp path] (7.239,1.478)--(7.239,1.658);
\draw[gp path] (7.239,8.287)--(7.239,8.107);
\node[gp node center] at (7.239,1.016) {$0.5$};
\draw[gp path] (8.717,1.478)--(8.717,1.658);
\draw[gp path] (8.717,8.287)--(8.717,8.107);
\node[gp node center] at (8.717,1.016) {$0.6$};
\draw[gp path] (10.194,1.478)--(10.194,1.658);
\draw[gp path] (10.194,8.287)--(10.194,8.107);
\node[gp node center] at (10.194,1.016) {$0.7$};
\draw[gp path] (11.671,1.478)--(11.671,1.658);
\draw[gp path] (11.671,8.287)--(11.671,8.107);
\node[gp node center] at (11.671,1.016) {$0.8$};
\draw[gp path] (2.808,8.287)--(2.808,1.478)--(11.671,1.478)--(11.671,8.287)--cycle;
\node[gp node left,font={\fontsize{13.0pt}{15.6pt}\selectfont}] at (3.547,3.180) {$\sqrt{S_{ep}}=45$ GeV};
\node[gp node left,font={\fontsize{13.0pt}{15.6pt}\selectfont}] at (3.547,2.329) {$Q^2=10$ GeV${}^2$};
\node[gp node left,font={\fontsize{13.0pt}{15.6pt}\selectfont}] at (6.796,3.180) {$x_{B}=0.005$};
\node[gp node left,font={\fontsize{13.0pt}{15.6pt}\selectfont}] at (6.796,2.329) {$P_{J/\psi}^{\perp}=2$ GeV};
\node[gp node left,font={\fontsize{20.0pt}{24.0pt}\selectfont}] at (6.501,7.095) {Model 2};
\node[gp node center,rotate=-270,font={\fontsize{18.0pt}{21.6pt}\selectfont}] at (0.438,4.882) {${\cal F}_2/2\sigma^{\rm U}_1$};
\node[gp node center,font={\fontsize{22.0pt}{26.4pt}\selectfont}] at (7.239,0.323) {$z_f$};
\node[gp node right,font={\fontsize{15.0pt}{18.0pt}\selectfont}] at (5.472,7.876) {$N^1$};
\gpcolor{rgb color={0.580,0.000,0.827}}
\gpsetdashtype{gp dt 1}
\gpsetlinewidth{3.00}
\draw[gp path] (3.360,7.876)--(4.644,7.876);
\draw[gp path] (2.808,4.950)--(2.956,4.958)--(3.103,4.966)--(3.251,4.974)--(3.399,4.983)%
  --(3.547,4.992)--(3.694,5.001)--(3.842,5.011)--(3.990,5.021)--(4.137,5.031)--(4.285,5.042)%
  --(4.433,5.053)--(4.581,5.064)--(4.728,5.076)--(4.876,5.087)--(5.024,5.099)--(5.171,5.111)%
  --(5.319,5.124)--(5.467,5.136)--(5.615,5.149)--(5.762,5.163)--(5.910,5.176)--(6.058,5.190)%
  --(6.205,5.203)--(6.353,5.218)--(6.501,5.232)--(6.649,5.246)--(6.796,5.261)--(6.944,5.276)%
  --(7.092,5.291)--(7.239,5.306)--(7.387,5.322)--(7.535,5.338)--(7.683,5.354)--(7.830,5.370)%
  --(7.978,5.386)--(8.126,5.403)--(8.274,5.419)--(8.421,5.436)--(8.569,5.453)--(8.717,5.470)%
  --(8.864,5.487)--(9.012,5.504)--(9.160,5.521)--(9.308,5.538)--(9.455,5.555)--(9.603,5.573)%
  --(9.751,5.590)--(9.898,5.607)--(10.046,5.624)--(10.194,5.641)--(10.342,5.658)--(10.489,5.675)%
  --(10.637,5.691)--(10.785,5.707)--(10.932,5.724)--(11.080,5.740)--(11.228,5.756)--(11.376,5.771)%
  --(11.523,5.787)--(11.671,5.802);
\gpcolor{color=gp lt color border}
\node[gp node right,font={\fontsize{15.0pt}{18.0pt}\selectfont}] at (5.472,7.414) {$N^2$};
\gpcolor{rgb color={0.000,0.620,0.451}}
\gpsetdashtype{gp dt 2}
\draw[gp path] (3.360,7.414)--(4.644,7.414);
\draw[gp path] (2.808,4.782)--(2.956,4.770)--(3.103,4.758)--(3.251,4.746)--(3.399,4.733)%
  --(3.547,4.720)--(3.694,4.707)--(3.842,4.694)--(3.990,4.680)--(4.137,4.667)--(4.285,4.653)%
  --(4.433,4.640)--(4.581,4.627)--(4.728,4.614)--(4.876,4.601)--(5.024,4.588)--(5.171,4.576)%
  --(5.319,4.564)--(5.467,4.553)--(5.615,4.542)--(5.762,4.532)--(5.910,4.522)--(6.058,4.513)%
  --(6.205,4.505)--(6.353,4.498)--(6.501,4.492)--(6.649,4.487)--(6.796,4.483)--(6.944,4.480)%
  --(7.092,4.478)--(7.239,4.478)--(7.387,4.479)--(7.535,4.481)--(7.683,4.484)--(7.830,4.490)%
  --(7.978,4.496)--(8.126,4.505)--(8.274,4.514)--(8.421,4.526)--(8.569,4.539)--(8.717,4.554)%
  --(8.864,4.570)--(9.012,4.588)--(9.160,4.608)--(9.308,4.629)--(9.455,4.651)--(9.603,4.675)%
  --(9.751,4.700)--(9.898,4.725)--(10.046,4.752)--(10.194,4.779)--(10.342,4.806)--(10.489,4.833)%
  --(10.637,4.859)--(10.785,4.884)--(10.932,4.908)--(11.080,4.930)--(11.228,4.950)--(11.376,4.966)%
  --(11.523,4.979)--(11.671,4.987);
\gpcolor{color=gp lt color border}
\node[gp node right,font={\fontsize{15.0pt}{18.0pt}\selectfont}] at (5.472,6.952) {$N^3$};
\gpcolor{rgb color={0.337,0.706,0.914}}
\gpsetdashtype{gp dt 3}
\draw[gp path] (3.360,6.952)--(4.644,6.952);
\draw[gp path] (2.808,4.789)--(2.956,4.780)--(3.103,4.770)--(3.251,4.761)--(3.399,4.750)%
  --(3.547,4.740)--(3.694,4.728)--(3.842,4.717)--(3.990,4.705)--(4.137,4.692)--(4.285,4.679)%
  --(4.433,4.665)--(4.581,4.651)--(4.728,4.636)--(4.876,4.621)--(5.024,4.605)--(5.171,4.588)%
  --(5.319,4.571)--(5.467,4.552)--(5.615,4.533)--(5.762,4.514)--(5.910,4.493)--(6.058,4.471)%
  --(6.205,4.448)--(6.353,4.425)--(6.501,4.400)--(6.649,4.374)--(6.796,4.346)--(6.944,4.318)%
  --(7.092,4.288)--(7.239,4.256)--(7.387,4.223)--(7.535,4.188)--(7.683,4.152)--(7.830,4.114)%
  --(7.978,4.074)--(8.126,4.033)--(8.274,3.989)--(8.421,3.944)--(8.569,3.896)--(8.717,3.847)%
  --(8.864,3.796)--(9.012,3.743)--(9.160,3.688)--(9.308,3.632)--(9.455,3.574)--(9.603,3.515)%
  --(9.751,3.454)--(9.898,3.392)--(10.046,3.330)--(10.194,3.268)--(10.342,3.205)--(10.489,3.143)%
  --(10.637,3.082)--(10.785,3.023)--(10.932,2.966)--(11.080,2.912)--(11.228,2.861)--(11.376,2.816)%
  --(11.523,2.775)--(11.671,2.740);
\gpcolor{color=gp lt color border}
\node[gp node right,font={\fontsize{15.0pt}{18.0pt}\selectfont}] at (5.472,6.490) {$N^4$};
\gpcolor{rgb color={0.902,0.624,0.000}}
\gpsetdashtype{gp dt 4}
\draw[gp path] (3.360,6.490)--(4.644,6.490);
\draw[gp path] (2.808,4.941)--(2.956,4.951)--(3.103,4.960)--(3.251,4.971)--(3.399,4.983)%
  --(3.547,4.995)--(3.694,5.008)--(3.842,5.022)--(3.990,5.036)--(4.137,5.052)--(4.285,5.068)%
  --(4.433,5.085)--(4.581,5.104)--(4.728,5.123)--(4.876,5.143)--(5.024,5.164)--(5.171,5.187)%
  --(5.319,5.210)--(5.467,5.235)--(5.615,5.260)--(5.762,5.288)--(5.910,5.316)--(6.058,5.346)%
  --(6.205,5.377)--(6.353,5.409)--(6.501,5.443)--(6.649,5.479)--(6.796,5.516)--(6.944,5.554)%
  --(7.092,5.595)--(7.239,5.637)--(7.387,5.681)--(7.535,5.726)--(7.683,5.774)--(7.830,5.823)%
  --(7.978,5.874)--(8.126,5.927)--(8.274,5.981)--(8.421,6.038)--(8.569,6.096)--(8.717,6.156)%
  --(8.864,6.218)--(9.012,6.281)--(9.160,6.345)--(9.308,6.411)--(9.455,6.478)--(9.603,6.546)%
  --(9.751,6.614)--(9.898,6.682)--(10.046,6.750)--(10.194,6.818)--(10.342,6.885)--(10.489,6.950)%
  --(10.637,7.013)--(10.785,7.074)--(10.932,7.131)--(11.080,7.184)--(11.228,7.232)--(11.376,7.275)%
  --(11.523,7.311)--(11.671,7.340);
\gpcolor{color=gp lt color border}
\node[gp node right,font={\fontsize{15.0pt}{18.0pt}\selectfont}] at (5.472,6.028) {$N^5$};
\gpcolor{rgb color={0.941,0.894,0.259}}
\gpsetdashtype{gp dt 5}
\draw[gp path] (3.360,6.028)--(4.644,6.028);
\draw[gp path] (2.808,4.906)--(2.956,4.909)--(3.103,4.911)--(3.251,4.914)--(3.399,4.917)%
  --(3.547,4.919)--(3.694,4.922)--(3.842,4.925)--(3.990,4.927)--(4.137,4.930)--(4.285,4.932)%
  --(4.433,4.935)--(4.581,4.937)--(4.728,4.940)--(4.876,4.942)--(5.024,4.944)--(5.171,4.946)%
  --(5.319,4.948)--(5.467,4.949)--(5.615,4.951)--(5.762,4.952)--(5.910,4.954)--(6.058,4.955)%
  --(6.205,4.956)--(6.353,4.957)--(6.501,4.957)--(6.649,4.958)--(6.796,4.958)--(6.944,4.959)%
  --(7.092,4.959)--(7.239,4.959)--(7.387,4.959)--(7.535,4.960)--(7.683,4.960)--(7.830,4.960)%
  --(7.978,4.960)--(8.126,4.960)--(8.274,4.960)--(8.421,4.961)--(8.569,4.961)--(8.717,4.962)%
  --(8.864,4.963)--(9.012,4.964)--(9.160,4.965)--(9.308,4.967)--(9.455,4.969)--(9.603,4.972)%
  --(9.751,4.974)--(9.898,4.978)--(10.046,4.981)--(10.194,4.985)--(10.342,4.989)--(10.489,4.994)%
  --(10.637,4.999)--(10.785,5.004)--(10.932,5.010)--(11.080,5.015)--(11.228,5.021)--(11.376,5.027)%
  --(11.523,5.033)--(11.671,5.038);
\gpcolor{color=gp lt color border}
\gpsetdashtype{gp dt solid}
\gpsetlinewidth{1.00}
\draw[gp path] (2.808,8.287)--(2.808,1.478)--(11.671,1.478)--(11.671,8.287)--cycle;
%% coordinates of the plot area
\gpdefrectangularnode{gp plot 1}{\pgfpoint{2.808cm}{1.478cm}}{\pgfpoint{11.671cm}{8.287cm}}
\end{tikzpicture}
}
%% gnuplot variables

%% file: F3_model2.tex
\scalebox{0.65}[0.65]{
\begin{tikzpicture}[gnuplot]
%% generated with GNUPLOT 5.4p5 (Lua 5.4; terminal rev. Jun 2020, script rev. 115)
%% 6/20/2023 12:26:18 PM
\tikzset{every node/.append style={font={\fontsize{15.0pt}{18.0pt}\selectfont}}}
\path (0.000,0.000) rectangle (12.500,8.750);
\gpcolor{color=gp lt color border}
\gpsetlinetype{gp lt border}
\gpsetdashtype{gp dt solid}
\gpsetlinewidth{1.00}
\draw[gp path] (3.084,2.235)--(3.264,2.235);
\draw[gp path] (11.671,2.235)--(11.491,2.235);
\node[gp node right] at (2.808,2.235) {$-0.0004$};
\draw[gp path] (3.084,3.748)--(3.264,3.748);
\draw[gp path] (11.671,3.748)--(11.491,3.748);
\node[gp node right] at (2.808,3.748) {$-0.0002$};
\draw[gp path] (3.084,5.261)--(3.264,5.261);
\draw[gp path] (11.671,5.261)--(11.491,5.261);
\node[gp node right] at (2.808,5.261) {$0$};
\draw[gp path] (3.084,6.774)--(3.264,6.774);
\draw[gp path] (11.671,6.774)--(11.491,6.774);
\node[gp node right] at (2.808,6.774) {$0.0002$};
\draw[gp path] (3.084,8.287)--(3.264,8.287);
\draw[gp path] (11.671,8.287)--(11.491,8.287);
\node[gp node right] at (2.808,8.287) {$0.0004$};
\draw[gp path] (3.084,1.478)--(3.084,1.658);
\draw[gp path] (3.084,8.287)--(3.084,8.107);
\node[gp node center] at (3.084,1.016) {$0.2$};
\draw[gp path] (4.515,1.478)--(4.515,1.658);
\draw[gp path] (4.515,8.287)--(4.515,8.107);
\node[gp node center] at (4.515,1.016) {$0.3$};
\draw[gp path] (5.946,1.478)--(5.946,1.658);
\draw[gp path] (5.946,8.287)--(5.946,8.107);
\node[gp node center] at (5.946,1.016) {$0.4$};
\draw[gp path] (7.377,1.478)--(7.377,1.658);
\draw[gp path] (7.377,8.287)--(7.377,8.107);
\node[gp node center] at (7.377,1.016) {$0.5$};
\draw[gp path] (8.809,1.478)--(8.809,1.658);
\draw[gp path] (8.809,8.287)--(8.809,8.107);
\node[gp node center] at (8.809,1.016) {$0.6$};
\draw[gp path] (10.240,1.478)--(10.240,1.658);
\draw[gp path] (10.240,8.287)--(10.240,8.107);
\node[gp node center] at (10.240,1.016) {$0.7$};
\draw[gp path] (11.671,1.478)--(11.671,1.658);
\draw[gp path] (11.671,8.287)--(11.671,8.107);
\node[gp node center] at (11.671,1.016) {$0.8$};
\draw[gp path] (3.084,8.287)--(3.084,1.478)--(11.671,1.478)--(11.671,8.287)--cycle;
\node[gp node left,font={\fontsize{13.0pt}{15.6pt}\selectfont}] at (3.370,2.991) {$\sqrt{S_{ep}}=45$ GeV};
\node[gp node left,font={\fontsize{13.0pt}{15.6pt}\selectfont}] at (3.370,2.235) {$Q^2=10$ GeV${}^2$};
\node[gp node left,font={\fontsize{13.0pt}{15.6pt}\selectfont}] at (6.662,2.991) {$x_{B}=0.005$};
\node[gp node left,font={\fontsize{13.0pt}{15.6pt}\selectfont}] at (6.662,2.235) {$P_{J/\psi}^{\perp}=2$ GeV};
\node[gp node left,font={\fontsize{20.0pt}{24.0pt}\selectfont}] at (6.662,7.530) {Model 2};
\node[gp node center,rotate=-270,font={\fontsize{18.0pt}{21.6pt}\selectfont}] at (0.438,4.882) {${\cal F}_3/2\sigma^{\rm U}_1$};
\node[gp node center,font={\fontsize{22.0pt}{26.4pt}\selectfont}] at (7.377,0.323) {$z_f$};
\node[gp node right,font={\fontsize{15.0pt}{18.0pt}\selectfont}] at (5.748,7.876) {$N^1$};
\gpcolor{rgb color={0.580,0.000,0.827}}
\gpsetdashtype{gp dt 1}
\gpsetlinewidth{3.00}
\draw[gp path] (3.636,7.876)--(4.920,7.876);
\draw[gp path] (3.084,5.269)--(3.227,5.268)--(3.370,5.267)--(3.513,5.265)--(3.656,5.262)%
  --(3.800,5.259)--(3.943,5.256)--(4.086,5.252)--(4.229,5.247)--(4.372,5.242)--(4.515,5.235)%
  --(4.658,5.229)--(4.801,5.221)--(4.945,5.212)--(5.088,5.203)--(5.231,5.193)--(5.374,5.182)%
  --(5.517,5.170)--(5.660,5.157)--(5.803,5.143)--(5.946,5.128)--(6.089,5.113)--(6.233,5.096)%
  --(6.376,5.079)--(6.519,5.060)--(6.662,5.041)--(6.805,5.021)--(6.948,5.000)--(7.091,4.978)%
  --(7.234,4.956)--(7.377,4.933)--(7.521,4.910)--(7.664,4.886)--(7.807,4.861)--(7.950,4.837)%
  --(8.093,4.812)--(8.236,4.787)--(8.379,4.763)--(8.522,4.738)--(8.666,4.714)--(8.809,4.691)%
  --(8.952,4.668)--(9.095,4.646)--(9.238,4.626)--(9.381,4.606)--(9.524,4.588)--(9.667,4.571)%
  --(9.810,4.556)--(9.954,4.543)--(10.097,4.532)--(10.240,4.523)--(10.383,4.517)--(10.526,4.513)%
  --(10.669,4.511)--(10.812,4.512)--(10.955,4.516)--(11.099,4.523)--(11.242,4.532)--(11.385,4.544)%
  --(11.528,4.558)--(11.671,4.576);
\gpcolor{color=gp lt color border}
\node[gp node right,font={\fontsize{15.0pt}{18.0pt}\selectfont}] at (5.748,7.414) {$N^2$};
\gpcolor{rgb color={0.000,0.620,0.451}}
\gpsetdashtype{gp dt 2}
\draw[gp path] (3.636,7.414)--(4.920,7.414);
\draw[gp path] (3.084,5.231)--(3.227,5.229)--(3.370,5.228)--(3.513,5.226)--(3.656,5.225)%
  --(3.800,5.223)--(3.943,5.222)--(4.086,5.222)--(4.229,5.221)--(4.372,5.221)--(4.515,5.221)%
  --(4.658,5.221)--(4.801,5.222)--(4.945,5.223)--(5.088,5.224)--(5.231,5.226)--(5.374,5.228)%
  --(5.517,5.231)--(5.660,5.233)--(5.803,5.237)--(5.946,5.240)--(6.089,5.245)--(6.233,5.249)%
  --(6.376,5.254)--(6.519,5.260)--(6.662,5.265)--(6.805,5.272)--(6.948,5.278)--(7.091,5.286)%
  --(7.234,5.293)--(7.377,5.301)--(7.521,5.310)--(7.664,5.319)--(7.807,5.328)--(7.950,5.338)%
  --(8.093,5.348)--(8.236,5.359)--(8.379,5.371)--(8.522,5.382)--(8.666,5.395)--(8.809,5.407)%
  --(8.952,5.420)--(9.095,5.434)--(9.238,5.448)--(9.381,5.463)--(9.524,5.478)--(9.667,5.494)%
  --(9.810,5.511)--(9.954,5.528)--(10.097,5.545)--(10.240,5.563)--(10.383,5.582)--(10.526,5.601)%
  --(10.669,5.620)--(10.812,5.640)--(10.955,5.660)--(11.099,5.679)--(11.242,5.699)--(11.385,5.719)%
  --(11.528,5.737)--(11.671,5.755);
\gpcolor{color=gp lt color border}
\node[gp node right,font={\fontsize{15.0pt}{18.0pt}\selectfont}] at (5.748,6.952) {$N^3$};
\gpcolor{rgb color={0.337,0.706,0.914}}
\gpsetdashtype{gp dt 3}
\draw[gp path] (3.636,6.952)--(4.920,6.952);
\draw[gp path] (3.084,5.154)--(3.227,5.151)--(3.370,5.148)--(3.513,5.147)--(3.656,5.146)%
  --(3.800,5.147)--(3.943,5.149)--(4.086,5.152)--(4.229,5.156)--(4.372,5.162)--(4.515,5.170)%
  --(4.658,5.179)--(4.801,5.191)--(4.945,5.204)--(5.088,5.219)--(5.231,5.236)--(5.374,5.256)%
  --(5.517,5.278)--(5.660,5.303)--(5.803,5.330)--(5.946,5.360)--(6.089,5.393)--(6.233,5.428)%
  --(6.376,5.467)--(6.519,5.509)--(6.662,5.554)--(6.805,5.602)--(6.948,5.653)--(7.091,5.708)%
  --(7.234,5.765)--(7.377,5.826)--(7.521,5.891)--(7.664,5.958)--(7.807,6.028)--(7.950,6.101)%
  --(8.093,6.177)--(8.236,6.255)--(8.379,6.335)--(8.522,6.417)--(8.666,6.500)--(8.809,6.585)%
  --(8.952,6.669)--(9.095,6.754)--(9.238,6.839)--(9.381,6.922)--(9.524,7.003)--(9.667,7.082)%
  --(9.810,7.157)--(9.954,7.228)--(10.097,7.293)--(10.240,7.353)--(10.383,7.405)--(10.526,7.450)%
  --(10.669,7.486)--(10.812,7.512)--(10.955,7.527)--(11.099,7.531)--(11.242,7.523)--(11.385,7.502)%
  --(11.528,7.468)--(11.671,7.421);
\gpcolor{color=gp lt color border}
\node[gp node right,font={\fontsize{15.0pt}{18.0pt}\selectfont}] at (5.748,6.490) {$N^4$};
\gpcolor{rgb color={0.902,0.624,0.000}}
\gpsetdashtype{gp dt 4}
\draw[gp path] (3.636,6.490)--(4.920,6.490);
\draw[gp path] (3.084,5.071)--(3.227,5.053)--(3.370,5.034)--(3.513,5.015)--(3.656,4.994)%
  --(3.800,4.972)--(3.943,4.950)--(4.086,4.926)--(4.229,4.901)--(4.372,4.875)--(4.515,4.848)%
  --(4.658,4.819)--(4.801,4.790)--(4.945,4.759)--(5.088,4.727)--(5.231,4.694)--(5.374,4.659)%
  --(5.517,4.624)--(5.660,4.587)--(5.803,4.548)--(5.946,4.509)--(6.089,4.469)--(6.233,4.427)%
  --(6.376,4.384)--(6.519,4.340)--(6.662,4.295)--(6.805,4.249)--(6.948,4.202)--(7.091,4.154)%
  --(7.234,4.106)--(7.377,4.057)--(7.521,4.007)--(7.664,3.957)--(7.807,3.907)--(7.950,3.857)%
  --(8.093,3.807)--(8.236,3.758)--(8.379,3.709)--(8.522,3.661)--(8.666,3.614)--(8.809,3.569)%
  --(8.952,3.525)--(9.095,3.484)--(9.238,3.445)--(9.381,3.409)--(9.524,3.377)--(9.667,3.348)%
  --(9.810,3.323)--(9.954,3.303)--(10.097,3.288)--(10.240,3.278)--(10.383,3.274)--(10.526,3.277)%
  --(10.669,3.286)--(10.812,3.302)--(10.955,3.326)--(11.099,3.357)--(11.242,3.396)--(11.385,3.443)%
  --(11.528,3.498)--(11.671,3.560);
\gpcolor{color=gp lt color border}
\node[gp node right,font={\fontsize{15.0pt}{18.0pt}\selectfont}] at (5.748,6.028) {$N^5$};
\gpcolor{rgb color={0.941,0.894,0.259}}
\gpsetdashtype{gp dt 5}
\draw[gp path] (3.636,6.028)--(4.920,6.028);
\draw[gp path] (3.084,5.223)--(3.227,5.217)--(3.370,5.211)--(3.513,5.204)--(3.656,5.196)%
  --(3.800,5.188)--(3.943,5.179)--(4.086,5.169)--(4.229,5.158)--(4.372,5.147)--(4.515,5.135)%
  --(4.658,5.121)--(4.801,5.108)--(4.945,5.093)--(5.088,5.077)--(5.231,5.060)--(5.374,5.043)%
  --(5.517,5.024)--(5.660,5.005)--(5.803,4.984)--(5.946,4.963)--(6.089,4.940)--(6.233,4.917)%
  --(6.376,4.892)--(6.519,4.867)--(6.662,4.840)--(6.805,4.813)--(6.948,4.785)--(7.091,4.756)%
  --(7.234,4.726)--(7.377,4.695)--(7.521,4.664)--(7.664,4.632)--(7.807,4.599)--(7.950,4.566)%
  --(8.093,4.533)--(8.236,4.500)--(8.379,4.467)--(8.522,4.434)--(8.666,4.401)--(8.809,4.368)%
  --(8.952,4.337)--(9.095,4.306)--(9.238,4.277)--(9.381,4.249)--(9.524,4.223)--(9.667,4.199)%
  --(9.810,4.177)--(9.954,4.158)--(10.097,4.141)--(10.240,4.129)--(10.383,4.119)--(10.526,4.114)%
  --(10.669,4.112)--(10.812,4.115)--(10.955,4.123)--(11.099,4.136)--(11.242,4.153)--(11.385,4.176)%
  --(11.528,4.204)--(11.671,4.237);
\gpcolor{color=gp lt color border}
\gpsetdashtype{gp dt solid}
\gpsetlinewidth{1.00}
\draw[gp path] (3.084,8.287)--(3.084,1.478)--(11.671,1.478)--(11.671,8.287)--cycle;
%% coordinates of the plot area
\gpdefrectangularnode{gp plot 1}{\pgfpoint{3.084cm}{1.478cm}}{\pgfpoint{11.671cm}{8.287cm}}
\end{tikzpicture}
}
%% gnuplot variables

%% file: F4_model2.tex
\scalebox{0.65}[0.65]{
\begin{tikzpicture}[gnuplot]
%% generated with GNUPLOT 5.4p5 (Lua 5.4; terminal rev. Jun 2020, script rev. 115)
%% 6/20/2023 12:33:05 PM
\tikzset{every node/.append style={font={\fontsize{15.0pt}{18.0pt}\selectfont}}}
\path (0.000,0.000) rectangle (12.500,8.750);
\gpcolor{color=gp lt color border}
\gpsetlinetype{gp lt border}
\gpsetdashtype{gp dt solid}
\gpsetlinewidth{1.00}
\draw[gp path] (2.808,1.478)--(2.988,1.478);
\draw[gp path] (11.671,1.478)--(11.491,1.478);
\node[gp node right] at (2.532,1.478) {$-0.004$};
\draw[gp path] (2.808,2.911)--(2.988,2.911);
\draw[gp path] (11.671,2.911)--(11.491,2.911);
\node[gp node right] at (2.532,2.911) {$-0.002$};
\draw[gp path] (2.808,4.345)--(2.988,4.345);
\draw[gp path] (11.671,4.345)--(11.491,4.345);
\node[gp node right] at (2.532,4.345) {$0$};
\draw[gp path] (2.808,5.778)--(2.988,5.778);
\draw[gp path] (11.671,5.778)--(11.491,5.778);
\node[gp node right] at (2.532,5.778) {$0.002$};
\draw[gp path] (2.808,7.212)--(2.988,7.212);
\draw[gp path] (11.671,7.212)--(11.491,7.212);
\node[gp node right] at (2.532,7.212) {$0.004$};
\draw[gp path] (2.808,1.478)--(2.808,1.658);
\draw[gp path] (2.808,8.287)--(2.808,8.107);
\node[gp node center] at (2.808,1.016) {$0.2$};
\draw[gp path] (4.285,1.478)--(4.285,1.658);
\draw[gp path] (4.285,8.287)--(4.285,8.107);
\node[gp node center] at (4.285,1.016) {$0.3$};
\draw[gp path] (5.762,1.478)--(5.762,1.658);
\draw[gp path] (5.762,8.287)--(5.762,8.107);
\node[gp node center] at (5.762,1.016) {$0.4$};
\draw[gp path] (7.239,1.478)--(7.239,1.658);
\draw[gp path] (7.239,8.287)--(7.239,8.107);
\node[gp node center] at (7.239,1.016) {$0.5$};
\draw[gp path] (8.717,1.478)--(8.717,1.658);
\draw[gp path] (8.717,8.287)--(8.717,8.107);
\node[gp node center] at (8.717,1.016) {$0.6$};
\draw[gp path] (10.194,1.478)--(10.194,1.658);
\draw[gp path] (10.194,8.287)--(10.194,8.107);
\node[gp node center] at (10.194,1.016) {$0.7$};
\draw[gp path] (11.671,1.478)--(11.671,1.658);
\draw[gp path] (11.671,8.287)--(11.671,8.107);
\node[gp node center] at (11.671,1.016) {$0.8$};
\draw[gp path] (2.808,8.287)--(2.808,1.478)--(11.671,1.478)--(11.671,8.287)--cycle;
\node[gp node left,font={\fontsize{13.0pt}{15.6pt}\selectfont}] at (3.399,2.911) {$\sqrt{S_{ep}}=45$ GeV};
\node[gp node left,font={\fontsize{13.0pt}{15.6pt}\selectfont}] at (3.399,2.195) {$Q^2=10$ GeV${}^2$};
\node[gp node left,font={\fontsize{13.0pt}{15.6pt}\selectfont}] at (6.501,2.911) {$x_{B}=0.005$};
\node[gp node left,font={\fontsize{13.0pt}{15.6pt}\selectfont}] at (6.501,2.195) {$P_{J/\psi}^{\perp}=2$ GeV};
\node[gp node left,font={\fontsize{20.0pt}{24.0pt}\selectfont}] at (6.501,7.212) {Model 2};
\node[gp node center,rotate=-270,font={\fontsize{18.0pt}{21.6pt}\selectfont}] at (0.438,4.882) {${\cal F}_4/2\sigma^{\rm U}_1$};
\node[gp node center,font={\fontsize{22.0pt}{26.4pt}\selectfont}] at (7.239,0.323) {$z_f$};
\node[gp node right,font={\fontsize{15.0pt}{18.0pt}\selectfont}] at (5.472,7.876) {$N^1$};
\gpcolor{rgb color={0.580,0.000,0.827}}
\gpsetdashtype{gp dt 1}
\gpsetlinewidth{3.00}
\draw[gp path] (3.360,7.876)--(4.644,7.876);
\draw[gp path] (2.808,4.176)--(2.956,4.164)--(3.103,4.151)--(3.251,4.138)--(3.399,4.125)%
  --(3.547,4.112)--(3.694,4.098)--(3.842,4.084)--(3.990,4.069)--(4.137,4.054)--(4.285,4.039)%
  --(4.433,4.023)--(4.581,4.007)--(4.728,3.991)--(4.876,3.974)--(5.024,3.957)--(5.171,3.939)%
  --(5.319,3.921)--(5.467,3.902)--(5.615,3.883)--(5.762,3.863)--(5.910,3.842)--(6.058,3.821)%
  --(6.205,3.800)--(6.353,3.778)--(6.501,3.755)--(6.649,3.732)--(6.796,3.708)--(6.944,3.683)%
  --(7.092,3.657)--(7.239,3.631)--(7.387,3.605)--(7.535,3.577)--(7.683,3.549)--(7.830,3.520)%
  --(7.978,3.490)--(8.126,3.460)--(8.274,3.429)--(8.421,3.398)--(8.569,3.366)--(8.717,3.333)%
  --(8.864,3.300)--(9.012,3.267)--(9.160,3.234)--(9.308,3.200)--(9.455,3.166)--(9.603,3.133)%
  --(9.751,3.099)--(9.898,3.067)--(10.046,3.034)--(10.194,3.003)--(10.342,2.973)--(10.489,2.944)%
  --(10.637,2.917)--(10.785,2.892)--(10.932,2.870)--(11.080,2.850)--(11.228,2.833)--(11.376,2.819)%
  --(11.523,2.809)--(11.671,2.804);
\gpcolor{color=gp lt color border}
\node[gp node right,font={\fontsize{15.0pt}{18.0pt}\selectfont}] at (5.472,7.414) {$N^2$};
\gpcolor{rgb color={0.000,0.620,0.451}}
\gpsetdashtype{gp dt 2}
\draw[gp path] (3.360,7.414)--(4.644,7.414);
\draw[gp path] (2.808,4.846)--(2.956,4.881)--(3.103,4.916)--(3.251,4.951)--(3.399,4.987)%
  --(3.547,5.023)--(3.694,5.060)--(3.842,5.097)--(3.990,5.135)--(4.137,5.173)--(4.285,5.211)%
  --(4.433,5.250)--(4.581,5.290)--(4.728,5.330)--(4.876,5.371)--(5.024,5.412)--(5.171,5.454)%
  --(5.319,5.497)--(5.467,5.541)--(5.615,5.586)--(5.762,5.631)--(5.910,5.678)--(6.058,5.725)%
  --(6.205,5.774)--(6.353,5.824)--(6.501,5.875)--(6.649,5.927)--(6.796,5.981)--(6.944,6.036)%
  --(7.092,6.092)--(7.239,6.150)--(7.387,6.209)--(7.535,6.270)--(7.683,6.332)--(7.830,6.395)%
  --(7.978,6.460)--(8.126,6.526)--(8.274,6.593)--(8.421,6.662)--(8.569,6.731)--(8.717,6.801)%
  --(8.864,6.871)--(9.012,6.942)--(9.160,7.013)--(9.308,7.083)--(9.455,7.152)--(9.603,7.220)%
  --(9.751,7.286)--(9.898,7.350)--(10.046,7.411)--(10.194,7.468)--(10.342,7.521)--(10.489,7.568)%
  --(10.637,7.610)--(10.785,7.644)--(10.932,7.671)--(11.080,7.689)--(11.228,7.697)--(11.376,7.695)%
  --(11.523,7.682)--(11.671,7.656);
\gpcolor{color=gp lt color border}
\node[gp node right,font={\fontsize{15.0pt}{18.0pt}\selectfont}] at (5.472,6.952) {$N^3$};
\gpcolor{rgb color={0.337,0.706,0.914}}
\gpsetdashtype{gp dt 3}
\draw[gp path] (3.360,6.952)--(4.644,6.952);
\draw[gp path] (2.808,4.291)--(2.956,4.288)--(3.103,4.285)--(3.251,4.282)--(3.399,4.278)%
  --(3.547,4.275)--(3.694,4.272)--(3.842,4.270)--(3.990,4.267)--(4.137,4.264)--(4.285,4.261)%
  --(4.433,4.258)--(4.581,4.255)--(4.728,4.253)--(4.876,4.250)--(5.024,4.248)--(5.171,4.245)%
  --(5.319,4.243)--(5.467,4.241)--(5.615,4.239)--(5.762,4.237)--(5.910,4.235)--(6.058,4.233)%
  --(6.205,4.231)--(6.353,4.229)--(6.501,4.228)--(6.649,4.227)--(6.796,4.225)--(6.944,4.224)%
  --(7.092,4.223)--(7.239,4.223)--(7.387,4.222)--(7.535,4.222)--(7.683,4.222)--(7.830,4.222)%
  --(7.978,4.222)--(8.126,4.222)--(8.274,4.223)--(8.421,4.224)--(8.569,4.225)--(8.717,4.226)%
  --(8.864,4.228)--(9.012,4.230)--(9.160,4.232)--(9.308,4.234)--(9.455,4.236)--(9.603,4.239)%
  --(9.751,4.242)--(9.898,4.245)--(10.046,4.249)--(10.194,4.252)--(10.342,4.256)--(10.489,4.260)%
  --(10.637,4.264)--(10.785,4.269)--(10.932,4.273)--(11.080,4.278)--(11.228,4.283)--(11.376,4.287)%
  --(11.523,4.292)--(11.671,4.297);
\gpcolor{color=gp lt color border}
\node[gp node right,font={\fontsize{15.0pt}{18.0pt}\selectfont}] at (5.472,6.490) {$N^4$};
\gpcolor{rgb color={0.902,0.624,0.000}}
\gpsetdashtype{gp dt 4}
\draw[gp path] (3.360,6.490)--(4.644,6.490);
\draw[gp path] (2.808,4.129)--(2.956,4.115)--(3.103,4.101)--(3.251,4.087)--(3.399,4.072)%
  --(3.547,4.058)--(3.694,4.044)--(3.842,4.029)--(3.990,4.015)--(4.137,4.000)--(4.285,3.986)%
  --(4.433,3.971)--(4.581,3.957)--(4.728,3.942)--(4.876,3.928)--(5.024,3.913)--(5.171,3.898)%
  --(5.319,3.884)--(5.467,3.869)--(5.615,3.854)--(5.762,3.839)--(5.910,3.823)--(6.058,3.808)%
  --(6.205,3.792)--(6.353,3.777)--(6.501,3.761)--(6.649,3.744)--(6.796,3.728)--(6.944,3.711)%
  --(7.092,3.693)--(7.239,3.675)--(7.387,3.657)--(7.535,3.638)--(7.683,3.618)--(7.830,3.598)%
  --(7.978,3.578)--(8.126,3.556)--(8.274,3.534)--(8.421,3.511)--(8.569,3.487)--(8.717,3.462)%
  --(8.864,3.437)--(9.012,3.411)--(9.160,3.383)--(9.308,3.355)--(9.455,3.327)--(9.603,3.297)%
  --(9.751,3.267)--(9.898,3.237)--(10.046,3.206)--(10.194,3.175)--(10.342,3.144)--(10.489,3.114)%
  --(10.637,3.084)--(10.785,3.055)--(10.932,3.027)--(11.080,3.001)--(11.228,2.977)--(11.376,2.956)%
  --(11.523,2.937)--(11.671,2.922);
\gpcolor{color=gp lt color border}
\node[gp node right,font={\fontsize{15.0pt}{18.0pt}\selectfont}] at (5.472,6.028) {$N^5$};
\gpcolor{rgb color={0.941,0.894,0.259}}
\gpsetdashtype{gp dt 5}
\draw[gp path] (3.360,6.028)--(4.644,6.028);
\draw[gp path] (2.808,4.183)--(2.956,4.172)--(3.103,4.161)--(3.251,4.150)--(3.399,4.139)%
  --(3.547,4.128)--(3.694,4.116)--(3.842,4.105)--(3.990,4.093)--(4.137,4.082)--(4.285,4.070)%
  --(4.433,4.058)--(4.581,4.046)--(4.728,4.034)--(4.876,4.022)--(5.024,4.010)--(5.171,3.998)%
  --(5.319,3.986)--(5.467,3.973)--(5.615,3.960)--(5.762,3.947)--(5.910,3.934)--(6.058,3.920)%
  --(6.205,3.906)--(6.353,3.892)--(6.501,3.878)--(6.649,3.863)--(6.796,3.847)--(6.944,3.831)%
  --(7.092,3.815)--(7.239,3.797)--(7.387,3.780)--(7.535,3.761)--(7.683,3.742)--(7.830,3.722)%
  --(7.978,3.701)--(8.126,3.679)--(8.274,3.656)--(8.421,3.632)--(8.569,3.607)--(8.717,3.581)%
  --(8.864,3.554)--(9.012,3.526)--(9.160,3.497)--(9.308,3.467)--(9.455,3.435)--(9.603,3.403)%
  --(9.751,3.370)--(9.898,3.337)--(10.046,3.302)--(10.194,3.268)--(10.342,3.233)--(10.489,3.199)%
  --(10.637,3.164)--(10.785,3.131)--(10.932,3.099)--(11.080,3.068)--(11.228,3.039)--(11.376,3.013)%
  --(11.523,2.990)--(11.671,2.971);
\gpcolor{color=gp lt color border}
\gpsetdashtype{gp dt solid}
\gpsetlinewidth{1.00}
\draw[gp path] (2.808,8.287)--(2.808,1.478)--(11.671,1.478)--(11.671,8.287)--cycle;
%% coordinates of the plot area
\gpdefrectangularnode{gp plot 1}{\pgfpoint{2.808cm}{1.478cm}}{\pgfpoint{11.671cm}{8.287cm}}
\end{tikzpicture}
}
%% gnuplot variables

%% file: F5_model2.tex
\scalebox{0.7}[0.7]{
\begin{tikzpicture}[gnuplot]
%% generated with GNUPLOT 5.4p5 (Lua 5.4; terminal rev. Jun 2020, script rev. 115)
%% 6/20/2023 12:39:02 PM
\tikzset{every node/.append style={font={\fontsize{15.0pt}{18.0pt}\selectfont}}}
\path (0.000,0.000) rectangle (12.500,8.750);
\gpcolor{color=gp lt color border}
\gpsetlinetype{gp lt border}
\gpsetdashtype{gp dt solid}
\gpsetlinewidth{1.00}
\draw[gp path] (3.084,1.478)--(3.264,1.478);
\draw[gp path] (11.671,1.478)--(11.491,1.478);
\node[gp node right] at (2.808,1.478) {$-0.0004$};
\draw[gp path] (3.084,3.080)--(3.264,3.080);
\draw[gp path] (11.671,3.080)--(11.491,3.080);
\node[gp node right] at (2.808,3.080) {$-0.0002$};
\draw[gp path] (3.084,4.682)--(3.264,4.682);
\draw[gp path] (11.671,4.682)--(11.491,4.682);
\node[gp node right] at (2.808,4.682) {$0$};
\draw[gp path] (3.084,6.284)--(3.264,6.284);
\draw[gp path] (11.671,6.284)--(11.491,6.284);
\node[gp node right] at (2.808,6.284) {$0.0002$};
\draw[gp path] (3.084,7.886)--(3.264,7.886);
\draw[gp path] (11.671,7.886)--(11.491,7.886);
\node[gp node right] at (2.808,7.886) {$0.0004$};
\draw[gp path] (3.084,1.478)--(3.084,1.658);
\draw[gp path] (3.084,8.287)--(3.084,8.107);
\node[gp node center] at (3.084,1.016) {$0.2$};
\draw[gp path] (4.515,1.478)--(4.515,1.658);
\draw[gp path] (4.515,8.287)--(4.515,8.107);
\node[gp node center] at (4.515,1.016) {$0.3$};
\draw[gp path] (5.946,1.478)--(5.946,1.658);
\draw[gp path] (5.946,8.287)--(5.946,8.107);
\node[gp node center] at (5.946,1.016) {$0.4$};
\draw[gp path] (7.377,1.478)--(7.377,1.658);
\draw[gp path] (7.377,8.287)--(7.377,8.107);
\node[gp node center] at (7.377,1.016) {$0.5$};
\draw[gp path] (8.809,1.478)--(8.809,1.658);
\draw[gp path] (8.809,8.287)--(8.809,8.107);
\node[gp node center] at (8.809,1.016) {$0.6$};
\draw[gp path] (10.240,1.478)--(10.240,1.658);
\draw[gp path] (10.240,8.287)--(10.240,8.107);
\node[gp node center] at (10.240,1.016) {$0.7$};
\draw[gp path] (11.671,1.478)--(11.671,1.658);
\draw[gp path] (11.671,8.287)--(11.671,8.107);
\node[gp node center] at (11.671,1.016) {$0.8$};
\draw[gp path] (3.084,8.287)--(3.084,1.478)--(11.671,1.478)--(11.671,8.287)--cycle;
\node[gp node left,font={\fontsize{13.0pt}{15.6pt}\selectfont}] at (3.800,3.080) {$\sqrt{S_{ep}}=45$ GeV};
\node[gp node left,font={\fontsize{13.0pt}{15.6pt}\selectfont}] at (3.800,2.279) {$Q^2=10$ GeV${}^2$};
\node[gp node left,font={\fontsize{13.0pt}{15.6pt}\selectfont}] at (6.948,3.080) {$x_{B}=0.005$};
\node[gp node left,font={\fontsize{13.0pt}{15.6pt}\selectfont}] at (6.948,2.279) {$P_{J/\psi}^{\perp}=2$ GeV};
\node[gp node left,font={\fontsize{20.0pt}{24.0pt}\selectfont}] at (6.662,7.085) {Model 2};
\node[gp node center,rotate=-270,font={\fontsize{18.0pt}{21.6pt}\selectfont}] at (0.438,4.882) {${\cal F}_5/2\sigma^{\rm U}_1$};
\node[gp node center,font={\fontsize{22.0pt}{26.4pt}\selectfont}] at (7.377,0.323) {$z_f$};
\node[gp node right,font={\fontsize{15.0pt}{18.0pt}\selectfont}] at (5.748,7.876) {$N^1$};
\gpcolor{rgb color={0.580,0.000,0.827}}
\gpsetdashtype{gp dt 1}
\gpsetlinewidth{3.00}
\draw[gp path] (3.636,7.876)--(4.920,7.876);
\draw[gp path] (3.084,4.005)--(3.227,3.996)--(3.370,3.989)--(3.513,3.983)--(3.656,3.977)%
  --(3.800,3.973)--(3.943,3.970)--(4.086,3.967)--(4.229,3.966)--(4.372,3.966)--(4.515,3.966)%
  --(4.658,3.968)--(4.801,3.971)--(4.945,3.975)--(5.088,3.979)--(5.231,3.985)--(5.374,3.992)%
  --(5.517,4.000)--(5.660,4.009)--(5.803,4.020)--(5.946,4.031)--(6.089,4.043)--(6.233,4.057)%
  --(6.376,4.072)--(6.519,4.088)--(6.662,4.105)--(6.805,4.123)--(6.948,4.142)--(7.091,4.163)%
  --(7.234,4.184)--(7.377,4.207)--(7.521,4.231)--(7.664,4.256)--(7.807,4.281)--(7.950,4.308)%
  --(8.093,4.336)--(8.236,4.364)--(8.379,4.393)--(8.522,4.423)--(8.666,4.454)--(8.809,4.484)%
  --(8.952,4.516)--(9.095,4.547)--(9.238,4.579)--(9.381,4.610)--(9.524,4.642)--(9.667,4.673)%
  --(9.810,4.703)--(9.954,4.733)--(10.097,4.761)--(10.240,4.789)--(10.383,4.815)--(10.526,4.840)%
  --(10.669,4.863)--(10.812,4.884)--(10.955,4.902)--(11.099,4.919)--(11.242,4.932)--(11.385,4.943)%
  --(11.528,4.952)--(11.671,4.957);
\gpcolor{color=gp lt color border}
\node[gp node right,font={\fontsize{15.0pt}{18.0pt}\selectfont}] at (5.748,7.414) {$N^2$};
\gpcolor{rgb color={0.000,0.620,0.451}}
\gpsetdashtype{gp dt 2}
\draw[gp path] (3.636,7.414)--(4.920,7.414);
\draw[gp path] (3.084,4.580)--(3.227,4.566)--(3.370,4.551)--(3.513,4.536)--(3.656,4.519)%
  --(3.800,4.501)--(3.943,4.482)--(4.086,4.462)--(4.229,4.441)--(4.372,4.419)--(4.515,4.395)%
  --(4.658,4.371)--(4.801,4.346)--(4.945,4.319)--(5.088,4.292)--(5.231,4.264)--(5.374,4.235)%
  --(5.517,4.204)--(5.660,4.173)--(5.803,4.142)--(5.946,4.109)--(6.089,4.076)--(6.233,4.042)%
  --(6.376,4.008)--(6.519,3.973)--(6.662,3.938)--(6.805,3.903)--(6.948,3.868)--(7.091,3.832)%
  --(7.234,3.797)--(7.377,3.762)--(7.521,3.728)--(7.664,3.694)--(7.807,3.661)--(7.950,3.629)%
  --(8.093,3.599)--(8.236,3.569)--(8.379,3.541)--(8.522,3.515)--(8.666,3.491)--(8.809,3.469)%
  --(8.952,3.450)--(9.095,3.433)--(9.238,3.420)--(9.381,3.409)--(9.524,3.403)--(9.667,3.400)%
  --(9.810,3.400)--(9.954,3.405)--(10.097,3.414)--(10.240,3.428)--(10.383,3.447)--(10.526,3.470)%
  --(10.669,3.498)--(10.812,3.530)--(10.955,3.568)--(11.099,3.610)--(11.242,3.657)--(11.385,3.707)%
  --(11.528,3.762)--(11.671,3.820);
\gpcolor{color=gp lt color border}
\node[gp node right,font={\fontsize{15.0pt}{18.0pt}\selectfont}] at (5.748,6.952) {$N^3$};
\gpcolor{rgb color={0.337,0.706,0.914}}
\gpsetdashtype{gp dt 3}
\draw[gp path] (3.636,6.952)--(4.920,6.952);
\draw[gp path] (3.084,5.482)--(3.227,5.501)--(3.370,5.521)--(3.513,5.540)--(3.656,5.559)%
  --(3.800,5.578)--(3.943,5.597)--(4.086,5.616)--(4.229,5.635)--(4.372,5.654)--(4.515,5.673)%
  --(4.658,5.693)--(4.801,5.712)--(4.945,5.732)--(5.088,5.751)--(5.231,5.771)--(5.374,5.791)%
  --(5.517,5.811)--(5.660,5.831)--(5.803,5.851)--(5.946,5.871)--(6.089,5.891)--(6.233,5.911)%
  --(6.376,5.931)--(6.519,5.951)--(6.662,5.971)--(6.805,5.990)--(6.948,6.010)--(7.091,6.029)%
  --(7.234,6.048)--(7.377,6.067)--(7.521,6.085)--(7.664,6.103)--(7.807,6.120)--(7.950,6.137)%
  --(8.093,6.153)--(8.236,6.168)--(8.379,6.182)--(8.522,6.195)--(8.666,6.207)--(8.809,6.217)%
  --(8.952,6.226)--(9.095,6.234)--(9.238,6.240)--(9.381,6.244)--(9.524,6.246)--(9.667,6.246)%
  --(9.810,6.244)--(9.954,6.239)--(10.097,6.231)--(10.240,6.221)--(10.383,6.208)--(10.526,6.192)%
  --(10.669,6.172)--(10.812,6.150)--(10.955,6.124)--(11.099,6.094)--(11.242,6.061)--(11.385,6.025)%
  --(11.528,5.985)--(11.671,5.941);
\gpcolor{color=gp lt color border}
\node[gp node right,font={\fontsize{15.0pt}{18.0pt}\selectfont}] at (5.748,6.490) {$N^4$};
\gpcolor{rgb color={0.902,0.624,0.000}}
\gpsetdashtype{gp dt 4}
\draw[gp path] (3.636,6.490)--(4.920,6.490);
\draw[gp path] (3.084,4.785)--(3.227,4.796)--(3.370,4.808)--(3.513,4.820)--(3.656,4.832)%
  --(3.800,4.846)--(3.943,4.859)--(4.086,4.874)--(4.229,4.889)--(4.372,4.904)--(4.515,4.920)%
  --(4.658,4.936)--(4.801,4.953)--(4.945,4.971)--(5.088,4.989)--(5.231,5.007)--(5.374,5.026)%
  --(5.517,5.045)--(5.660,5.064)--(5.803,5.084)--(5.946,5.103)--(6.089,5.124)--(6.233,5.144)%
  --(6.376,5.164)--(6.519,5.185)--(6.662,5.205)--(6.805,5.225)--(6.948,5.245)--(7.091,5.265)%
  --(7.234,5.285)--(7.377,5.304)--(7.521,5.323)--(7.664,5.342)--(7.807,5.359)--(7.950,5.376)%
  --(8.093,5.393)--(8.236,5.408)--(8.379,5.422)--(8.522,5.435)--(8.666,5.447)--(8.809,5.458)%
  --(8.952,5.467)--(9.095,5.475)--(9.238,5.481)--(9.381,5.485)--(9.524,5.488)--(9.667,5.488)%
  --(9.810,5.487)--(9.954,5.483)--(10.097,5.478)--(10.240,5.470)--(10.383,5.460)--(10.526,5.447)%
  --(10.669,5.432)--(10.812,5.415)--(10.955,5.396)--(11.099,5.374)--(11.242,5.350)--(11.385,5.324)%
  --(11.528,5.296)--(11.671,5.266);
\gpcolor{color=gp lt color border}
\node[gp node right,font={\fontsize{15.0pt}{18.0pt}\selectfont}] at (5.748,6.028) {$N^5$};
\gpcolor{rgb color={0.941,0.894,0.259}}
\gpsetdashtype{gp dt 5}
\draw[gp path] (3.636,6.028)--(4.920,6.028);
\draw[gp path] (3.084,4.729)--(3.227,4.735)--(3.370,4.741)--(3.513,4.747)--(3.656,4.754)%
  --(3.800,4.761)--(3.943,4.769)--(4.086,4.777)--(4.229,4.785)--(4.372,4.794)--(4.515,4.802)%
  --(4.658,4.812)--(4.801,4.821)--(4.945,4.831)--(5.088,4.840)--(5.231,4.851)--(5.374,4.861)%
  --(5.517,4.871)--(5.660,4.882)--(5.803,4.893)--(5.946,4.904)--(6.089,4.914)--(6.233,4.925)%
  --(6.376,4.936)--(6.519,4.947)--(6.662,4.958)--(6.805,4.968)--(6.948,4.979)--(7.091,4.989)%
  --(7.234,4.999)--(7.377,5.008)--(7.521,5.017)--(7.664,5.026)--(7.807,5.034)--(7.950,5.041)%
  --(8.093,5.048)--(8.236,5.055)--(8.379,5.060)--(8.522,5.065)--(8.666,5.069)--(8.809,5.072)%
  --(8.952,5.074)--(9.095,5.076)--(9.238,5.076)--(9.381,5.075)--(9.524,5.073)--(9.667,5.070)%
  --(9.810,5.066)--(9.954,5.061)--(10.097,5.055)--(10.240,5.048)--(10.383,5.039)--(10.526,5.030)%
  --(10.669,5.020)--(10.812,5.008)--(10.955,4.996)--(11.099,4.983)--(11.242,4.969)--(11.385,4.955)%
  --(11.528,4.940)--(11.671,4.924);
\gpcolor{color=gp lt color border}
\gpsetdashtype{gp dt solid}
\gpsetlinewidth{1.00}
\draw[gp path] (3.084,8.287)--(3.084,1.478)--(11.671,1.478)--(11.671,8.287)--cycle;
%% coordinates of the plot area
\gpdefrectangularnode{gp plot 1}{\pgfpoint{3.084cm}{1.478cm}}{\pgfpoint{11.671cm}{8.287cm}}
\end{tikzpicture}
}
%% gnuplot variables